\documentclass[acmsmall,nonacm,screen]{acmart}

\AtBeginDocument{%
  }

\usepackage{array}
\usepackage{amsmath}
\usepackage{booktabs}
\usepackage{xcolor}
\usepackage{ulem}
\usepackage{multirow, lineno, pifont}
\usepackage{colortbl}
\usepackage{tabularx}
\usepackage{cleveref}
\usepackage{ragged2e}
\usepackage[ruled, vlined, linesnumbered]{algorithm2e}

\usepackage{hyperref}

\usepackage[edges]{forest}
\usepackage{graphicx}

\usepackage{tikz}
\usepackage[disable]{todonotes}
\usetikzlibrary{fadings}
\usetikzlibrary{mindmap,trees}
\usetikzlibrary{arrows,automata,shapes,positioning,shadows,trees}
\usetikzlibrary{shapes,snakes,shadows}
\usetikzlibrary{shapes.arrows}
\usetikzlibrary{calc,shapes, positioning}
\usetikzlibrary{decorations.text}
\usetikzlibrary{matrix,chains,positioning,decorations.pathreplacing,arrows}
\usetikzlibrary{bayesnet}
\usepackage[edges]{forest}
\usetikzlibrary{shadows.blur}
\usetikzlibrary{shapes.geometric}

\usepackage{pgfplots}
\usepackage{pgfplotstable}
\usepackage{pgfmath,pgffor}
\usepgfplotslibrary{colorbrewer}
\pgfplotsset{compat = 1.14, cycle list/Set1-8}
\usetikzlibrary{pgfplots.statistics, pgfplots.colorbrewer}
\pgfplotsset{compat=1.8}

\tikzstyle{edge}=[-latex',draw=black!90,shorten <=1pt,shorten >=1pt]
\tikzstyle{redge}=[latex'-,draw=black!90,shorten <=1pt,shorten >=1pt]
\tikzstyle{dedge}=[latex'-latex',draw=black!90,shorten <=1pt,shorten >=1pt]

\tikzstyle{block}=[draw, text width=5em,align=center,shape=rectangle, rounded corners, , align=center]
\tikzstyle{nobox}=[align=center]
\definecolor{emb}{RGB}{209,228,252}
\definecolor{hidden-blue}{RGB}{194,232,247}
\definecolor{hidden-orange}{RGB}{224,224,224}
\definecolor{hidden-yellow}{RGB}{242,244,193}
\definecolor{output-purple}{RGB}{219,203,231}
\definecolor{output-green}{RGB}{204,231,207}
\definecolor{output-black}{RGB}{0,0,0}
\definecolor{output-white}{RGB}{255,255,255}
\definecolor{myorange}{RGB}{255,208,153}
\definecolor{mygreen}{RGB}{166,207,152}

\definecolor{hiddendraw}{RGB}{85,124,85}

\tikzstyle{emb-purple}=[
    rectangle,
    draw=output-purple!50!purple,
    fill=output-purple,
    text opacity=1,
    minimum height=1.5em,
    minimum width=1.5em,
    inner sep=0pt,
    align=center,
    fill opacity=.5,
    ]

\tikzstyle{emb-blue}=[
    rectangle,
    draw=emb!50!blue,
    fill=emb,
    text opacity=1,
    minimum height=1.5em,
    minimum width=1.5em,
    inner sep=0pt,
    align=center,
    fill opacity=.5,
]

\definecolor{colortwo}{rgb}{0.4,0.77,0.17}
\definecolor{colorthree}{rgb}{0.01,0.51,0.93}

\newenvironment{newcontentblock}{\begingroup}{\endgroup}

\definecolor{ratinglow}{HTML}{B56B4B}
\definecolor{ratingmedium}{HTML}{B38B36}
\definecolor{ratinghigh}{HTML}{36877E}
\definecolor{ratingveryhigh}{HTML}{386A98}
\definecolor{ratinglowbg}{HTML}{FAF4F0}
\definecolor{ratingmediumbg}{HTML}{FAF7EE}
\definecolor{ratinghighbg}{HTML}{F2F7F5}
\definecolor{ratingveryhighbg}{HTML}{F1F5FA}
\colorlet{ratinglowmediumbg}{ratinglowbg!50!ratingmediumbg}
\colorlet{ratingmediumhighbg}{ratingmediumbg!50!ratinghighbg}
\definecolor{ratingheader}{HTML}{F6F7F8}
\definecolor{ratingrule}{HTML}{A9B0B5}
\definecolor{ratingcite}{HTML}{536B80}
\newcommand{\ratingsymbol}[1]{%
  \tikz[baseline=-0.42ex,x=1pt,y=1pt,scale=0.70,rounded corners=0pt]{%
    \ifcsname ratingshape#1\endcsname\csname ratingshape#1\endcsname\fi}}

\usetikzlibrary{backgrounds}
\newcommand{\decoderbackgroundtop}{3.5pt}
\newcommand{\decoderbackgroundbottom}{1.5pt}
\newcommand{\decodercontentheight}{37pt}
\newcommand{\decodercell}[2]{%
  \tikz[baseline=(card.north)]{\node[rounded corners=4pt,
    inner xsep=3pt,inner ysep=3pt,
    text width=\dimexpr\linewidth-6pt\relax,align=left] (card)
    {\parbox[t][\decodercontentheight][t]{\dimexpr\linewidth-6pt\relax}{\RaggedRight #2}};
    \begin{scope}[on background layer]
      \path[overlay,fill=#1,draw=none,rounded corners=4pt]
        ([yshift=\decoderbackgroundtop]card.north west) rectangle ([yshift=\decoderbackgroundbottom]card.south east);
    \end{scope}}}
\newcommand{\ratingcell}[4]{%
  \decodercell{#1}{\ratingsymbol{#2}\enspace\textbf{#3}%
    \if\relax\detokenize{#4}\relax\else\par\vspace{3pt}#4\fi}}
\newcommand{\ratingkey}[3]{%
  \tikz[baseline=(key.base)]{\node[rounded corners=3pt,fill=#1,
    inner xsep=7pt,inner ysep=4pt] (key)
    {\ratingsymbol{#2}\enspace\textbf{#3}};}}
  
\NewDocumentCommand{\delbyzy}{m O{ZY}}{
    \sout{#1}
    \textsuperscript{\color{purple}[#2]}
}

\newcommand{\Clifford}{\mathcal{C}}
\newcommand{\Pauli}{\mathcal{P}}

\newcommand{\nc}{\newcommand}
\nc{\rnc}{\renewcommand}
\nc{\lbar}[1]{\overline{#1}}
\nc{\bra}[1]{\langle#1|}
\nc{\ket}[1]{|#1\rangle}
\nc{\ketbra}[2]{|#1\rangle\!\langle#2|}
\nc{\braket}[2]{\langle#1|#2\rangle}

\begin{document}

\title{Quantum Compiler Design for Fault-Tolerant Quantum Computing}

\author{Chenghong Zhu}
\authornote{Co-first Authors.}
\affiliation{%
  \institution{QudeLeap Research}
  \city{Shanghai}
  \country{China}}
\affiliation{%
  \institution{The Hong Kong University of Science and Technology (Guangzhou)}
  \city{Guangzhou}
  \country{China}}

\author{Jiahan Chen}
\authornotemark[1]
\affiliation{%
\institution{QudeLeap Research}
  \city{Shanghai}
  \country{China}}
\affiliation{\institution{The Hong Kong University of Science and Technology (Guangzhou)}
  \city{Guangzhou}
  \country{China}
  }

\author{Keming He}
\affiliation{%
\institution{QudeLeap Research}
  \city{Shanghai}
  \country{China}}
\affiliation{%
  \institution{The Hong Kong University of Science and Technology (Guangzhou)}
  \city{Guangzhou}
  \country{China}
  }

\author{Hongshun Yao}
\affiliation{%
\institution{QudeLeap Research}
  \city{Shanghai}
  \country{China}}
\affiliation{%
  \institution{The Hong Kong University of Science and Technology (Guangzhou)}
  \city{Guangzhou}
  \country{China}
  }
\author{Zhaohui Yang}
\affiliation{%
  \institution{The Hong Kong University of Science and Technology}
  \city{Hong Kong}
  \country{Hong Kong}}
\author{Jin-Guo Liu}
\affiliation{%
  \institution{The Hong Kong University of Science and Technology (Guangzhou)}
  \city{Guangzhou}
  \country{China}
  }
\author{Anbang Wu}
\affiliation{%
  \institution{Shanghai Jiao Tong University}
  \city{Shanghai}
  \country{China}}
\author{Xiaotong Ni}
\affiliation{%
\institution{Hangzhou MatriQ Computing Co., Ltd}
  \city{Hangzhou}
  \country{China}
}

\author{Xingsheng Luan}
\affiliation{%
\institution{Atomqubic Quantum Technology Co., Ltd}
  \city{Shanghai}
  \country{China}}
\affiliation{%
  \institution{State Key Laboratory of Quantum Optics Technologies and Devices, and Institute of Opto-Electronics, Shanxi University}
  \city{Taiyuan}
  \country{China}
  }
\author{Zhuo Fu}
\affiliation{%
\institution{CAS Cold Atom Technology (Wuhan) Co., Ltd.}
  \city{Wuhan}
  \country{China}
}
  
\author{Shenggen Zheng}
\affiliation{%
  \institution{Quantum Science Center of Guangdong-Hong Kong-Macao Greater Bay Area}
  \city{Shenzhen}
  \country{China}}
\author{Xin Wang}
\authornote{Corresponding Author: \href{mailto:felixxinwang@hkust-gz.edu.cn}{felixxinwang@hkust-gz.edu.cn}.}
\affiliation{%
  \institution{The Hong Kong University of Science and Technology (Guangzhou)}
  \city{Guangzhou}
  \country{China}}
\email{felixxinwang@hkust-gz.edu.cn}

\begin{abstract}

Scalable quantum computation is expected to rely on fault-tolerant quantum computation (FTQC), in which quantum error correction (QEC) suppresses physical errors sufficiently to support reliable logical operations. This requires quantum compilation to move beyond general-purpose circuit optimization toward encoding-aware and protocol-structured compilation across the full stack of fault-tolerant quantum computers. Beyond circuit synthesis and hardware mapping, an FTQC compiler must lower algorithm-level operations into the logical gate set supported by the chosen code, coordinate encoded data and ancilla resources, realize logical operations together with repeated syndrome extraction under hardware constraints, and provide the resulting measurement stream to real-time decoding and frame-based feedback for continuous error correction.

This survey presents a full-stack view of compiler design for QEC-protected quantum computation. We organize existing work into three interacting layers: logical-level QEC compilation, physical-level QEC realization, and decoder runtime integration. At the logical level, we review surface-code lattice-surgery compilers, beyond-surface-code code-surgery frameworks including emerging qLDPC approaches, and compilation support for non-Clifford operations such as magic-state distillation and code switching. At the physical level, we survey hardware-aware QEC realization on superconducting, trapped-ion, and neutral-atom platforms. We further examine decoder models, real-time decoding systems, and frame-management mechanisms that close the feedback loop during fault-tolerant execution. Finally, we identify open challenges in cross-layer optimization, qLDPC compilation, compiler-decoder co-design, runtime adaptivity, and the development of integrated and benchmarkable FTQC compilation stacks. An actively maintained paper list is available at: \href{https://github.com/chenghongz/QEC-compiler-design}{github.com/chenghongz/QEC-compiler-design}.

\end{abstract}





\authorsaddresses{}

\maketitle

\tableofcontents

\section{Introduction}

Quantum computing has emerged as a new computational paradigm and a specialized accelerator for problems in cryptography, quantum chemistry, and machine learning~\cite{shor1994algorithms,aspuru2005simulated,biamonte2017quantum}. This promise is grounded in quantum algorithms that outperform the best-known classical algorithms for factoring and discrete logarithms, and achieve a provable quadratic query advantage for unstructured search~\cite{shor1994algorithms,grover1996fast}. Experimentally, random-circuit-sampling and boson-sampling experiments have provided early demonstrations of quantum advantage~\cite{arute2019quantum,zhong2020quantum}, while recent certified-randomness experiments have shown that quantum processors can generate verifiable randomness beyond classical simulation assumptions~\cite{liu2025certified}.

These advances have stimulated rapid progress in quantum hardware. Several leading platforms have emerged, including superconducting circuits, trapped ions, and neutral atoms~\cite{bruzewicz2019trapped,henriet2020quantum,kjaergaard2020superconducting}. Improvements in device fabrication, control, and trapping have pushed these platforms to increasingly large programmable system sizes. Superconducting circuits have crossed the thousand-qubit scale in announced processors, exemplified by IBM's 1121-qubit Condor chip~\cite{castelvecchi2023ibm}. Trapped-ion systems have reached nearly one hundred physical qubits while preserving high-fidelity, all-to-all control in QCCD architectures~\cite{ransford2025helios98qubittrappedionquantum}. Neutral-atom tweezer arrays have demonstrated registers with more than 6100 highly coherent atomic qubits across nearly 12000 trapping sites~\cite{Manetsch_2025}. At the level of raw trapped-atom count, a metasurface-generated tweezer array has trapped 11,000 individual atoms, the largest reported count to date~\cite{wang2026trapping11000}. Yet useful large-scale quantum computation remains limited by noise. Decoherence, imperfect gates, and faulty measurements all accumulate with circuit size. Quantum error correction (QEC)~\cite{terhal2015quantum} is therefore the central mechanism for scaling quantum computers from noisy devices to fault-tolerant machines.

The field is now entering a stage where QEC is no longer only a theoretical abstraction. Superconducting processors have demonstrated below-threshold surface-code memories with integrated real-time decoding~\cite{acharya2024quantum}. Reconfigurable neutral-atom arrays have demonstrated logical processors and mechanisms for universal fault-tolerant operation~\cite{bluvstein2024logical,bluvstein2026faulttolerant,rines2025demonstration}. Atom Computing has performed up to 90 toric-code syndrome-extraction cycles with mid-circuit measurement and replacement of lost atoms~\cite{atomcomputing2026toric}. Trapped-ion processors have also demonstrated real-time fault-tolerant QEC and repeated logical-qubit protection, showing that encoded operations can outperform corresponding physical baselines~\cite{reichardt2024demonstration,tham2026breakeven,dasu2026computing}. Beyond individual experimental demonstrations, recent fault-tolerant processor blueprints have begun to integrate QEC codes, logical operations, decoding, and hardware architecture into end-to-end system designs. Representative examples include IBM's bivariate-bicycle-code architecture, trapped-ion proposals such as the Walking Cat architecture, and other emerging high-rate qLDPC processor designs~\cite{yoder2025tour,tripier2026faulttolerantquantumcomputingtrapped,bhardwaj2026high,bravyi2024high,haner2026computing256bitellipticcurve}. Together, these experimental and architectural developments show that practical QEC is not determined by code parameters alone. The effective performance of a fault-tolerant system depends on how algorithmic operations are lowered into logical fault-tolerant procedures, how these procedures and syndrome-extraction circuits are mapped and scheduled on hardware, and whether the resulting measurement stream can be decoded fast enough to sustain continuous error correction. These coupled requirements make QEC compilation an inherently full-stack problem.

\begin{figure}[h]
    \centering
    \includegraphics[width=1\linewidth]{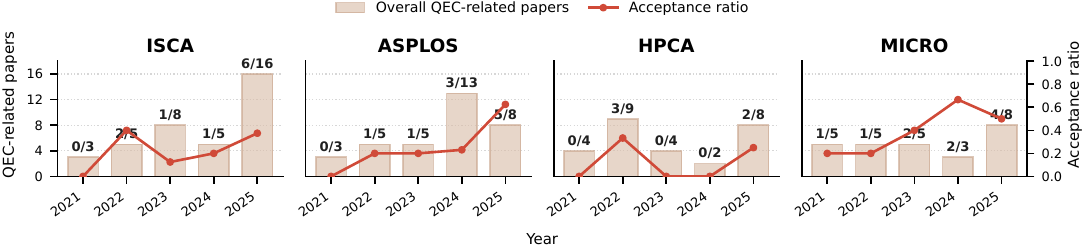}
    \caption{Trends in QEC related publications at premier computer architecture conferences (ISCA, ASPLOS, HPCA, and MICRO) from 2021 to 2025.}
    \label{fig:qec_accepted_paper}
\end{figure}

This systems view also changes the compiler abstraction. In NISQ compilation, a quantum program is typically lowered to a native gate sequence and optimized for device-level costs such as depth and routing overhead~\cite{zhu2025quantumcompilerdesignqubit}. In QEC-protected computation, the compiler operates on encoded objects and fault-tolerant procedures rather than individual physical gates. A logical operation may be realized through mechanisms such as lattice surgery or magic-state preparation, while each encoded block is maintained by repeated syndrome extraction. The resulting measurement record becomes part of the classical decoding and feedback loop. Compiler decisions therefore affect both quantum execution and classical control, so QEC-aware compilation has to reason jointly about reliability and resource cost across logical operations, hardware realization, and decoding.

The importance of this compiler perspective is increasingly visible in recent publication trends. Quantum software and architecture have become active research areas, and premier computer architecture venues now regularly publish work on QEC-aware compilation and fault-tolerant system design. As shown in Fig.~\ref{fig:qec_accepted_paper}, QEC-related papers at ISCA, ASPLOS, HPCA, and MICRO have grown from sparse acceptances in 2021 to substantially higher counts by 2025. The corresponding acceptance-ratio curves also rise in the later years, suggesting growing receptivity to QEC-related contributions. These trends indicate that QEC has become a recurring theme in computer architecture, reflecting the broader movement from noisy intermediate-scale demonstrations toward fault-tolerant quantum computing.

\begin{figure}[t]
    \centering
    \includegraphics[width=\linewidth]{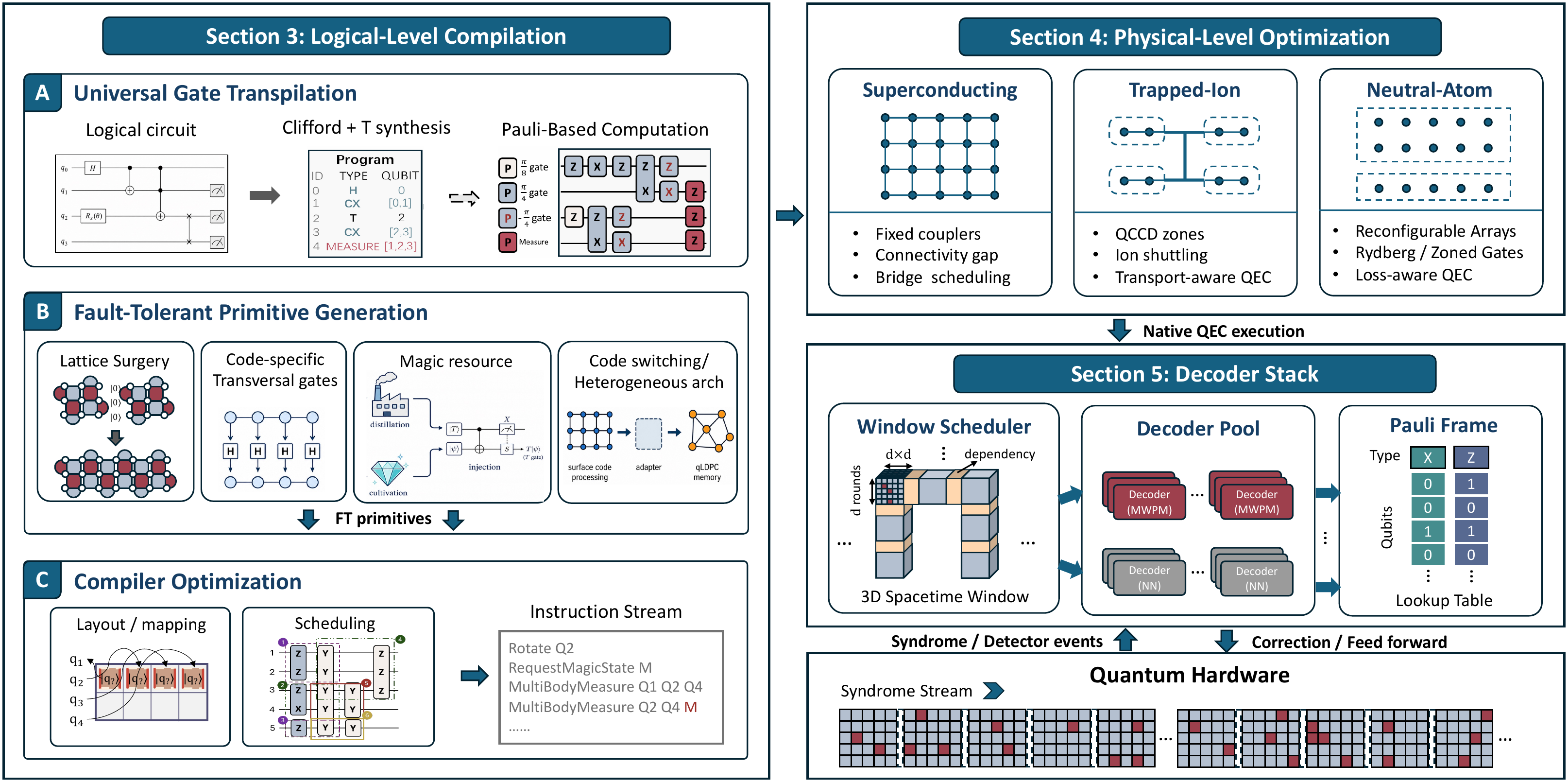}
    \caption{Overall picture of the QEC-protected quantum computing stack: logical-level compilation transpiles and lowers a logical circuit into fault-tolerant primitives and an optimized instruction stream; physical-level optimization compiles it to native operations across superconducting, trapped-ion, and neutral-atom platforms; and the decoder stack processes the resulting syndrome stream to feed corrections back via the Pauli frame.}
    \label{fig:main_figure}
\end{figure}

These trends motivate a stack-oriented view of QEC-aware compilation. At the logical level, the compiler selects fault-tolerant procedures for encoded operations, including mechanisms such as lattice surgery and magic-state preparation. At the physical level, these procedures are realized under platform-specific constraints, where connectivity, movement, native gates, and syndrome-extraction timing determine the execution cost. At the runtime level, the resulting measurement stream is decoded and incorporated into frame updates fast enough to support continuous error correction. Fig.~\ref{fig:main_figure} summarizes this full-stack view. The rest of this survey follows the same organization, reviewing logical-level QEC optimization, physical-level QEC implementation, and decoder/runtime support.

\textbf{Contributions.} In summary, the contributions of this survey paper are as follows:
\begin{itemize}
    \item We present a full-stack taxonomy of QEC compiler design, connecting logical-operation synthesis, physical-level syndrome-extraction and hardware realization, and decoder/runtime support.
    \item We survey logical-level QEC compilation techniques, including surface-code lattice-surgery compilers, beyond-surface-code and qLDPC code-surgery frameworks, magic-state resource scheduling, code switching, and early fault-tolerant execution models.
    \item We review physical-level QEC implementation across superconducting circuits, trapped-ion QCCD systems, and neutral-atom arrays, emphasizing the platform-specific constraints that shape compiler design.
    \item We discuss decoder-aware compilation and real-time runtime integration, including detector error models, representative decoder algorithms, Pauli and Clifford frames, window decoding, and noise- or side-information-aware decoding.
    \item We identify open problems in end-to-end FTQC compilation, qLDPC compiler abstractions, compiler-decoder co-design, runtime adaptivity, and benchmark standardization.
\end{itemize}

\textbf{Survey Scope.}
This survey covers literature available through June 2026, with no lower publication-year cutoff so that foundational methods can be retained. We consider peer-reviewed conference and journal papers, together with technically substantive preprints, from quantum computing and information, computer architecture and systems, programming languages and compilers, and electronic design automation. Papers devoted solely to QEC-code construction, asymptotic or threshold analysis, decoder theory, or experimental demonstrations are excluded from the compiler taxonomy. Selected examples are cited only when they establish code properties, hardware constraints, or runtime requirements relevant to compilation.

\textbf{Organization.} The remainder of this survey is organized as follows. Section~\ref{sec:pre} introduces the background on quantum computation, QEC codes, fault-tolerant gate sets, quantum compilation, and representative hardware platforms. Section~\ref{sec:logical_level_qec_optimization} reviews logical-level QEC optimization, including lattice surgery, magic-state distillation, early fault-tolerant compilation, and code switching. Section~\ref{sec:physical_level_qec_implementation} discusses physical-level QEC implementation on superconducting circuits, trapped ions, and neutral-atom arrays, with an emphasis on hardware-aware mapping and scheduling. Section~\ref{sec:decoder} surveys decoder models, decoder implementations, Pauli-frame tracking, and window decoding. Finally, Section~\ref{sec:conclusion} concludes the survey and discusses the need for more integrated end-to-end FTQC compilation.

\section{Background}
\label{sec:pre}

\subsection{Quantum Computation}\label{subsec:quantumcomputation}
\subsubsection{\textbf{Qubit.}} In quantum computing, the basic unit of quantum information is the quantum bit or qubit, whose state is described and evolves in accordance with the framework of quantum mechanics~\cite{nielsen2010quantum}. A single-qubit pure state is described by a unit vector in the Hilbert space $\mathcal{H}(\mathbb{C}^2)$, which is commonly written in Dirac notation $\ket{\psi} = \alpha\ket{0} + \beta\ket{1}$, with $\ket{0} = (1,0)^T$, $\ket{1} =(0,1)^T$ and $\alpha,\beta\in \mathbb{C}$ subject to $|\alpha|^2 + |\beta|^2 = 1$. The complex conjugate of $\ket{\psi}$ is denoted as $\bra{\psi} = \ket{\psi}^\dagger$. The $N$-qubit Hilbert space is formed by the tensor product ``$\otimes$'' of $N$ single-qubit state spaces with dimension $d=2^N$. For an $N$-qubit state $\ket{\psi}\in \mathcal{H}(\mathbb{C}^{d})$, it can be written as $\ket{\psi} = \sum_{i=0}^{d-1} a_i \ket{i}$, with $\sum_{i=0}^{d-1} \lvert a_i\rvert^2 = 1$ and $\{\ket{i}\}_{i=0}^{d-1}$ form a set of orthogonal basis vectors, each $\ket{i}$ having 1 in the $i$-th element and 0 elsewhere.

\subsubsection{\textbf{Quantum Operation.}} In the representation of quantum circuit model~\cite{nielsen2010quantum}, quantum computing performs by altering the states of qubits through quantum operations (or gates). Similar to classical computing, QC systems generally support a universal gate set usually composed of basic single-qubit rotations and one or more two-qubit gates. These gates are capable of acting on one or two qubits simultaneously and expressing any quantum program~\cite{Barenco_1995}. Common single-qubit rotation gates include $R_x(\theta)=e^{-i\frac{\theta}{2}X}$, $R_y(\theta)=e^{-i\frac{\theta}{2}Y}$, $R_z(\theta)=e^{-i\frac{\theta}{2}Z}$, which are in the matrix exponential form of Pauli matrices,
\begin{equation}
    X = \begin{pmatrix}
        0 & 1 \\ 1 & 0
    \end{pmatrix},\quad
    Y = \begin{pmatrix}
        0 & -i \\ i & 0 \\
    \end{pmatrix},\quad
    Z = \begin{pmatrix}
        1 & 0 \\ 0 & -1 \\
    \end{pmatrix}.
\end{equation}
Common two-qubit gates in superconducting devices include controlled-X gate $\text{CNOT} = \ketbra{0}{0} \otimes I + \ketbra{1}{1} \otimes X$  and controlled-Z gate $\text{CZ}= \ketbra{0}{0} \otimes I + \ketbra{1}{1} \otimes Z$, which can generate maximal entanglement among qubits. Parametrized $XX(\theta)=e^{-i\frac{\theta}{2} X\otimes X}$ gates and CZ gate are the most frequently used two-qubit gates in trapped-ion devices and neutral atom devices, respectively. SWAP gate swaps the states of two qubits and can be decomposed into three CNOT gates.  Quantum programs are typically expressed using high-level languages with gate sets that are not directly compatible with the requirements of QEC codes, such as the surface code, which often compile logical algorithms into a Clifford+$T$ gate set. This incompatibility necessitates the use of gate synthesis techniques such as the Solovay-Kitaev algorithm~\cite{dawson2005solovaykitaevalgorithm} or more advanced methods~\cite{ross2016optimalancillafreecliffordtapproximation} to translate arbitrary gates into sequences composed of Clifford and $T$ gates, while minimizing the number of costly $T$ gates.

\subsection{Quantum Error Correction} 

Below, we briefly introduce the basic concepts of quantum error-correction codes.

\subsubsection{\textbf{Stabilizer Code}} A stabilizer code is a quantum error-correcting code that defines a protected logical subspace via a set of commuting Pauli constraints. Let $\mathcal{P}_n$ denote the $n$-qubit Pauli group. A stabilizer code is specified by an Abelian subgroup $\mathcal{S} \subset \mathcal{P}_n$ with $-I \notin \mathcal{S}$, called the stabilizer group. The codespace $\mathcal{C}$ is the simultaneous $+1$ eigenspace of all stabilizers,
$\mathcal{C} \;=\; \left\{\,\ket{\psi} : s\ket{\psi}=\ket{\psi},\ \forall s\in\mathcal{S} \,\right\}$. The normalizer $\mathcal{N}(\mathcal{S})$ is the set of Pauli operators that commute with every element of $\mathcal{S}$. Logical Pauli operators correspond to elements of $\mathcal{N}(\mathcal{S})\setminus\mathcal{S}$ that act nontrivially on $\mathcal{C}$. If $\mathcal{S}$ has $n-k$ independent generators, the code encodes $k$ logical qubits using $n$ physical qubits and is denoted $[[n,k,d]]$, where the distance $d$ is the minimum weight of any nontrivial logical operator. Operationally, larger distance typically yields better logical error suppression at the expense of higher physical resource requirements, reflecting a basic reliability resource tradeoff in fault-tolerant quantum computing.

Errors of a stabilizer code are modeled as Pauli operators $E\in \mathcal{P}_n$. Measuring the stabilizer generators yields a binary syndrome indicating which checks anticommute with $E$. A decoder maps the observed syndrome to a recovery operator $R$. Decoding succeeds if $RE$ acts trivially on the logical subspace up to a stabilizer; otherwise $RE\in \mathcal{N}(\mathcal{S}) \setminus\mathcal{S}$ and a logical error has occurred. Many stabilizer codes such as surface code, are degenerate, where distinct Pauli errors that differ by a stabilizer produce the same syndrome. Consequently, decoding amounts to selecting a likely error coset rather than identifying a unique error.

\subsubsection{\textbf{Universal Gate Set}}

A universal gate set is a collection of quantum operations from which any unitary on a finite number of qubits can be approximated to arbitrary accuracy. In fault-tolerant quantum computing, universality is typically obtained by combining a set of efficiently implementable Clifford operations with at least one non-Clifford operation. The Clifford group is generated by the single-qubit Hadamard gate $H$, phase $S$, and CNOT gate. Clifford circuits map Pauli operators to Pauli operators under conjugation, which makes them well matched to stabilizer codes and enables efficient classical tracking of Pauli-frame updates.

Clifford gates alone are not universal, so an additional non-Clifford resource is required. A common choice is the $T$ gate, $T = \text{diag}(1, e^{i\pi/4})$, which together with Clifford gates forms the widely used Clifford + $T$ gate set. Clifford + $T$ is a standard baseline in QEC and architectural studies because it yields a simple resource model in terms of $T$-count, $T$-depth, and the magic state factory throughput, which connects algorithm-level compilation to fault-tolerant resource estimates. Other universal choices are also used, such as Clifford combined with CCZ or CCX gates, or Clifford combined with small-angle $R_z$ rotations synthesized from distilled resource states.

In quantum error correction, fault-tolerant logical gates can be implemented using several mechanisms. A particularly simple approach is a transversal gate, where the logical gate is realized by applying the corresponding physical gate to each qubit in a qubitwise manner. Many stabilizer codes admit transversal implementations for various logical gates, and particularly CSS codes can support transversal CNOT gate. However, the Eastin--Knill theorem implies that no quantum error correcting code can realize a universal set of logical gates using only transversal unitary operations~\cite{eastin2009restrictions}. As a consequence, universality requires additional non-transversal resources and procedures. In Clifford + $T$ setting, non-Clifford logical gates are commonly implemented via state injection and gate teleportation. These protocols consume prepared resource states and apply a fixed sequence of Clifford operations and measurements, followed by a classically conditioned Pauli or Clifford correction. Typical examples include the $\ket{Y} = (\ket{0} + i\ket{1})/\sqrt{2}$ for implementing phase gate and $\ket{T} = T\ket{+} = (\ket{0} + e^{i\pi /4}\ket{1})/\sqrt{2}$ for implementing $T$ gate. This introduces an additional requirement to produce high-fidelity resource states, motivating magic state distillation as one of the major overhead components in fault-tolerant architectures. Lattice surgery and code switching give further ways to build logical gates, and we return to both in Section~\ref{sec:logical_level_qec_optimization}.

%
%
%
\definecolor{xred}{RGB}{214,39,40}     
\definecolor{zblue}{RGB}{31,119,180}   
\definecolor{ccgreen}{RGB}{44,160,44}  
\colorlet{Xcol}{xred!60}
\colorlet{Zcol}{zblue!50}
\colorlet{Gcol}{ccgreen!55}
\providecommand{\Rcc}{0.5}                 
\providecommand{\hexcc}[3]{
  \fill[#3,draw=black,line width=0.4pt]
    ($(#1,#2)+(0:\Rcc)$)   -- ($(#1,#2)+(60:\Rcc)$)  -- ($(#1,#2)+(120:\Rcc)$) --
    ($(#1,#2)+(180:\Rcc)$) -- ($(#1,#2)+(240:\Rcc)$) -- ($(#1,#2)+(300:\Rcc)$) -- cycle;}

\begin{figure}[t]
  \centering
  \providecommand{\panelht}{2.7cm}
  \setlength{\tabcolsep}{0.45cm}%
  \begin{tabular}{@{}ccc@{}}
  \resizebox{!}{\panelht}{%
  \begin{tikzpicture}[scale=0.70]
    \foreach \x in {0,1,2,3}{
      \foreach \y in {0,1,2,3}{
        \pgfmathtruncatemacro{\s}{mod(\x+\y,2)}
        \ifnum\s=0 \fill[Zcol] (\x,\y) rectangle ++(1,1);
        \else      \fill[Xcol] (\x,\y) rectangle ++(1,1);\fi
      }
    }
    \fill[Xcol,draw=black,line width=0.4pt] (1,0) arc (0:-180:0.5)  -- cycle;
    \fill[Xcol,draw=black,line width=0.4pt] (3,0) arc (0:-180:0.5)  -- cycle;
    \fill[Xcol,draw=black,line width=0.4pt] (1,4) arc (180:0:0.5)   -- cycle;
    \fill[Xcol,draw=black,line width=0.4pt] (3,4) arc (180:0:0.5)   -- cycle;
    \fill[Zcol,draw=black,line width=0.4pt] (0,2) arc (90:270:0.5)  -- cycle;
    \fill[Zcol,draw=black,line width=0.4pt] (0,4) arc (90:270:0.5)  -- cycle;
    \fill[Zcol,draw=black,line width=0.4pt] (4,1) arc (90:-90:0.5)  -- cycle;
    \fill[Zcol,draw=black,line width=0.4pt] (4,3) arc (90:-90:0.5)  -- cycle;
    \draw[gray!55,line width=0.4pt] (0,0) grid (4,4);
    \foreach \i in {0,...,4}\foreach \j in {0,...,4}{\fill (\i,\j) circle (2.6pt);}
    \node at (0.5,0.5){\large $Z$};
    \node at (1.5,0.5){\large $X$};
  \end{tikzpicture}}
  &
  \resizebox{!}{\panelht}{%
  \begin{tikzpicture}[scale=1.35]
    \coordinate (h0) at (0,0);          \coordinate (h1) at (0.75,0.433);
    \coordinate (h2) at (0,0.866);      \coordinate (h3) at (-0.75,0.433);
    \coordinate (h4) at (-0.75,-0.433); \coordinate (h5) at (0,-0.866);
    \coordinate (h6) at (0.75,-0.433);
    \hexcc{0}{0}{Xcol}
    \hexcc{0.75}{0.433}{Gcol}   \hexcc{0}{0.866}{Zcol}
    \hexcc{-0.75}{0.433}{Gcol}  \hexcc{-0.75}{-0.433}{Zcol}
    \hexcc{0}{-0.866}{Gcol}     \hexcc{0.75}{-0.433}{Zcol}
    \foreach \c in {h0,h1,h2,h3,h4,h5,h6}{
      \foreach \a in {0,60,120,180,240,300}{\fill ($(\c)+(\a:\Rcc)$) circle (1.3pt);}}
  \end{tikzpicture}}
  &
  \resizebox{!}{\panelht}{%
  \begin{tikzpicture}[scale=0.82,
     qubit/.style={circle,draw=black,fill=white,minimum size=13pt,inner sep=0pt,line width=0.7pt},
     xcheck/.style={rectangle,draw=black,fill=xred!75,minimum size=12pt,inner sep=0pt,line width=0.7pt},
     zcheck/.style={rectangle,draw=black,fill=zblue!70,minimum size=12pt,inner sep=0pt,line width=0.7pt}]
    \node[qubit] (q0) at (0,0){};    \node[qubit] (q1) at (0.85,0){};
    \node[qubit] (q2) at (1.70,0){}; \node at (2.50,0){$\cdots$};
    \node[qubit] (q3) at (3.30,0){}; \node[qubit] (q4) at (4.15,0){};
    \node[qubit] (q5) at (5.00,0){};
    \node[xcheck] (x0) at (0.65,1.9){}; \node[xcheck] (x1) at (2.50,1.9){}; \node[xcheck] (x2) at (4.35,1.9){};
    \node[zcheck] (z0) at (0.65,-1.9){}; \node[zcheck] (z1) at (2.50,-1.9){}; \node[zcheck] (z2) at (4.35,-1.9){};
    \draw (x0)--(q0); \draw (x0)--(q1); \draw (x0)--(q3);
    \draw (x1)--(q1); \draw (x1)--(q2); \draw (x1)--(q4);
    \draw (x2)--(q2); \draw (x2)--(q4); \draw (x2)--(q5);
    \draw (z0)--(q0); \draw (z0)--(q2); \draw (z0)--(q4);
    \draw (z1)--(q1); \draw (z1)--(q3); \draw (z1)--(q5);
    \draw (z2)--(q2); \draw (z2)--(q3); \draw (z2)--(q5);
  \end{tikzpicture}}
  \\[3pt]
  {\small (a) Surface code} & {\small (b) Color code} & {\small (c) qLDPC code} \\
  \end{tabular}

  \caption{Three representative families of quantum error-correcting codes.
  (a)~A distance-5 rotated \emph{surface code}: data qubits (dots) on a 2D lattice
  with weight-4 $X$ (red) and $Z$ (blue) plaquette stabilizers and weight-2
  boundary stabilizers. (b)~A patch of the $6.6.6$ (hexagonal) \emph{color code}:
  a data qubit sits on every vertex of a trivalent lattice whose faces carry the
  standard red/green/blue three-coloring, with each face hosting one $X$- and one
  $Z$-type stabilizer. (c)~A \emph{qLDPC code} shown as a Tanner graph: data qubits
  (circles) connected to sparse $X$- (red) and $Z$-type (blue) checks (squares),
  generally with non-local connectivity.}
  \label{fig:qec_code_families}
\end{figure}
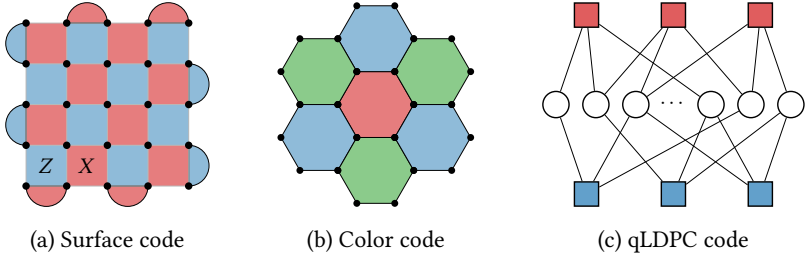

\subsubsection{\textbf{Surface Code}}

Surface codes are among the most extensively studied quantum error-correcting codes for fault-tolerant quantum computation~\cite{dennis2002topological, fowler2012surface}. They are widely used as a reference code family in QEC architecture studies because they combine local stabilizer measurements, compatibility with two-dimensional hardware layouts, and high thresholds under commonly studied circuit-level noise models. Many resource-estimation and fault-tolerant architecture frameworks therefore use surface code patches as the basic unit for representing protected logical qubits.

Surface codes are a family of two-dimensional topological stabilizer codes defined on lattices with local stabilizer checks. In the standard construction, data qubits are placed on the edges of a two-dimensional cellulation, while vertex and plaquette operators define the two types of stabilizer checks. The logical Pauli operators of a surface code are represented by string-like operators connecting appropriate boundaries of the patch. Since a nontrivial logical operator must cross the patch, the code distance is determined by the minimum length of such boundary-connecting operators. This geometric characterization provides a direct relation between the physical layout and the protection strength of the code.

Surface codes are commonly adopted as a baseline because several components of the fault-tolerant stack have been systematically developed for this code family. The regular lattice structure gives a clear physical layout, the stabilizer checks lead to standard syndrome extraction circuits. Under standard phenomenological or circuit-level Pauli noise models, the associated decoding problem can be mapped to a graph-matching problem, enabling the efficient use of minimum-weight perfect matching and related decoders. These features make the surface code a useful reference point for comparing QEC codes, decoders, compilation methods, and architecture-level resource estimates.

Fault-tolerant logical operations in commonly used surface code architectures are implemented through geometric procedures such as code deformation, lattice surgery,  and state injection. Lattice surgery allows logical qubits represented by different patches to interact through joint measurements, providing a structured way to compile logical Clifford operations. In this sense, the surface code is both a code construction and a concrete architectural model for organizing fault-tolerant quantum computation.

Despite these practical advantages, the surface code has a low encoding rate. A single distance $d$ planar surface code patch encodes one logical qubit while using $O(d^2)$ physical qubits. Increasing the distance improves the suppression of logical errors, but it also increases the number of physical qubits and the number of syndrome measurement rounds required for reliable fault-tolerant computation. This tradeoff is a major source of space-time overhead in surface code-based architectures.

\subsubsection{\textbf{Color Codes}}

Color codes are another important family of topological stabilizer codes for fault-tolerant quantum computation~\cite{bombin2006topological}. In two dimensions, color codes are defined on lattices whose faces are three-colorable, meaning that each face can be assigned one of three colors such that adjacent faces have different colors. A standard construction uses a trivalent lattice, and the triangular 6.6.6 color code based on a hexagonal tiling is a commonly studied example. This structure gives color codes local stabilizer measurements and makes them compatible with two-dimensional hardware layouts, similarly to surface codes.

Like other two-dimensional topological stabilizer codes, color codes are constrained by geometric trade-offs between code distance, encoding rate, and the total number of physical qubits. For two-dimensional local stabilizer codes, the Bravyi--Poulin--Terhal bound implies a trade-off of the form $k d^2 = O(n)$~\cite{bravyi2010tradeoffs}. Consequently, two-dimensional color-code families cannot simultaneously achieve constant encoding rate and distance growing as a power of the block length. Increasing the distance therefore still requires a growing number of physical qubits, which limits their asymptotic overhead compared with more general qLDPC code families.

Color codes are nevertheless useful in fault-tolerant quantum computation because of their transversal-gate structure. In particular, certain triangular 2D color codes on suitably chosen lattices admit useful transversal Clifford operations~\cite{landahl2011fault}. Beyond Clifford gates, color-code and related constructions also play an important role in protocols based on magic-state distillation, code switching, and gauge fixing. These protocols allow color-code schemes to exploit different code structures to access or prepare non-Clifford resources with reduced overhead.

\subsubsection{\textbf{qLDPC Codes}}

Two-dimensional topological codes such as surface codes and color codes provide hardware-friendly routes to fault-tolerant quantum computation, but their encoding rate is limited by geometric constraints. Surface-code architectures illustrate this trade-off clearly. A distance-$d$ rotated surface-code patch encodes a single logical qubit using $n=d^2$ data qubits and approximately $n-1$ ancillary qubits for syndrome extraction, resulting in a low encoding rate when all physical qubits are counted. As the target logical error rate becomes more stringent, a larger code distance is required, which quickly increases the physical-qubit count and the space-time cost of fault-tolerant computation. This motivates the study of quantum low-density parity-check (qLDPC) codes as a route toward reducing the spatial overhead of QEC.

A qLDPC code is a family of stabilizer codes in which each stabilizer generator acts on only a bounded number of physical qubits, and each physical qubit participates in only a bounded number of stabilizer checks~\cite{breuckmann2021quantum}. This sparsity condition is the quantum analogue of the low-density parity-check structure in classical coding theory. In the QEC setting, sparsity matters for both the code definition and syndrome extraction. Bounded-weight checks and bounded qubit degree provide a basis for controlling the depth, gate count, and number of fault locations in repeated measurement circuits.

Kasai and collaborators construct regular girth-eight CSS qLDPC codes across several rates and degrees while reconciling orthogonality with favorable Tanner graphs~\cite{kasai2026orthogonality,okada2026highgirth,okada2026twobranch,okada2026ratetwothirds}. Their pair-partition method also makes orthogonality and fixed-instance distance bounds directly verifiable~\cite{okada2026pairpartition}.

To enable low-overhead fault-tolerant quantum computation, a practically useful qLDPC code should satisfy several desirable properties: (1). \textit{Large code distance and high encoding rate:} The code should support a large distance $d$ to suppress logical errors, while maintaining a high encoding rate $r = k/n$, where $k$ is the number of logical qubits. In particular, constant-rate qLDPC codes are attractive because they can reduce the spatial overhead of logical memory and data blocks~\cite{gottesman2013fault}. (2). \textit{Short-depth syndrome measurement circuits:} The stabilizer measurements should be implementable with shallow and low-gate-count circuits, so that the errors introduced during repeated QEC cycles do not grow significantly with the code size. (3). \textit{Low-overhead logical operations and logical qubit addressing:} The architecture should support initialization, readout, and fault-tolerant logical operations on individual or selected logical qubits. This requirement is important for compiling algorithm-level circuits into executable fault-tolerant procedures.

Although qLDPC codes have sparse stabilizer checks, many constructions involve nonlocal connectivity when embedded into a two-dimensional hardware layout. Implementing such checks may require long-range couplers, qubit movement, modular architectures, or layered interaction structures. For platforms with mainly local interactions, such as superconducting qubits, it is therefore useful to identify qLDPC constructions with approximately planar or otherwise hardware-compatible layouts.

Overall, qLDPC codes offer the possibility of reducing the qubit overhead of fault-tolerant quantum computation through better encoding rates and sparse checks. At the same time, practical qLDPC-based architectures must address syndrome measurement scheduling, decoder design, logical qubit addressing, and fault-tolerant logical operations. These issues make qLDPC codes an important target for full-stack QEC compiler design.

\subsubsection{\textbf{Bivariate Bicycle Codes}}

Bivariate bicycle (BB) codes are a practically relevant class of qLDPC codes constructed from polynomial generators, first introduced in~\cite{kovalev2013quantum} and further investigated in recent architecture-oriented studies~\cite{bravyi2024high}. They provide concrete examples of sparse quantum codes with higher encoding rates than surface-code patches, while still admitting structured syndrome measurement circuits. Both the toric code and tensor products of cyclic codes can be viewed as instances of BB codes~\cite{eberhardt2024logical}.

Let $S_\ell$ denote the $\ell \times \ell$ binary cyclic shift matrix, where each row has a single 1 in the position one to the right modulo $\ell$. Then define the operators:
\begin{equation}
x = S_\ell \otimes I_m, \quad y = I_\ell \otimes S_m,
\end{equation}
where $I_\ell$ and $I_m$ are identity matrices of size $\ell$ and $m$ respectively. A BB code is defined by binary matrices $A$ and $B$ expressed as sums of powers of $x$ and $y$:
\begin{equation}
A = A_1 + A_2 + A_3, \quad B = B_1 + B_2 + B_3,
\end{equation}
where each $A_i$ and $B_j$ is a monomial in either $x$ or $y$, and all $A_i$ are distinct among themselves, as are all $B_j$. Note that each row and each column of $A$ and $B$ contains exactly three nonzero entries, and moreover, any $A_i$ commutes with any $B_j$ due to the commutativity of $x$ and $y$. \\
This pair $(A, B)$ defines a CSS qLDPC code with $X$- and $Z$-stabilizer check matrices given by:
\begin{equation}
H^X = [A \mid B], \quad H^Z = [B^T \mid A^T],
\end{equation}
where $[\cdot \mid \cdot]$ denotes horizontal concatenation and $(\cdot)^T$ denotes matrix transposition. These check matrices satisfy the commutativity condition $H^X (H^Z)^T = 0$. The resulting quantum code has block length $n = 2 \ell m$. The dimension $k$ and distance $d$ of the code are given by~\cite{bravyi2024high}:
\begin{equation}
k = 2 \cdot \dim (\ker A \cap \ker B), \quad
d = \min \left\{ \|v\| : v \in \ker(H^X) \setminus \mathrm{rs}(H^Z) \right\},
\end{equation}
where $\ker$ denotes the kernel (null space) and $\mathrm{rs}$ denotes the row space over $\mathbb{F}_2$.

Several BB code instances have been proposed with strong practical potential. In particular, the $[[144,12,12]]$ and $[[288,12,18]]$ codes, known respectively as the gross code and two-Gross code~\cite{yoder2025tour}, offer several-fold reductions in qubit overhead while achieving comparable logical error suppression~\cite{bravyi2024high}. These instances also support structured syndrome measurement circuits, including depth-7 circuits in the constructions studied in~\cite{bravyi2024high}, and can be implemented using a two-layer planar layout. This makes them more compatible with hardware platforms that favor local or layered interactions.

Recent work has also studied fault-tolerant logical operations and compilation strategies for BB codes and related bicycle-code families, including proposals for neutral-atom platforms~\cite{viszlai2023matching}. Therefore, BB codes provide a useful case study for connecting qLDPC code parameters with syndrome measurement design, decoder implementation, and hardware-aware compilation.

\subsubsection{\textbf{Syndrome Measurement Circuit.}}

Syndrome measurement is the basic mechanism by which quantum error correction protects encoded quantum information during computation. Physical errors may occur on data qubits while a quantum state is stored, moved, or operated on. If these errors are not detected and corrected in time, they can accumulate and eventually induce logical failures. Syndrome measurement circuits are therefore used repeatedly to extract information about errors without directly measuring or destroying the encoded logical state. Most fault-tolerant gadgets, including logical memory, logical gate implementation, state preparation, and measurement, rely on such syndrome extraction procedures to keep the encoded state within the correctable regime.

In practice, syndrome extraction is implemented by coupling data qubits to ancilla qubits. The ancilla qubits are initialized, interact with selected data qubits through native gates, and are then measured to produce syndrome bits. A full syndrome measurement round consists of many such check measurements arranged in a schedule. This schedule must respect the fact that different checks may share data qubits, that hardware may only support certain qubit interactions, and that measurements and resets may take non-negligible time. Therefore, even when the stabilizer checks of a code are fixed, there can be many possible circuit-level implementations of the same syndrome extraction task.

The measurement protocol itself has to be fault tolerant. Faults occurring during syndrome extraction should not spread into data errors that are too large for the code to correct. For example, the ordering of two-qubit gates can affect the propagation of hook errors and hence the effective circuit-level distance. Additional ancilla qubits may also be introduced, such as flag ancillas~\cite{chao2018quantum, chamberland2018flag}, to detect dangerous fault propagation. More advanced protocols, such as single-shot syndrome extraction~\cite{campbell2019theory, lin2024single}, further modify the measurement structure so that reliable error information can be obtained with fewer repeated rounds under suitable code constructions. Syndrome measurement is therefore a protocol design problem in its own right. The same set of checks admits many circuits, and they differ in how faults spread.

From a practical compilation viewpoint, the goal is to generate
syndrome measurement circuits that are compatible with the target
hardware while preserving the intended fault-tolerance properties.
Circuit depth and ancilla count must be balanced against fault
propagation and circuit-level distance. The circuit also determines
the measurement record and detector structure available to the
decoder, so its design must account for decoding requirements.

\subsubsection{\textbf{Fault-Tolerant Architecture.}}

While syndrome measurement circuits describe how to protect quantum information at the level of a code block, fault-tolerant quantum computation requires a larger architectural organization. A fault-tolerant architecture specifies how logical qubits are laid out and operated on, and how they connect to classical control. Its job is to keep errors correctable at every step, whether a qubit is idle, inside a gate, or being measured.

A typical fault-tolerant architecture consists of several interacting components. On the quantum side it fixes the logical layout and the QEC cycle, together with the logical operations the layout supports. On the classical side it collects syndromes and decodes them in real time, then feeds the result back through the Pauli frame. A non-Clifford gate is needed on top for universality, and in most designs it comes from magic states that are distilled and then injected. These components together define the effective logical instruction set seen by a QEC-aware compiler.

The surface code architecture provides a representative example. In this setting, logical qubits can be represented by surface code patches, and logical operations can be implemented through patch deformation, lattice surgery, logical measurements, and magic-state injection. A logical program then runs as a space-time pattern over these patches. Patches are initialized and kept alive by repeated syndrome extraction. Merges, splits, and logical measurements act on them, and the classical frame is updated after each measurement. This makes the architecture naturally compatible with a compiler view, where the input is an algorithm-level logical circuit and the output is a scheduled sequence of fault-tolerant operations.

In designing a fault-tolerant architecture, several practical considerations commonly arise. The architecture is expected to reduce the logical error rate to the level required by the target algorithm, while keeping the associated space-time overhead within a feasible range. The overhead has a spatial part and a temporal part. The spatial part is physical qubits, including those spent on routing and on magic-state factories. The temporal part is repeated QEC cycles and logical measurements. The architecture should also be regular enough that compilation and decoding can be organized systematically. Hardware sets the outer limits through its native gate set and connectivity, and through how fast it can measure, reset, and feed a result back.

Therefore, a fault-tolerant architecture can be viewed as the target machine model for QEC compilation. It specifies the logical operations and resource abstractions available to the compiler, while also constraining the placement, scheduling, and execution of fault-tolerant procedures. In a full-stack QEC compiler, this architectural layer connects low-level syndrome measurement circuits with high-level logical circuit compilation, resource estimation, and hardware-aware optimization.

\subsection{Quantum Program Compilation}
\label{subsec:quantum_compilation}

\textit{Quantum program compilation} or quantum compilation is the process of translating a quantum program or quantum algorithm into operations that can be executed on a target quantum device. A typical quantum compilation flow consists of two closely related stages. (1).\textit{circuit synthesis} translates the algorithm into a gate set supported by the software stack and applies equivalence-preserving transformations to reduce the cost of the circuit. (2).\textit{qubit mapping and routing} adapts the synthesized circuit to the target hardware. At this hardware-facing stage, the compiler assigns program qubits to physical qubits, realizes required interactions through routing or movement, and schedules operations while respecting circuit dependencies and device timing constraints. This mapping and routing stage forms the central interface between the abstract circuit model and the physical architecture. In the NISQ era, the quality of quantum compilation is commonly evaluated by the overhead introduced during this transformation, including additional gates, increased circuit duration, estimated fidelity loss, and classical compilation cost~\cite{zhu2025quantumcompilerdesignqubit}.

The mapping and routing problem is highly platform dependent. In superconducting processors, the coupling graph restricts which two-qubit gates can be executed directly, so distant interactions must be implemented through local routing and scheduled under nearest-neighbor interference constraints. Trapped-ion processors expose more flexible connectivity within an interaction region, but scalable QCCD systems shift the main cost to ion transport and zone management. Neutral-atom arrays provide reconfigurable geometry through atom movement and Rydberg-mediated interactions, while making motion legality and addressability part of the compilation problem. Thus, even before QEC is introduced, quantum compilation is already architecture dependent: the same logical circuit can induce different routing overheads, timing profiles, and fidelity estimates on different devices.

Fault-tolerant quantum computation lifts this compilation problem to a higher level of abstraction. The compiler now works on encoded blocks. It lowers the algorithm into fault-tolerant procedures, keeps each block alive through syndrome extraction, and passes measurement results to the classical control stack. Correctness is also stronger than unitary equivalence. A QEC-aware transformation must preserve the fault-tolerance assumptions of the selected code and protocol, including the effective distance of the implemented circuit, the structure of propagated faults, and the detector information required by the decoder.

This distinction explains why QEC-aware compilation cannot be treated as a direct extension of ordinary qubit mapping and routing. Familiar goals such as reducing depth and communication remain important, but they now act inside fault-tolerant procedures. A routing decision may change the spacetime volume of lattice surgery. A syndrome-extraction schedule may change the detector error model. A hardware placement changes how logical errors accumulate, not just gate fidelity. Throughout this survey, we use \textit{QEC-aware compilation} to refer to the broader problem of lowering algorithm-level logical programs into fault-tolerant operations, hardware-realizable QEC cycles, and decoder-compatible runtime interfaces.

\subsection{Brief Introduction to Quantum Hardware}

Quantum hardware provides the physical substrate on which compiled quantum circuits are executed. In the abstract circuit model, gates are ideal and interactions can be specified without reference to a device topology. Real processors instead expose a limited set of native operations and a hardware-dependent notion of which qubits can interact efficiently. These restrictions make quantum compilation inherently architecture dependent. A circuit that is simple in the abstract model may require substantial routing, movement, or scheduling overhead on a real device. Following the cross-architectural compilation perspective of~\cite{zhu2025quantumcompilerdesignqubit}, we briefly summarize the hardware properties most relevant to qubit mapping and routing. 
In this survey, we focus on superconducting, trapped-ion, and neutral-atom platforms, whose QEC-aware compilation literature supports consistent cross-platform comparison. Photonic/fusion-based and semiconductor spin-qubit platforms are outside our scope.

\paragraph{Superconducting circuits.} Superconducting processors implement qubits using fabricated circuit elements and typically support relatively fast gate operations. Their main compilation constraint is the fixed coupling graph defined during fabrication. Two-qubit gates are usually native only between connected qubits, so an interaction between distant program qubits must be realized through routing on the coupling graph. This makes placement and SWAP insertion central backend tasks. Scheduling is also device dependent, since parallel operations can be limited by effects such as crosstalk and frequency crowding.

\paragraph{Trapped-ion.} Trapped-ion processors provide a different communication model. Within a single interaction region, ions can often be coupled with much more flexibility than in nearest-neighbor solid-state devices. This reduces the pressure of static connectivity in small systems, but scalable trapped-ion architectures introduce a different backend problem. In Quantum Charge-Coupled Device (QCCD) systems~\cite{pino2021demonstration,kaushal2020shuttling}, ions are transported between zones so that gates, storage, and measurement can be performed in appropriate regions. Compilation must therefore coordinate gate execution with transport operations such as shuttling and local reordering~\cite{murali2020architecting_iontrap,schoenberger2024shuttling,schoenberger2024using,zhu2025s,ruan2025trapsimdsimdawarecompileroptimization,wu2025muss}. The dominant cost is no longer only the number of two-qubit gates, but also the latency, congestion, and fidelity impact introduced by ion motion.

\paragraph{Neutral-atom arrays} Neutral-atom processors offer a more reconfigurable model. Atoms trapped in optical tweezers can be arranged in programmable geometries, and Rydberg-mediated interactions can entangle atoms over distances beyond strict nearest-neighbor layouts. The compiler can therefore rearrange the register mid-execution to make later interactions cheaper. The same flexibility introduces new legality constraints, since atom motion must avoid collisions and gates must respect addressing and blockade conditions~\cite{baker2021exploiting_atom,tan2022qubit_atom,tan2023compiling_atom,wang2024atomique,lin2024reuse,ruan2024powermove}. Neutral-atom compilation is therefore often organized around alternating stages of movement and entangling operations. Compared with superconducting and trapped-ion systems, its main backend question is how to use reconfigurability without allowing motion planning and control constraints to dominate the execution cost.

\section{Logical-level QEC optimization}
\label{sec:logical_level_qec_optimization}

Once qubits are encoded, an algorithmic gate can no longer be treated as a primitive physical operation. It must be realized through a fault-tolerant logical procedure supported by the chosen code and architecture. Logical-level compilation therefore selects and organizes these procedures, resolving how logical operations are implemented and coordinating their execution to reduce space--time and resource overhead. The first question is thus what logical operations the encoding can support fault tolerantly, and through which implementation mechanisms.

The fault-tolerant logical instruction set is constrained by both error propagation and locality. 
Transversal gates are attractive because they limit fault propagation within a code block, but 
the Eastin-Knill theorem rules out universality using transversal gates alone under its assumptions~\cite{eastin2009restrictions}. 
Locality-preserving logical operations are further restricted by the Bravyi--K\"onig bound for geometrically local topological stabilizer codes~\cite{bravyi2013classification}, while intrinsic locality dimension extends such locality-based constraints beyond fixed Euclidean geometries to more general code-connectivity structures~\cite{lu2026intrinsiclocalitydimensionquantum}. 
Universal fault-tolerant computation therefore generally combines complementary mechanisms, including code-specific protected gates, logical measurements and code surgery, resource-state injection, and code or gauge switching~\cite{horsman2012surface,paetznick2013universal,anderson2014fault,bombin2015gauge}.

These implementation choices directly determine the compiler problem. In surface-code architectures based on lattice surgery, logical interactions are expressed through joint logical Pauli measurements between code patches~\cite{horsman2012surface,fowler2018low}. The compiler must consequently arrange patches, allocate routing or ancillary regions, and schedule joint measurements while preserving sufficient parallelism. Beyond the surface code, there is no single corresponding patch-level abstraction. Depending on the code family, logical operations may exploit transversal or fold-transversal transformations and qubit permutations, or may instead be realized through code-surgery gadgets, bridge systems, and adapters~\cite{eberhardt2024logical,cohen2022low,cross2024improved,swaroop2026universal}. The compiler must therefore reason about code-specific logical access, ancillary resources, inter-block interactions, and operation scheduling rather than assuming one uniform routing model. Non-Clifford implementations introduce another form of resource coordination: resource-state injection and distillation require preparation, storage, and delivery of ancillary states, while code switching introduces explicit transitions between encodings.

We organize this section according to these logical-operation requirements. We first consider Clifford-operation implementation, beginning with surface-code lattice surgery and its compiler stack before extending to beyond-surface-code and qLDPC code-surgery frameworks. We then consider the mechanisms used to complete universal fault-tolerant computation, focusing on magic-state distillation, code switching and heterogeneous encodings, and early fault-tolerant compilation based on injected rotation resources.

\subsection{Clifford Operation Implementation}

Logical Clifford operations constitute a central component of fault-tolerant quantum computation, providing the entangling and basis-changing operations required throughout a logical circuit. Among them, the logical CNOT provides a representative case: it is a generating entangling operation of the Clifford group and exposes many of the placement, communication, and scheduling constraints encountered by logical-level compilers. Although Clifford operations are often treated as comparatively inexpensive relative to non-Clifford operations such as the $T$ gate, their realization can still contribute substantially to the space-time cost of large computations, particularly when logical interactions are non-local or highly concurrent. Efficient implementation of logical Clifford operations is therefore an important compilation objective.

A logical CNOT can be realized through several fault-tolerant mechanisms, including transversal coupling between code blocks, defect braiding, and lattice surgery~\cite{fowler2012surface,horsman2012surface}. In a transversal implementation, corresponding physical qubits in two encoded blocks interact pairwise, limiting the propagation of a single fault to at most one qubit within each block. Its practical cost, however, depends strongly on the availability of inter-block connectivity or on the routing required to realize these pairwise interactions. Lattice surgery instead implements logical interactions through joint logical Pauli measurements, typically involving operators such as $X_LX_L$ and $Z_LZ_L$, and can realize a logical CNOT using an auxiliary code patch~\cite{horsman2012surface,fowler2018low}. Because these operations can be implemented through local boundary measurements between neighboring patches, lattice surgery has become a standard execution abstraction for surface-code logical compilation.

When lattice surgery is adopted as the logical execution model, compilation becomes a space--time optimization problem over encoded regions and logical measurements. The compiler must assign logical qubits to code patches, allocate ancillary or routing regions for joint measurements, and schedule operations subject to spatial conflicts, data dependencies, and available parallelism. We therefore begin with the merge and split primitives and the logical CNOT constructed from them, before surveying surface-code lattice-surgery compilers and their solver-based, heuristic, and resource-aware optimization strategies. We then consider extensions beyond the surface code, where color codes, folded constructions, and qLDPC codes expose different transversal, code-surgery, adapter, and inter-block interaction mechanisms and consequently require more general compiler abstractions.



\subsubsection{\textbf{Lattice Surgery Introduction}} 

Lattice surgery is a measurement-based method for implementing logical operations on planar surface-code patches. Instead of moving defects through a code or relying on transversal coupling between distant blocks, it modifies the stabilizers measured along patch boundaries so as to read out a joint logical Pauli operator, while preserving locality in a two-dimensional nearest-neighbor layout~\cite{horsman2012surface}. This locality is why lattice surgery became the standard abstraction for surface-code compilation. Logical computation turns into a geometric problem: where to put each patch, and when to measure which pair of boundaries.

The two basic primitives are \textit{merge} and \textit{split}, a pair of inverse deformations of the code geometry. A merge fuses two patches into one, while a split divides one patch into two. Neither is an ordinary unitary gate acting on a fixed code block. Both are measurement-driven changes of geometry whose intrinsically random outcomes must be exposed to the classical control stack.

\textit{Merge.} Two neighboring patches are joined along compatible boundaries by activating an intermediate strip of qubits, equivalently by switching on new stabilizer checks across the shared boundary. These checks span both patches, and their combined outcome measures a joint logical Pauli product. The measured operator is the logical string that runs parallel to the seam. When the patches are joined so that a logical-$Z$ string of each lies along the seam, those strings become equivalent in the fused patch and the merge reads out $Z_L Z_L$ (Fig.~\ref{fig:lattice_surgery}). The complementary arrangement, with the logical-$X$ strings along the seam, reads out $X_L X_L$. For a distance-$d$ code, the modified stabilizers are measured for $d$ rounds, so that faults occurring during the deformation are detected and the joint parity is inferred reliably~\cite{horsman2012surface}. The parity outcome is random, so it is treated as a classical byproduct, either absorbed into the Pauli frame or used to reinterpret subsequent measurements.

\textit{Split.} A split reverses this deformation. A selected strip of data qubits is measured individually in a single-qubit Pauli basis, which severs the stabilizer connections across the strip and opens two new patch boundaries. After the required $d$ rounds of syndrome extraction, the single patch has become two encoded patches. Depending on the measurement basis and boundary type, a split can distribute the original logical state across the two new patches as an entangled state. For example, a patch in $\alpha\ket{0}_L + \beta\ket{1}_L$ splits into the correlated pair $\alpha\ket{00}_L + \beta\ket{11}_L$, up to Pauli-frame corrections~\cite{horsman2012surface}.

%
\providecommand{\lsbulk}[4]{\foreach \xx in {#1,...,#2}{\foreach \yy in {#3,...,#4}{\pgfmathtruncatemacro{\pp}{mod(\xx+\yy,2)}\ifnum\pp=0 \fill[Zcol] (\xx,\yy) rectangle ++(1,1);\else \fill[Xcol] (\xx,\yy) rectangle ++(1,1);\fi}}}
\providecommand{\lssgrid}[4]{\draw[gray!55,line width=0.4pt] (#1,#3) grid (#2,#4);\foreach \xx in {#1,...,#2}{\foreach \yy in {#3,...,#4}{\fill (\xx,\yy) circle (1.5pt);}}}
\providecommand{\lstabU}[3]{\fill[#3,draw=black,line width=0.4pt] (#1-0.5,#2) arc (180:0:0.5) -- cycle;}
\providecommand{\lstabD}[3]{\fill[#3,draw=black,line width=0.4pt] (#1+0.5,#2) arc (0:-180:0.5) -- cycle;}
\providecommand{\lstabL}[3]{\fill[#3,draw=black,line width=0.4pt] (#1,#2+0.5) arc (90:270:0.5) -- cycle;}
\providecommand{\lstabR}[3]{\fill[#3,draw=black,line width=0.4pt] (#1,#2+0.5) arc (90:-90:0.5) -- cycle;}
\providecommand{\lstwo}{%
  \lsbulk{0}{1}{0}{1}\lsbulk{0}{1}{4}{5}%
  \lstabD{0.5}{0}{Xcol}\lstabU{1.5}{2}{Xcol}\lstabL{0}{1.5}{Zcol}\lstabR{2}{0.5}{Zcol}%
  \lstabD{0.5}{4}{Xcol}\lstabU{1.5}{6}{Xcol}\lstabL{0}{5.5}{Zcol}\lstabR{2}{4.5}{Zcol}%
  \lssgrid{0}{2}{0}{2}\lssgrid{0}{2}{4}{6}%
  \draw[logop,line width=1.4pt] (0,2)--(2,2);\draw[logop,line width=1.4pt] (0,4)--(2,4);%
  \node[font=\scriptsize,text=logop] at (-0.72,2){$\bar Z_1$};\node[font=\scriptsize,text=logop] at (-0.72,4){$\bar Z_2$};%
}

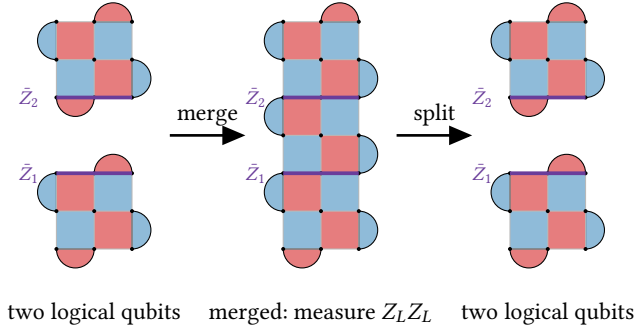
\begin{figure}[t]
  \centering
  \definecolor{xred}{RGB}{214,39,40}\definecolor{zblue}{RGB}{31,119,180}\definecolor{logop}{RGB}{106,61,154}%
  \colorlet{Xcol}{xred!60}\colorlet{Zcol}{zblue!50}%
  \begin{tikzpicture}[scale=0.5]
    \begin{scope}[xshift=0cm]\lstwo\node[font=\small] at (1,-1.7){two logical qubits};\end{scope}
    \draw[->,line width=1pt] (3.0,3) -- (5.0,3) node[midway,above,font=\small]{merge};
    \begin{scope}[xshift=6cm]
      \lsbulk{0}{1}{0}{5}
      \lstabD{0.5}{0}{Xcol}\lstabU{1.5}{6}{Xcol}
      \lstabL{0}{1.5}{Zcol}\lstabL{0}{3.5}{Zcol}\lstabL{0}{5.5}{Zcol}
      \lstabR{2}{0.5}{Zcol}\lstabR{2}{2.5}{Zcol}\lstabR{2}{4.5}{Zcol}
      \lssgrid{0}{2}{0}{6}
      \draw[logop,line width=1.4pt] (0,2)--(2,2);\draw[logop,line width=1.4pt] (0,4)--(2,4);
      \node[font=\scriptsize,text=logop] at (-0.72,2){$\bar Z_1$};\node[font=\scriptsize,text=logop] at (-0.72,4){$\bar Z_2$};
      \node[font=\small] at (1,-1.7){merged: measure $Z_LZ_L$};
    \end{scope}
    \draw[->,line width=1pt] (9.0,3) -- (11.0,3) node[midway,above,font=\small]{split};
    \begin{scope}[xshift=12cm]\lstwo\node[font=\small] at (1,-1.7){two logical qubits};\end{scope}
  \end{tikzpicture}
  \caption{Lattice surgery: merging and then splitting two logical qubits. Two
  distance-3 rotated surface-code patches encode logical qubits whose logical-$Z$
  operators $\bar Z_1$ and $\bar Z_2$ (purple) run horizontally. Merging the patches along those boundaries fuses them into a single patch (middle), whose seam reads out the joint logical measurement $Z_LZ_L$; a subsequent split restores the two qubits. $X$-type stabilizers are red and $Z$-type blue, following Fig.~\ref{fig:qec_code_families}.}
  \label{fig:lattice_surgery}
\end{figure}

\textit{Logical Operations from Surgery Primitives.}
Together, these primitives are sufficient to express the Clifford interactions required by many surface-code circuits. A lattice-surgery CNOT, for instance, can be assembled from joint $XX$ and $ZZ$ measurements among a control patch, a target patch, and an ancilla patch~\cite{horsman2012surface}. Non-Clifford operations still require additional resources, such as magic-state injection, but lattice surgery supplies the geometric substrate on which those resources are consumed. The compiler can therefore treat lattice surgery as an instruction set with explicit space-time costs: patch placement determines which joint measurements are local, routing determines when auxiliary area is needed, and scheduling determines how many operations run in parallel.

\subsubsection{\textbf{Surface Code Lattice Surgery Compilers}}

\tikzstyle{leaf}=[draw=hiddendraw,
    rounded corners,minimum height=1em,
    fill=mygreen!40,text opacity=1, align=center,
    fill opacity=.5,  text=black,align=left,font=\scriptsize,
    inner xsep=3pt,
    inner ysep=1pt,
    ]
\tikzstyle{middle}=[draw=hiddendraw,
    rounded corners,minimum height=1em,
    fill=output-white!40,text opacity=1, align=center,
    fill opacity=.5,  text=black,align=left,font=\scriptsize,
    inner xsep=3pt,
    inner ysep=1pt,
    ]
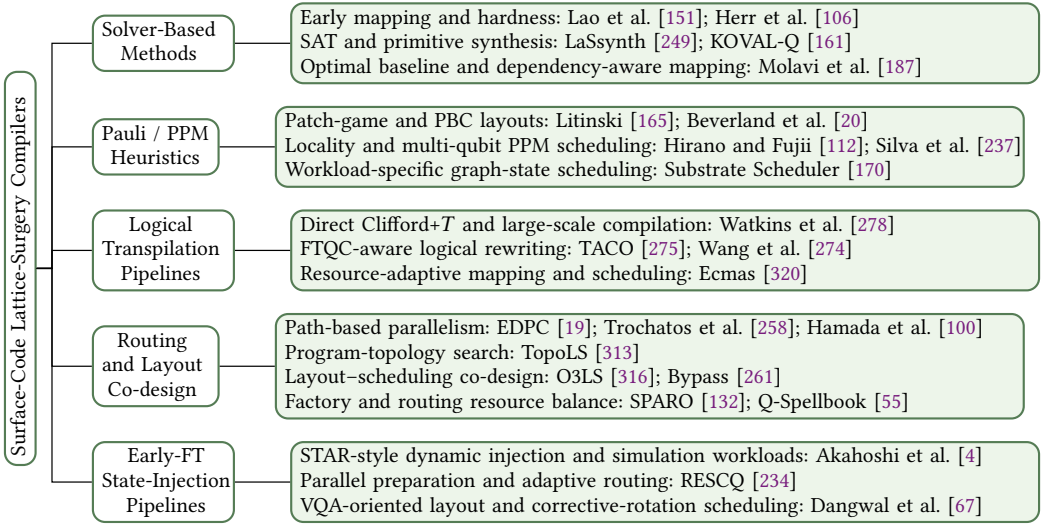
\begin{figure}[!ht]
\centering
\resizebox{\textwidth}{!}{
\begin{forest}
for tree={
  forked edges,
  grow=east,
  reversed=true,
  anchor=base west,
  parent anchor=east,
  child anchor=west,
  base=middle,
  font=\scriptsize,
  rectangle,
  line width=0.7pt,
  draw=output-black,
  rounded corners,
  align=left,
  minimum width=2em,
  l sep=18pt,
  s sep=7pt,
  inner xsep=4pt,
  inner ysep=2pt,
},
where level=1{text width=4.0em,align=center,font=\scriptsize}{},
where level=2{text width=23.8em,font=\scriptsize}{}
[{Surface-Code Lattice-Surgery Compilers}, middle, rotate=90, anchor=north, edge=output-black, text width=12.2em
  [{Solver-Based \\Methods}, middle, edge=output-black
    [{Early mapping and hardness: Lao et al.~\cite{lao2018mapping}; Herr et al.~\cite{herr2017optimization}\\
      SAT and primitive synthesis: LaSsynth~\cite{Tan2024SATScalpel}; KOVAL-Q~\cite{liao2026design}\\
      Optimal baseline and dependency-aware mapping: Molavi et al.~\cite{molavi2025dependency}},
      leaf, edge=output-black]
  ]
  [{Pauli / PPM\\Heuristics}, middle, edge=output-black, text width=3.5em
    [{Patch-game and PBC layouts: Litinski~\cite{litinski2019game}; Beverland et al.~\cite{beverland2022assessing}\\
      Locality and multi-qubit PPM scheduling: Hirano and Fujii~\cite{hirano2025localityaware}; Silva et al.~\cite{silva2024multi}\\
      Workload-specific graph-state scheduling: Substrate Scheduler~\cite{Liu2023SubstrateScheduler}},
      leaf, edge=output-black]
  ]
  [{Logical\\Transpilation\\Pipelines}, middle, edge=output-black
    [{Direct Clifford+\ensuremath{T} and large-scale compilation: Watkins et al.~\cite{Watkins2024highperformance}\\
      FTQC-aware logical rewriting: TACO~\cite{wang2024optimizingftqcprogramsqec}; Wang et al.~\cite{wang2025tableau}\\
      Resource-adaptive mapping and scheduling: Ecmas~\cite{zhu2024ecmas}},
      leaf, edge=output-black]
  ]
  [{ Routing \\ and Layout\\ Co-design}, middle, edge=output-black, text width=3.5em
    [{Path-based parallelism: EDPC~\cite{beverland2022SurfaceCodeCompilation}; Trochatos et al.~\cite{trochatos2025trace}; Hamada et al.~\cite{hamada2024efficient}\\
      Program-topology search: TopoLS~\cite{zhou2026topols}\\
      Layout--scheduling co-design: O3LS~\cite{zhu2026o3ls}; Bypass~\cite{ueno2024highperformancescalable}\\
      Factory and routing resource balance: SPARO~\cite{kan2025sparo}; Q-Spellbook~\cite{chatterjee2025qspellbook}},
      leaf, edge=output-black]
  ]
  [{Early-FT\\State-Injection\\Pipelines}, middle, edge=output-black
    [{STAR-style dynamic injection and simulation workloads: Akahoshi et al.~\cite{akahoshi2025compilation}\\
      Parallel preparation and adaptive routing: RESCQ~\cite{sethi2025rescq}\\
      VQA-oriented layout and corrective-rotation scheduling: Dangwal et al.~\cite{dangwal2025variational}},
      leaf, edge=output-black]
  ]
]
\end{forest}
}
\caption{A taxonomy of surface-code lattice-surgery compilers.}
\label{fig:taxonomy_of_surface_code_ls_compilers}
\end{figure}

\begin{figure}[t]
    \centering
    \includegraphics[width=1\linewidth]{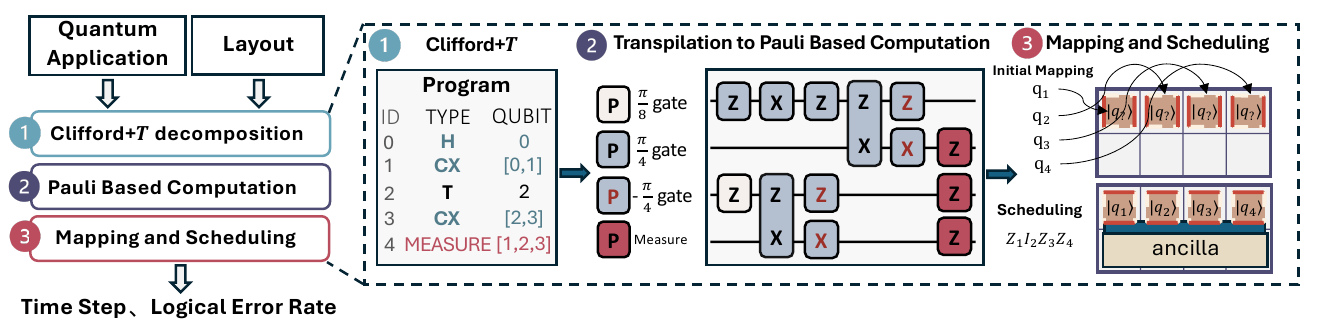}
    \caption{Pipeline for executing logical circuits on surface code quantum computers. First, decompose the logical circuit into the Clifford plus T instruction set. Optionally, transpile the circuit to a Pauli based computation form. Finally, apply mapping and scheduling to realize the instructions on the target hardware.}
    \label{fig:LS_pipeline}
\end{figure}

\paragraph{Background and Typical Pipeline.}
Fig.~\ref{fig:LS_pipeline} illustrates a representative top-down workflow for compiling logical operations for surface-code-based quantum computers, following common practices in~\cite{beverland2022assessing,hirano2025localityaware}. To place this workflow within the broader research landscape, Fig.~\ref{fig:taxonomy_of_surface_code_ls_compilers} presents a taxonomy of existing surface-code lattice-surgery compilers and their principal design choices. Given a quantum application and a specified logical-qubit layout, the workflow progressively lowers the application into a scheduled sequence of lattice-surgery operations and estimates the resulting execution time and logical error rate. We describe each compilation stage below.

\textit{Step {1}: Clifford+$T$ Decomposition.} The program begins with a quantum algorithm written in a high-level language (Fig.~\ref{fig:LS_pipeline}.1). Since such algorithms are not expressed in the Clifford+$T$ gate set required for fault-tolerant execution, gate synthesis becomes necessary. This is typically done using the Solovay-Kitaev algorithm~\cite{dawson2005solovaykitaevalgorithm} or more advanced techniques~\cite{ross2016optimalancillafreecliffordtapproximation, li2025noncliffordfusiontgateoptimization}.

\textit{Step {2}: Transpilation to Pauli-Based Computation.} The decomposed Clifford+$T$ circuits are subsequently transpiled into Pauli product rotations (Fig.~\ref{fig:LS_pipeline}.2). We use the Pauli-product-rotation convention $P_{\theta} = \exp(-i\theta P)$, where $P$ is a multi-qubit Pauli operator and global phases are ignored. Thus $S = Z_{\pi/4}$ and $T = Z_{\pi/8}$ under this convention. Equivalently, the angle $\theta$ is half of the rotation angle in the common notation $R_P(\phi)=\exp(-i\phi P/2)$. The standard decompositions are then given as $H = Z_{\pi/4}X_{\pi/4}Z_{\pi/4}$ and $CNOT = (Z\otimes X)_{\pi/4} (I\otimes X)_{-\pi/4} (Z\otimes I)_{-\pi/4}$. There are several rules for simplifying circuits based on the commutation relations of Pauli operators.

If $P$ and $P'$ commute i.e. $PP' - P'P = 0$, then $P_{\pi/4}$ can be moved past $P'_{\theta}$. If $P$ and $P'$ anti-commute i.e. $PP' + P'P = 0$, $P'_{\theta}$ turns into $(iPP')_{\theta}$ when passing $P_{\pi/4}$. 
Clifford gates can be commuted through the circuit and absorbed into final measurements, as they map Pauli operators to Pauli operators. Transpiling to Pauli-based computation is optional. It changes the
operations presented to the scheduler, and its benefits depend on
the resulting measurement supports and the target architecture.
The compilers reviewed below make different choices about how
extensively to propagate Clifford operations.

\textit{Step {3}: Logical Level Mapping and Scheduling.} After transpilation of Pauli product rotations, the compiler realizes these instructions by mapping and scheduling them according to the rules required by lattice surgery (Fig.~\ref{fig:LS_pipeline}.3). The instructions are initially mapped by assigning logical qubits to different patches of data layout, with the goal of maximizing opportunities for simultaneous multi-patch measurements while minimizing time costs. 

\paragraph{Solver-Based and Optimization-Based Methods. } 
The first line of the surface code lattice surgery compilers formulates the problem as explicit optimization problems. \citet{lao2018mapping} present an early end-to-end mapping flow for lattice-surgery-based fault-tolerant circuits. The first stage performs scheduling to minimize total latency while preserving data dependencies. It uses the fact that many CNOTs commute to reorder them and increase parallelism in the rest of the circuit. After scheduling, the mapper solves an initial placement problem formulated as a quadratic assignment problem (QAP), based on the insight that frequently interacting logical qubits should be placed close together. The placement cost is driven by Manhattan distance weighted by interaction frequency to reduce future communication. Finally, the router applies a sliding-window (buffer)-based heuristic over the scheduled instruction stream, inserting movement operations as needed and locally rescheduling within the window to satisfy architectural constraints while limiting overhead. \citet{molavi2025dependency} also proposes a SAT-based optimal scheduling algorithm that minimizes the number of time steps. Given a fixed data layout and an input circuit, it encodes execution into Boolean variables subject to constraints that (i) preserve the circuit’s logical dependencies, (ii) ensure routing paths are pairwise disjoint, and (iii) connect the qubit pairs participating in each CNOT at the time of execution.

LaSsynth~\cite{Tan2024SATScalpel} moves to a lower-level subroutine-synthesis problem. It introduces LaSre, a native lattice-surgery representation based on three-dimensional pipe diagrams over a discretized spacetime grid, and encodes the existence of a valid lattice-surgery construction within a bounded spacetime volume as a SAT instance. By repeatedly querying the SAT solver while shrinking the allowed volume, LaSsynth can exhaustively search the bounded design space and prove volume optimality within that formulation. It is meant for synthesizing frequently reused subroutines. For example, it reports volume reductions for 15-to-1 $T$-factory designs relative to human-designed baselines. Recent SAT-based EDA kernels further broaden this subroutine-synthesis direction. KOVAL-Q~\cite{liao2026design} describes logical operations in terms of stabilizer faces, which admits more surface-code encodings as targets and intermediate configurations than pipe diagrams do. It separates lattice-surgery legality, functional correctness, and fault-tolerance checks into SAT-based subkernels, and uses them to optimize and verify fundamental logical operations such as CNOTs and patch rotations. Like LaSsynth, KOVAL-Q is an EDA kernel: it discovers and optimizes primitives that a program-level compiler can later invoke.

These solver-based methods expose different levels of the lattice-surgery design stack. Lao et al. target early end-to-end mapping and routing and Molavi et al. optimize whole-program surface-code mapping and routing while using SAT as an optimal baseline. LaSsynth and KOVAL-Q synthesize or verify small but important lattice-surgery subroutines. Their limited scalability is consistent with the known hardness of lattice-surgery optimization~\cite{herr2017optimization}: exact formulations are valuable for proving optimality on bounded instances and for producing high-quality primitive libraries, while large-scale applications generally require heuristic, graph-based, or layout-aware compilation strategies. Below, we highlight representative lattice-surgery compilers that rely on heuristic strategies.

\paragraph{Heuristic-Based and Scalable Methods. } Toward scalable compilation for surface-code lattice surgery, a substantial body of work has proposed heuristic techniques. These methods primarily differ in whether they adopt Pauli-based computation or compile circuits directly in the Clifford+T model. In both cases, the central goal is to maximize logical-level parallelism and thereby minimize overall latency, typically measured in the number of time steps or QEC cycles. Algorithm.~\ref{algo:LS_scheduling} presents a general heuristic scheduling framework. The input is a gate sequence describing the target quantum circuit, and the procedure is agnostic to whether the computation is expressed using Pauli-based techniques or compiled directly in the Clifford+T model. Given the input circuit and an initial data layout, the scheduler iterates through the remaining operations and checks whether the next operation is executable under the current layout and constraints. If it is not, the scheduler inserts the required lattice-surgery and patch-management actions (e.g., moves, merges, splits, or rotates) until the operation becomes feasible. It then schedules the operation, updates the data layout accordingly, and repeats this process until the program completes. Below, we review prior work on heuristic lattice-surgery scheduling frameworks.

The most widely used starting point is Litinski's ``game of surface codes''~\cite{litinski2019game}, which states lattice-surgery compilation as an optimization problem with a small set of moves. In this formulation, lattice-surgery operations are expressed as a set of primitive instructions (e.g., patch moves, rotations, splits, and merges), each associated with an explicit time cost measured in code cycles. For heuristic optimization,~\cite{litinski2019game} uses the commutation properties of Pauli operators to eliminate explicit Clifford gates by pushing them to the right and absorbing their effects into the final measurements. Litinski also proposes several standard data-layout templates, including the compact, intermediate, and fast layout. The compact data layout uses $1.5n + 3$ tiles to store $n$ qubits, the intermediate data layouts use $2n + 4$ tiles and the fast data layouts uses $2n +\sqrt{8n} + 1$ tiles. These layouts trade space for time in magic-state consumption: the fast layout achieves the lowest latency for consuming magic states, but it incurs the highest space overhead. A Python implementation aimed at smaller-scale studies and supporting Pauli-based computation is presented in~\cite{leblond2023realistic}, with the code available on GitHub~\cite{latticesurgery_compiler_github}.

Several works refine this Pauli-based view by adding locality, parallelism, or workload-specific structure. \citet{hirano2025localityaware} propose locality-aware Pauli-based computation (LAPBC) to increase parallelism on both standard and sparse layouts, which use approximately $2.25N$ and $4N$ data patches, respectively. LAPBC stops short of propagating every Clifford operation. Instead, LAPBC restricts propagation to single-qubit Clifford gates, so only a subset of qubits is affected at each step and more operations can be executed in parallel. Using this technique, LAPBC outperforms the approach of~\cite{litinski2019game} on random circuit samples and a 2D Ising Hamiltonian benchmark. \citet{silva2024multi} study the scheduling of multi-qubit lattice-surgery operations, focusing on how to schedule Pauli measurements that involve several logical patches. Substrate Scheduler~\cite{Liu2023SubstrateScheduler} addresses a more specialized workload that is compiling arbitrary graph states. It minimizes space-time volume and demonstrates compilation for graph states with thousands of vertices, making it a useful example of workload-specific lattice-surgery scheduling.

Transpiler-level optimizations can also reshape the workload before lattice-surgery scheduling. TACO~\cite{wang2024optimizingftqcprogramsqec} observes that Clifford gates can become a major FTQC bottleneck once $T$-gate costs are reduced, and introduces FTQC-aware circuit and architecture co-design to reduce Clifford overhead while preserving useful parallelism. TQC~\cite{wang2025tableau} takes a complementary tableau-based approach that uses commutativity, operation fusion, basis-aware reordering, and latency hiding to reduce runtime overhead in resource-efficient logical architectures. Both are transpilation and co-design methods. They reshape the operation stream before it reaches a lattice-surgery scheduler.

\begin{algorithm}[t]
    \caption{Heuristic Scheduling Framework}
    \label{trivial_scheduling}
    \KwIn{Logical Gate Sequence $\mathcal{S}$}
    \KwOut{Physical Gate Sequence $\mathcal{S}'$}
    $\mathcal{S}' \leftarrow \{\}$\;
    Get Data Layout $\mathcal{B}$\;
    \For{$g \in \mathcal{S}$}{
        \If{not $executable\_check(B,g)$}{
            $patch\_op \leftarrow patch\_scheduling(\mathcal{B}, g)$\;
            $\mathcal{S}'\leftarrow \mathcal{S}'+ patch\_op$\;            
            Update Data Layout $\mathcal{B}$\;
        }
        $path \leftarrow path\_routing(g)$\;
        $\mathcal{S}'\leftarrow \mathcal{S}'+ ls(path, g)$\;
    }
    \Return $\mathcal{S}'$\;
\label{algo:LS_scheduling}
\end{algorithm}

Another line of work compiles Clifford+$T$ programs more directly into lattice-surgery instructions. Watkins et al.~\cite{Watkins2024highperformance} introduce a high-performance compiler for very large surface-code computations. The compiler uses a lattice-surgery intermediate representation, supports customizable layouts, and emphasizes scalability: it can process very large logical circuits using a streaming pipeline. LeBlond et al.~\cite{leblond2023realistic} build on this open-source lattice-surgery compiler to provide a resource-estimation pipeline for practical Clifford+$T$ circuits. Their flow separates an abstract, layout-independent instruction stage from a local lattice-surgery allocation stage, enabling post-hoc analysis of magic-state production and storage requirements. These systems make lattice-surgery compilation runnable at application scale.

Resource-adaptive mapping and scheduling methods further optimize how a circuit uses the available surface-code layout. Ecmas and Ecmas+~\cite{zhu2024ecmas,zhu2025ecmas+} formalize mapping and scheduling for both double-defect and lattice-surgery models, introducing circuit and chip metrics to choose resource allocations adapted to each workload. Such methods are closer to traditional compiler mapping and scheduling, in the sense that they explicitly balance circuit parallelism against communication capacity and available physical resources.

\paragraph{Routing and Layout Co-design}
A second group of scalable lattice-surgery compilers focuses on the resources that surround each logical operation, including routing space, data-layout geometry, and auxiliary patches such as magic-state factories. In these methods, executing a single logical operation is not the main bottleneck. The harder problem is routing many long-range operations without congestion and without over-provisioning the surface-code layout.

Beverland et al.~\cite{beverland2022SurfaceCodeCompilation} address this problem through edge-disjoint paths compilation. Their approach models long-range lattice-surgery operations as the preparation of Bell pairs along edge-disjoint ancilla paths, allowing multiple Clifford operations to proceed in parallel when the layout provides sufficient routing capacity. Scheduling then becomes a graph-routing problem, and path conflicts set the achievable parallelism. Hamada et al.~\cite{hamada2024efficient} refine this routing-centric view by exploiting the internal structure of lattice-surgery instructions. Instead of treating each logical operation as an indivisible unit, they decompose operations into smaller components, such as Bell-state preparation and measurement, and schedule these components in a three-dimensional spacetime lattice. The third axis is time, so the scheduler can start parts of a future operation while the current one is still running and relieve congestion.

Recent compilers also optimize the topology of the lattice-surgery program itself. TopoLS~\cite{zhou2026topols} uses ZX diagrams as an intermediate representation and combines topology-preserving program transformations with Monte Carlo tree search for placement and routing in three-dimensional spacetime. This makes the compiler less tied to a fixed gate sequence and allows it to search over semantically equivalent lattice-surgery structures.

Layout choice is another major determinant of routing cost. O3LS~\cite{zhu2026o3ls} studies this layout-scheduling interaction directly. It observes that extra routing space cuts time steps but has a price. Patches move farther and rotate more often, and while they wait they accumulate logical error. O3LS therefore combines automatic layout search, loose scheduling, and circuit-synthesis optimizations. Its squeezed-layout search reduces space overhead, while loose scheduling allows patch functionality to be reassigned during execution to avoid redundant movements and operations. Bypass~\cite{ueno2024highperformancescalable} pushes the same idea closer to the architecture substrate. Rather than only improving routing on a fixed two-dimensional patch array, it identifies bottlenecks in lattice-surgery execution and proposes a 2.5D organization with dense and sparse qubit layers. This architecture-level change increases routing flexibility and reduces congestion for large-scale lattice-surgery workloads.

Finally, several systems balance the resources around the scheduler. Q-Spellbook~\cite{chatterjee2025qspellbook} studies the joint choice of data-block layouts and magic-state distillation protocols, exposing the tradeoff between tile count, distillation throughput, and execution time. SPARO~\cite{kan2025sparo} dynamically allocates surface-code resources among computation, routing, and magic-state factories using an active logical-error model. Pure Magic~\cite{hofmeyr2026puremagicdynamicschedulerlattice} considers a cultivation-based magic-state supply model and repurposes magic-state preparation resources for routing when beneficial. Together, these works show that scalable lattice-surgery compilation is a whole-computation resource-allocation problem. Routing paths, layout geometry, factory throughput, and architecture-level resource allocation all have to be coordinated.

\subsubsection{\textbf{Beyond Surface-Code and qLDPC Code-Surgery Compilers}}

Surface-code compilers rely on a mature patch abstraction. Each logical qubit is a two-dimensional patch, and compilation reduces to placing patches, routing ancilla regions, and scheduling merges and splits. This abstraction does not transfer directly to codes beyond the surface code. Color codes and folded surface codes differ in their transversal-gate support and boundary-access rules. High-rate qLDPC codes additionally encode multiple logical qubits per block, requiring selective logical access and code-specific inter-block operations, which may use surgery adapters or bridges. As a result, the compiler stack beyond the surface code is less mature. Many works provide logical-measurement primitives that a compiler may consume, while only a smaller set provide automated synthesis or program-level compilation pipelines.

The most direct attempts to generalize lattice-surgery compilation beyond the surface code introduce new abstractions for representing the code substrate and the surgery operation itself. Herzog et al.~\cite{herzog2025latticesurgerycompilation} propose a beyond-surface-code compilation framework based on a code substrate, which is a microscopic blueprint for implementing a topological code and its lattice-surgery rules. After abstracting from microscopic stabilizer details, the compilation problem becomes a mapping-and-routing problem on a macroscopic routing graph, possibly with substrate-specific constraints. This is the first framework that separates logical compilation from surface-code patch geometry, and it is demonstrated on color-code and folded-surface-code substrates.

For CSS codes, SSIP~\cite{cowtan2024ssip} provides a more algebraic form of automation. It implements Safe Surgery by Identifying Pushouts, an open-source Python package that automates internal and external surgery between CSS codeblocks using chain-complex pushouts. SSIP is a surgery synthesis tool. It outputs a valid surgery construction between two codes, and leaves placement and scheduling to the caller. Similarly, Engineering CSS Surgery~\cite{poirson2025engineering} introduces a systematic framework for designing and analyzing surgery protocols for arbitrary CSS codes. Its subcode formalism tracks both the physical data needed for surgery and the induced logical operation, and it gives a CNOT construction for arbitrary CSS codes. This gives a reusable surgery-design abstraction that a scheduler can build on. GeneCS~\cite{zhou2026genecs} is closer to a general code-surgery compiler. It treats code-surgery construction for arbitrary stabilizer codes as a graph-optimization problem, synthesizing resource-efficient protocols for both single-code and cross-code logical operations. Compared with purely theoretical surgery constructions, GeneCS explicitly optimizes the ancillary graph and check structure, incorporates degree constraints, and targets scalable compilation of surgery instances.

\paragraph{Architecture-specific qLDPC compilation pipelines.}
A second group of works develops qLDPC computation around a particular architectural interface. These systems are more compiler-like than pure logical primitives, but their compilation choices are tightly coupled to a proposed architecture. Extractor architectures~\cite{he2025extractors} provide an architectural bridge from qLDPC memory to logical computation. An extractor system augments a qLDPC memory block so that arbitrary logical Pauli operators can be measured through a fixed-connectivity auxiliary system, enabling Pauli-based computation across extractor-augmented computational blocks. This work defines a logical-computation architecture and a measurement interface that later compilers can target. The bicycle architecture of Yoder et al.~\cite{yoder2025tour} is more architecture-specific and compiler-facing. It builds a modular FTQC architecture around bivariate-bicycle codes, defines explicit logical instruction sets, estimates instruction-level logical error rates, and develops a compilation strategy from Pauli-based computation to the bicycle instruction set. Follow-up systems work by Liu et al.~\cite{liu2026assessing} further studies the compiler bottlenecks of early bicycle architectures, identifying inter-module communication from non-Clifford operations as a dominant cost and proposing optimizations such as factory-side synthesis and critical-path Clifford insertion.

\paragraph{Logical code-surgery primitives for qLDPC computation.}
Most qLDPC surgery papers fall into a third category. They define measurement primitives, adapters, or bridges that a compiler can later call. Cohen et al.~\cite{cohen2022low} established a foundational qLDPC computation model in which long-range connectivity enables low-overhead logical Pauli measurements. Cross et al.~\cite{cross2024improved} improve this direction through gauge-fixed qLDPC surgery, reducing ancilla overhead and introducing bridge systems for joint measurements between code families. Zhang and Li~\cite{zhang2025time} study time-efficient logical operations by measuring arbitrary commuting sets of logical Pauli operators in time independent of the number of measured operators, exposing a parallel logical-measurement primitive for qLDPC computation.

Several recent works formulate these primitives in more general algebraic language. Williamson and Yoder~\cite{williamson2026low} measure logical operators by treating them as symmetries and gauging them, giving a flexible logical-measurement method applicable beyond a single code family. Ide et al.~\cite{ide2025fault} introduce homological measurement, which views lattice surgery and related logical-measurement protocols as special cases of chain-complex operations and gives edge-expanded protocols for arbitrary qLDPC logical Paulis. Universal adapters~\cite{swaroop2026universal} connect qLDPC codeblocks through repetition-code adapters, allowing joint logical Pauli measurements within or between codeblocks independent of the specific LDPC codes involved.

Other primitives focus on parallelism and asymptotic overhead. Parallel code surgery~\cite{cowtan2026parallel} measures many logical Pauli products in parallel on qLDPC codes while preserving LDPC structure and fault distance. Fast surgery~\cite{baspin2025fast} reduces generalized qLDPC surgery to a constant number of syndrome-measurement rounds using homomorphic chain-complex constructions. Parsimonious qLDPC surgery~\cite{yuan2026parsimonious} further reduces the ancilla overhead for measuring arbitrary logical Paulis and can improve the overhead of several existing surgery schemes. Batched high-rate logical operations~\cite{xu2025batched} extend the same broader theme from high-rate memory to high-rate computation by applying shared operations across many qLDPC blocks in parallel. Taken together, they define the operation library and cost model that a qLDPC compiler will have to schedule.

\subsection{Non-Clifford Operation Implementation}

Non-Clifford operations complete the logical gate set required for universal fault-tolerant quantum computation, but they often introduce a qualitatively different compilation bottleneck from Clifford operations. Many Clifford operations can be realized through code-supported mechanisms such as transversal gates, logical Pauli measurements, or code deformation, so their execution cost is largely shaped by logical connectivity, placement, routing, and scheduling. Non-Clifford operations, in contrast, frequently require additional fault-tolerant resources whose preparation and delivery can be substantially more expensive than the logical operation itself. In architectures based on magic-state injection, for example, the production, storage, routing, and consumption of high-fidelity resource states can dominate space-time overhead or limit the sustained throughput of the computation~\cite{bravyi2005universal}. This cost asymmetry motivates treating non-Clifford implementation as a distinct compiler problem.

Broadly, non-Clifford operations can be supplied through two classes of mechanisms. The first exploits logical gates supported directly by the structure of a code. Recent homological and algebraic constructions use structures such as cup products, intersection properties, and code multiplication properties to obtain protected controlled-$Z$-type logical operations~\cite{breuckmann2026cups,golowich2025asymptotically}. Related constructions provide transversal multi-controlled-$Z$ gates for selected qLDPC and quantum locally testable code families~\cite{li2026transversal}. For a compiler, however, gate existence alone is insufficient: the logical operands must also be addressable, concurrent applications must be compatible, and the required connectivity and ancillary resources must be exposed by the architecture. Recent work on addressable and parallelizable transversal non-Clifford gates makes this distinction explicit~\cite{he2025asymptotically,guemard2026good}. Thus, the usefulness of a direct non-Clifford gate depends not only on whether the code admits it, but also on the execution model through which the compiler can invoke it.

When a required non-Clifford operation is not available through a suitable direct mechanism, the architecture must provide an indirect fault-tolerant route. The most established approach is resource-state injection, where specially prepared non-stabilizer states are consumed to realize logical non-Clifford gates. Achieving the required logical fidelity then introduces additional compilation problems in resource-state preparation, distillation, placement, buffering, routing, and consumption. Code switching provides a complementary strategy by transferring logical information between encodings with different protected gate sets~\cite{anderson2014fault}, while early fault-tolerant architectures explore injected continuous rotations and runtime correction to reduce the cost of fully distilled resources. These mechanisms turn non-Clifford compilation into a resource-flow and scheduling problem in addition to a logical gate-synthesis problem. We therefore first review magic-state distillation and its associated resource-management techniques, then discuss code switching and heterogeneous encodings, and finally consider early fault-tolerant compilation schemes based on dynamically injected rotations.

\subsubsection{Magic State Distillation}

\paragraph{Background.}
Magic state distillation (MSD) provides a standard route for preparing the high-fidelity resource states required by injected non-Clifford operations. ...

\subsubsection{\textbf{Magic-State Distillation and Cultivation}}

Fault-tolerant non-Clifford gates are commonly implemented by consuming high-fidelity magic states through state injection or gate teleportation. Magic-state distillation (MSD) prepares these resources by processing multiple noisy input states, producing fewer states with lower error rates. Canonical examples include 7-to-1 $\ket{Y}$-state distillation based on the $[[7,1,3]]$ Steane code and 15-to-1 $\ket{T}$-state distillation based on the $[[15,1,3]]$ quantum Reed-Muller code~\cite{Bravyi_2005,raussendorf2007topological,raussendorf2006fault}. These codes are also related to two- and three-dimensional color-code constructions~\cite{bombin2006topological}. Magic-state cultivation provides a different route to the same resource. It begins with a small encoded magic-state seed and gradually increases its protection through local code growth and repeated fault-detecting measurements~\cite{gidney2024magic,hirano2025efficient,vaknin2025magic,claes2025cultivating,sahay2025fold}.

\paragraph{Optimizing Magic State Distillation.} Beyond these canonical baselines, MSD protocols have been continuously optimized to reduce space time overhead and improve yield, with a prominent line of practical constructions based on triorthogonal and related code families~\cite{bravyi2012magic, haah2017magic, haah2018codes, wills2025constant}. Recently, physical-level distillation and preparation protocols have attracted significant attention because they aim to reduce the number of costly logical distillation rounds in practical FTQC. Zero-level distillation improves the effective quality of injected magic states using physical-level error detection, then teleports the resulting state into the target topological code~\cite{itogawa2025efficient, chamberland2020very}. Magic state cultivation further develops this direction by growing a high-fidelity logical magic state through local, measurement-driven operations, shifting part of the factory cost from multi-round distillation into local preparation steps \cite{gidney2024magic, hirano2025efficient, vaknin2025magic, claes2025cultivating, sahay2025fold}. Collectively, these approaches can reduce factory depth and qubit overhead and can relax architectural pressure on long-range connectivity, thereby improving the practicality of universal fault-tolerant computation.

To reduce the overhead of magic state distillation in resource-efficient FTQC architectures, \citet{ding2018magic} propose magic state functional units and a compilation flow that maps and schedules multi-level distillation circuits as modular services under limited hardware resources. Their mapping and scheduling account for the geometric constraints induced by braiding defects in the defect based surface code. \citet{holmes2019resource} further optimize factory resource usage by co-designing distillation layouts and execution schedules to reduce footprint and increase distilled magic state throughput in defect based surface code implementations, making explicit the tradeoffs among factory area, routing congestion, and delivered magic state rate.

On the other hand, lattice surgery based approaches provide an alternative architectural model for magic state distillation ~\cite{litinski2019game}. Within this framework, \citet{litinski2019magic} argues that magic state distillation can be implemented as compact, pipelined factories, reducing space time overhead by sustaining continuous output and exploiting efficient patch movement and merging. \citet{hirano2024leveraging} use zero-level distillation as a physical-level preprocessing stage that improves the effective quality of injected magic states and can reduce the number of logical-level distillation rounds required to reach a target error rate. \citet{hirano2024magicpool} address the operational impact of probabilistic distillation with Magicpool, which maintains a reservoir of distilled states and uses allocation and recovery policies to prevent distillation failures from causing global stalls. \citet{hofmeyr2026puremagicdynamicschedulerlattice} study scheduling in the presence of magic state cultivation under lattice surgery, showing how cultivation competes for lattice-surgery resources and motivating integrated schedulers that jointly plan computation and state preparation to sustain end-to-end throughput.

\subsubsection{\textbf{Code Switching}}

A central limitation of transversal fault-tolerant gates is captured by the Eastin-Knill theorem: no finite-dimensional quantum error-correcting code can provide a universal logical gate set using transversal gates alone. This forces architectures and compilers to combine transversal operations with additional mechanisms to form the universal gate set. In this context, \textit{code switching transfers a logical state between distinct stabilizer codes during execution}, allowing different algorithm segments to exploit complementary code properties, such as favorable transversal gates or lower-cost logical constructions. Compiler design must therefore decide where and when to switch codes, and account for the time, space, and logical-error overheads introduced by each switch.

\paragraph{Compiler Design for Code Switching}

\citet{weilandt2025minimizing} are motivated by the fact that code switching enables a universal fault tolerant gate set by moving logical information between different QECCs, but each switch is expensive in time and space and can increase the logical error rate, so the number of switches should be minimized. They formulate minimizing the required code switching operations for a given logical circuit as a polynomial time optimization by reducing it to a minimum cut problem on a graph derived from the circuit, and they extend the formulation with practical compilation knobs such as preferring switches during idle periods (to reduce depth overhead) and biasing toward one code when desired. Experimentally, they show that the approach scales efficiently and can be applied to large logical circuits, including instances with up to 1024 qubits.

\citet{wu2023enablingfullstackquantumcomputing}'s main setting is dynamic switching among different error corrected encodings of a logical qubit (that is, switching between distinct stabilizer codes) and they identify three full stack challenges: realizing such dynamic logical qubits on hardware, deciding when to change the encoding, and improving end to end performance across programs with different characteristics. To address this, they propose CECQ, a full stack framework that explores the FTQC design space for changeable logical qubits, and report experimental results on a variety of quantum programs showing that CECQ is effective at improving overall system performance relative to non adaptive, fixed encoding baselines.

\citet{stein2024architecture} proposes a heterogeneous QEC framework HetEC that aims to combine the respective advantages of the surface code and the gross code. Inspired by a von Neumann-style architecture, HetEC separates the system into a processing unit built from surface-code blocks, a memory unit built from gross-code blocks, and an ancilla bus that transfers quantum information between these blocks. The motivation for this design is that the surface code has high physical overhead, with resource costs growing roughly quadratically with the number of correctable errors. In contrast, qLDPC codes can offer more favorable scaling, but practical schemes for universal FT computation with these codes are still less developed, which motivates combining them in a heterogeneous framework. Given the limited number of surface-code tiles, HetEC constrains the weight of any Pauli-product measurement to not exceed the available tiles. It further schedules operations to maximize gate concurrency and minimize data movement between regions. Under the same physical error rate, HetEC achieves up to a 6.42× reduction in physical‑qubit overhead.

\subsubsection{\textbf{Early Fault-Tolerant Compiler}}

Executing fault-tolerant logical operations requires non-Clifford gates, which are challenging to implement on QEC codes such as the surface code. In particular, reliably performing the $T$ gate necessitates resource-intensive protocols like magic state distillation. However, magic state distillation incurs a substantial overhead, often requiring many physical qubits. The transition from NISQ-era devices to FTQC presents a significant gap, with several orders of magnitude difference in the number of qubits required. To explore the feasibility of early FTQC systems with around $10^3$ to $10^4$ qubits, recent work has investigated the compilation of continuous rotation gates that are not protected by QEC and are executed via ancilla state injection without relying on magic state distillation.

The rotation is implemented by injecting the state $m_\theta \equiv R_z(\theta)\ket{+} = \frac{1}{\sqrt{2}}(e^{-i\theta/2}\ket{0}+e^{i\theta/2}\ket{1})$, where $\theta$ can be any angle. However, this approach introduces additional compilation challenges because the implementation is inherently non‑deterministic. Specifically, the outcome may correspond to either $R_z(\theta)\ket{\psi}$ or $R_z(-\theta)\ket{\psi}$ with equal probability. When the undesired rotation occurs, an additional corrective rotation $R_z(2\theta)$ must be applied to the output state. The correction loop repeats until the expected result is reached, requiring \textit{real-time compilation} to ensure correct execution.

\begin{figure}[t]
    \centering
    \includegraphics[width=1\linewidth]{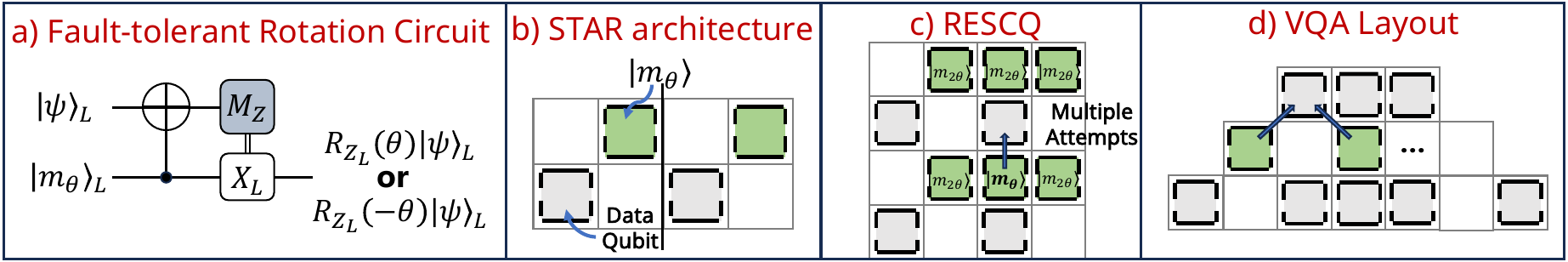}
    \caption{(a) Fault-tolerant rotation circuit (b) STAR Architecture~\cite{akahoshi2024partially} (c) RESCQ~\cite{sethi2025rescq} (d) VQA Layout~\cite{dangwal2025variational}.}
    \label{fig:eft_compiler_summary}
\end{figure}

In the proposed STAR architecture~\cite{akahoshi2024partially} (see Fig.~\ref{fig:eft_compiler_summary}b), a minimal structure is used to illustrate the technique. The layout is a $2\times2$ grid in which one tile hosts the data block, one tile hosts the resource state factory, and the remaining two tiles are ancilla used for routing. The protocol repeatedly attempts to prepare the resource state in the upper-right ancilla of the atomic patch. If preparation succeeds, state injection is performed. Otherwise, the preparation step is retried until success. For applications of the STAR architecture,~\cite{akahoshi2025compilation} focus on quantum simulation of the 2D Hubbard model. They insert fermionic SWAPs to modify the Jordan–Wigner (JW) mapping, placing interacting spin‑orbital pairs adjacently, which enables parallel protocol execution and minimizes overall runtime overhead.

However, these two compilers are tailored to a fixed pattern in which a single data patch is responsible for state preparation, missing opportunities for parallel preparation. In RESCQ~\cite{sethi2025rescq} (see Fig.~\ref{fig:eft_compiler_summary}c), multiple preparations are attempted in parallel. It also uses a dynamically recomputed minimum spanning tree for routing, yielding an average 2$\times$ reduction in time overhead compared with the STAR architecture. Subsequently,~\cite{dangwal2025variational} (see Fig.~\ref{fig:eft_compiler_summary}d) introduced a compiler that designs variational quantum algorithm (VQA) tailored data layouts to exploit the algorithmic parallelism inherent in VQAs. It further introduces a two-ancilla patch-shuttling scheme that prepares the $2\theta$ resource while consuming the $\theta$ resource. If gate teleportation implements $R_z(-\theta)$, the second resource supplies the $R_z(2\theta)$ correction; heralded preparation failures instead trigger preparation retries. Recent work has also extended STAR architecture through code-hardware co-design. Transversal STAR combines small-angle resource-state injection with transversal Clifford operations on neutral-atom hardware~\cite{ismail2026transversal}. High-rate STAR further uses bicycle-chain codes to support parallel state injection across multiple logical qubits, reducing the space overhead of early fault-tolerant quantum simulation~\cite{ismail2026fastparallelhighratestar}.


\section{Physical-level QEC Implementation}
\label{sec:physical_level_qec_implementation}

After logical operations have been scheduled, they must be lowered to executable operations on quantum hardware. This lowering proceeds in two stages. First, the compiler schedules the syndrome-extraction circuit. This step orders the stabilizer measurements and assigns ancillas to them, and it sets the timing of each QEC cycle. The schedule must not weaken the fault-tolerance properties of the code. This stage is largely independent of a particular hardware platform. Second, the scheduled circuit is realized on a concrete device under hardware-specific constraints. Superconducting processors are governed by fixed nearest-neighbor connectivity, trapped-ion processors require ion shuttling and chain reordering, and neutral-atom arrays expose atom transport and reconfiguration. Table~\ref{tab:physical_qec_system_comparison} summarizes representative physical-level QEC implementation works across these code families and hardware platforms. This section first reviews code-level syndrome-extraction scheduling and then examines how compilers address the platform-specific constraints of these three mainstream hardware technologies.

\subsection{Code-Level Syndrome-Extraction Scheduling}
\label{subsec:syndrome_measurement_scheduling_compiler_design}

\textbf{Task characterization.} Syndrome-extraction scheduling chooses a repeated measurement cycle for a stabilizer code or a CSS parity-check specification. A basic instance specifies the checks to be measured. It also fixes a measurement gadget, or a set of admissible gadgets, together with the available ancilla resources. The scheduler chooses the temporal order of elementary check interactions and the parallel tick structure of one QEC round. When ancillas are scarce, it also determines how ancillas are reused across checks. Its main artifact is a syndrome-extraction circuit $C_{\mathrm{SEC}}$, or a round schedule $\tau$, whose repetition produces the syndrome stream consumed by the decoder. Variants of the problem differ in what they add. Some add connectivity or timing constraints. Others fix a native measurement primitive, or bring in a noise model and a decoder so that schedules can be scored. This layer sits below logical lattice-surgery compilation, which schedules encoded patches and logical Pauli measurements, and above hardware realization, which places and routes the selected circuit on a concrete device.

The optimization criteria combine fault-tolerance and implementation goals. Low two-qubit depth and limited idling reduce exposure to circuit noise. The order of gates within a stabilizer measurement controls how ancilla and data faults propagate into correlated data errors. Surface-code schedules are often judged by hook-error orientation and circuit distance. qLDPC schedules are also judged by effective distance, harmful low-weight residual failures, and logical performance under circuit-level decoding. A compact abstraction is
\begin{equation}
(\mathcal{S},\mathcal{A},\Gamma,\mathcal{N},D)
  \mapsto
  (\tau,C_{\mathrm{SEC}},E),
\end{equation}
where $\mathcal{S}$ is the check specification, $\mathcal{A}$ describes ancilla resources and measurement gadgets, and $\Gamma$ denotes optional connectivity or timing constraints. The noise model $\mathcal{N}$ and decoder $D$ enter when the scheduler is evaluated or guided by circuit-level performance. The evaluation record $E$ depends on the formulation. Search and SMT-based tools may report depth or optimality certificates. Fault-tolerance analyses may report circuit distance, residual-error metrics, logical failure rate, or threshold. Resource-constrained formulations may also report physical-qubit count or hardware-realization cost. Detector construction and DEM generation sit downstream of the scheduler. A few schedulers optimize against them directly, and we note those cases below.

\textbf{Manual schedules and schedule primitives.} Before schedule search became common, most syndrome-extraction cycles were hand-written templates. A simple baseline is to order the check interactions by qubit or stabilizer indices, as in early experimental QEC demonstrations~\cite{ryananderson2021realtime,zhao2022realization}. A second family of baselines minimizes the total circuit depth, thereby reducing idle exposure and the time over which data and ancilla qubits can accumulate physical errors~\cite{kang2025quits,menon2026magic}. At a fixed or comparable depth, the interaction order also determines how correlated faults are oriented.

For the rotated surface code, local plaquette geometry permits repeated $N$- or $Z$-shaped schedules whose purpose is to orient hook errors away from logical operators. Recent diagonal surface-code schedules propose a globally uniform orientation that simplifies hook-error-safe scheduling under suitable timing assumptions~\cite{kishony2026surfacecodeoffthehookdiagonal}. Similarly, the bivariate-bicycle quantum-memory proposal of Bravyi et al.~\cite{bravyi2024high} uses a code-specific low-depth syndrome cycle on a degree-six connectivity graph, providing an important baseline for qLDPC memory implementations. Other schedule primitives relax or analyze the scheduling problem for specific code families. Tangling schedules use deliberately nonstandard extraction patterns to ease connectivity requirements~\cite{geher2024tangling}, while distance-preserving measurement results for hypergraph-product codes give correctness criteria for when low-depth measurement circuits preserve effective distance~\cite{manes2025distance}. Overall, manual schedules and schedule primitives define the constraints that syndrome-extraction compilers must preserve, including the orientation of correlated faults, achievable depth or connectivity targets, and preservation of the intended code distance. They are baselines and design rules that a schedule compiler has to respect.

\textbf{Automated syndrome-measurement schedulers and optimizers.} Recent work has begun to replace hand-designed extraction cycles with compiler-style schedule synthesis and circuit optimization. AlphaSyndrome~\cite{liu2026alphasyndrome} targets general commuting stabilizer codes, using Monte Carlo tree search guided by code structure and decoder feedback to explore both ordering and parallelism. PropHunt~\cite{viszlai2026prophunt} optimizes syndrome-measurement circuits for CSS codes at the level of individual CNOT orderings and schedules. Rather than applying generic gate-count or depth optimizations, it analyzes circuit-level fault propagation and decoding ambiguity, then iteratively applies verified reordering or rescheduling changes that improve the resulting syndrome-measurement circuit. For qLDPC deployment, Auto-Stabilizer-Check (ASC)~\cite{zhang2026optimal} formulates syndrome-extraction circuit construction as an SMT problem over the legality and timing constraints of stabilizer measurement. By exploiting the sparsity of qLDPC parity-check matrices and the commutativity of stabilizer-measurement subroutines, ASC returns depth-optimal circuits when the SMT instance is satisfiable, and near-optimal circuits when the solver times out. Complementing this solver-based approach, \cite{tan2026syndrome} exploits the algebraic product structure of practical CSS/qLDPC code families to construct near-optimal syndrome-extraction schedules. It focuses on quasi-Abelian lifted-product codes, including bivariate-bicycle, trivariate-tricycle, and hypergraph-product codes, and provides a constructive scheduling strategy whose CNOT depth is often optimal and otherwise within one layer of a fundamental lower bound.

A complementary line of work studies syndrome-extraction design through structured circuit constructions or explicit resource constraints. Left--right syndrome-extraction circuits provide a low-depth construction that staggers $X$- and $Z$-check measurements without interleaving their gates, generalizing earlier left--right constructions to arbitrary CSS codes~\cite{strikis2026high}. Given parity-check matrices, the framework analyzes and constructs syndrome-extraction circuits using residual-error metrics that capture idling, effective distance, and harmful low-weight failure mechanisms. Few-ancilla syndrome scheduling studies a resource-constrained variant in which fewer ancillas than stabilizers are available, so ancillas must be reused across checks and the scheduler must trade syndrome-cycle depth against total physical-qubit count and logical performance~\cite{sato2025scheduling}. Together with search- and solver-based optimizers, these works show that syndrome-measurement scheduling is emerging as a distinct compiler layer. It takes a code as input and produces a circuit as output, and its schedules are scored by simulation.

\subsection{Hardware-Aware QEC Realization on Superconducting Circuits}

\begin{table*}[t]
\centering
\begingroup
\caption{Representative physical-level QEC implementation works.}
\label{tab:physical_qec_system_comparison}
\definecolor{qecTableHeader}{RGB}{235,235,235}
\definecolor{qecTableRule}{RGB}{170,170,170}
\definecolor{qecSuperconducting}{RGB}{215,230,251}
\definecolor{qecTrappedIon}{RGB}{232,225,247}
\definecolor{qecNeutralAtom}{RGB}{225,241,226}
\renewcommand{\arraystretch}{1.18}
\setlength{\tabcolsep}{6pt}
\setlength{\aboverulesep}{0pt}
\setlength{\belowrulesep}{0pt}
\resizebox{\linewidth}{!}{%
\begin{tabular}{c|l|c|c|c|c}
\arrayrulecolor{black}
\toprule
\arrayrulecolor{qecTableRule}
\rowcolor{qecTableHeader}
\textbf{Target Hardware} & \multicolumn{1}{c|}{\textbf{Reference}} & \textbf{Year} & \textbf{Venue} & \textbf{QEC Code} & \textbf{Code Availability} \\[3pt]
\hline
\rowcolor{qecSuperconducting!55}
\cellcolor{qecSuperconducting} & Surf-Stitch~\cite{Wu2022synthesis} & 2022 & ISCA & Surface Code & - \\
\rowcolor{qecSuperconducting!20}
\cellcolor{qecSuperconducting} & QECC-Synth~\cite{yin2024qeccsynth} & 2025 & ASPLOS & Stabilizer Code & \href{https://zenodo.org/records/14061096}{Link} \\
\rowcolor{qecSuperconducting!55}
\cellcolor{qecSuperconducting} & \citet{vittal2024flag} & 2024 & MICRO & qLDPC Code & \href{https://zenodo.org/records/13820399}{Link} \\
\rowcolor{qecSuperconducting!20}
\cellcolor{qecSuperconducting} & Louvre~\cite{zhou2025louvre} & 2025 & arXiv & qLDPC Code & - \\
\rowcolor{qecSuperconducting!55}
\cellcolor{qecSuperconducting} & \citet{zhao2025simple} & 2025 & arXiv & Surface/qLDPC Code & - \\
\rowcolor{qecSuperconducting!20}
\cellcolor{qecSuperconducting} & HAL~\cite{mathews2026placing} & 2026 & npj QI & qLDPC Code & - \\
\rowcolor{qecSuperconducting!55}
\cellcolor{qecSuperconducting} & CaliQEC~\cite{fang2024caliscalpelinsitufinegrainedqubit} & 2025 & ISCA & Surface Code & - \\
\rowcolor{qecSuperconducting!20}
\cellcolor{qecSuperconducting}\multirow{-8}{*}{\textbf{Superconducting}} & Youtiao~\cite{tian2025youtiao} & 2025 & MICRO & Surface Code & - \\
\arrayrulecolor{black!60}\hline
\arrayrulecolor{qecTableRule}
\rowcolor{qecTrappedIon!55}
\cellcolor{qecTrappedIon} & TISCC~\cite{leblond2023tiscc} & 2023 & SC-W & Surface Code & \href{https://github.com/ORNL-QCI/TISCC}{Link} \\
\rowcolor{qecTrappedIon!20}
\cellcolor{qecTrappedIon} & Jones and Murali~\cite{jones2026architecting} & 2026 & ASPLOS & Surface Code & - \\
\rowcolor{qecTrappedIon!55}
\cellcolor{qecTrappedIon} & iSwitch~\cite{yin2025flexion} & 2026 & ASPLOS & Surface Code & - \\
\rowcolor{qecTrappedIon!20}
\cellcolor{qecTrappedIon} & Lee et al.~\cite{lee2026ion} & 2026 & PRA & Color Code & - \\
\rowcolor{qecTrappedIon!55}
\cellcolor{qecTrappedIon} & Moveless~\cite{khan2025movelessminimizingoverheadqccds} & 2025 & QCE & Stabilizer Code & \href{https://github.com/sahilkhan123/Moveless}{Link} \\
\rowcolor{qecTrappedIon!20}
\cellcolor{qecTrappedIon} & Cyclone~\cite{khan2025cyclone} & 2026 & HPCA & Stabilizer Code & - \\
\rowcolor{qecTrappedIon!55}
\cellcolor{qecTrappedIon}\multirow{-7}{*}{\textbf{Trapped-Ion}} & Walking Cat~\cite{tripier2026faulttolerantquantumcomputingtrapped} & 2026 & arXiv & qLDPC Code & - \\
\arrayrulecolor{black!60}\hline
\arrayrulecolor{qecTableRule}
\rowcolor{qecNeutralAtom!55}
\cellcolor{qecNeutralAtom} & Bluvstein et al.~\cite{bluvstein2024logical} & 2024 & Nature & Surface/Color Code & - \\
\rowcolor{qecNeutralAtom!20}
\cellcolor{qecNeutralAtom} & Viszlai et al.~\cite{viszlai2023architecture} & 2025 & HPCA & Surface Code & - \\
\rowcolor{qecNeutralAtom!55}
\cellcolor{qecNeutralAtom} & qSIEVE~\cite{viszlai2023matching} & 2026 & TQC & qLDPC Code & - \\
\rowcolor{qecNeutralAtom!20}
\cellcolor{qecNeutralAtom} & Pecorari et al.~\cite{pecorari2025high} & 2025 & Nat. Commun. & qLDPC Code & - \\
\rowcolor{qecNeutralAtom!55}
\cellcolor{qecNeutralAtom} & ConiQ~\cite{liu2025coniqenablingconcatenatedquantum} & 2025 & QCE & Concatenated Code & - \\
\rowcolor{qecNeutralAtom!20}
\cellcolor{qecNeutralAtom} & Stade et al.~\cite{stade2024abstract} & 2024 & QCE & CSS Code & - \\
\rowcolor{qecNeutralAtom!55}
\cellcolor{qecNeutralAtom} & Perrin et al.~\cite{perrin2025atomloss} & 2025 & Quantum & Surface Code & - \\
\rowcolor{qecNeutralAtom!20}
\cellcolor{qecNeutralAtom}\multirow{-8}{*}{\textbf{Neutral Atom}} & Pecorari et al.~\cite{pecorari2025lowdepth} & 2025 & PRL & Surface Code & - \\
\arrayrulecolor{black}
\bottomrule
\end{tabular}
}
\endgroup
\end{table*}

\subsubsection{\textbf{Challenges}} 

Superconducting circuits are among the leading platforms for near-term QEC because they provide fast gates, fast measurement and reset, and lithographically defined two-dimensional device layouts. Their main compiler challenge is that the physical coupling graph is fixed, sparse, and strongly device dependent. Each vendor fixes its own geometry, such as IBM's heavy-hex lattice~\cite{ibm2021heavyhex}, Google's Willow-class square grid~\cite{google2024willowSpec}, and Rigetti's Ankaa-class square lattice~\cite{aws2024ankaa2}. This creates a \textit{connectivity gap} between the interaction graph required by a syndrome-extraction circuit and the couplers supplied by the chip. Even surface-code implementations assume repeated local degree-four stabilizer measurements, while high-rate qLDPC memories, such as bivariate-bicycle (BB) codes, can require degree-six connectivity or a two-layer planar embedding~\cite{bravyi2024high}.

When the stabilizer-interaction graph does not match the device coupling graph, the compiler must introduce an indirect realization, such as routing through ancilla qubits or redesigning the measurement circuit around the available hardware. This closes the connectivity gap at the cost of changing the syndrome-cycle depth, error exposure, and fault-propagation structure. Superconducting QEC compilation is therefore a joint problem. Layout and routing are chosen together with the coupler budget and the calibration constraints, and the whole is judged by circuit-level logical performance.

\subsubsection{\textbf{Bridge and Flag Based Realization}}

One direct response to the connectivity gap is to preserve the target stabilizer measurements while adding ancilla qubits. Lao and Almudever~\cite{lao2020fault} combine flag fault tolerance with bridge qubits to implement stabilizer measurements with restricted connectivity. In this approach, flag qubits help detect dangerous fault propagation, while bridge qubits mediate interactions that are not directly supported by the device graph. Surf-Stitch~\cite{Wu2022synthesis} specializes bridge-based synthesis to surface-code implementations on superconducting processors. Its pipeline addresses three coupled tasks: selecting a geometric allocation of data qubits, constructing bridge structures that connect the data qubits participating in each stabilizer measurement, and scheduling the resulting syndrome-extraction operations. Bridge qubits sit inside the data-qubit regions, and compatible bridge paths are merged, which shortens each error-detection cycle on sparse hardware. QECC-Synth~\cite{yin2024qeccsynth} later generalizes this bridge-based view beyond a single surface-code layout. Instead of assuming a fixed bridge construction, it explores different ancilla-bridge choices and circuit transformations, formalizing the design space through a two-stage MaxSAT formulation and scalable heuristics. Compared with Surf-Stitch's specialized mapping, this makes QECC-Synth closer to a reusable layout synthesizer for diverse QEC codes and hardware graphs.

For qLDPC codes, Flag-Proxy Networks (FPNs)~\cite{vittal2024flag} combine architecture design, scheduling, and decoding. FPNs use flag and proxy qubits to reduce the native connectivity required by hyperbolic surface and hyperbolic color codes, pair this hardware abstraction with a greedy syndrome-extraction scheduler, and exploit flag information in decoding. This changes both the hardware abstraction and the decoding interface so that higher-connectivity qLDPC checks can be implemented on lower-degree superconducting layouts.

\subsubsection{\textbf{Expanded Instruction Sets}} 
A second class of approaches reduces static connectivity requirements by redesigning the syndrome-extraction circuit or by inserting routing operations into the extraction cycle. These methods do not remove the need for local two-qubit couplings when a gate is executed. Instead, they avoid requiring the original data-ancilla interaction graph to be embedded as a fixed subgraph of the device. Time-dynamic circuits and expanded-instruction-set approaches use operations such as SWAP or iSWAP, together with possible role exchange or gate reordering, so that the same syndrome information can be extracted with fewer static couplers or lower average coupling degree.

McEwen, Bacon, and Gidney~\cite{mcewen2023relaxing} introduce time-dynamic surface-code circuits designed directly in spacetime using detecting regions. Their constructions embed surface-code circuits on hexagonal connectivity, use native iSWAP-type gates, and swap data and measurement roles during execution. Louvre~\cite{zhou2025louvre} applies a related idea to generalized-bicycle and bivariate-bicycle codes. BB-code syndrome extraction naturally requires degree-six connectivity, which is challenging for superconducting hardware. Louvre also expands the available instruction set with iSWAP/SWAP-style routing operations inside the syndrome-extraction cycle. Its Louvre-7 variant reduces the average degree for BB codes while preserving the original syndrome-extraction depth, whereas Louvre-8 further reduces connectivity with a modest depth increase. Thus, Louvre trades temporal routing and native-gate support for a lower static coupler degree. Other routing-oriented constructions make this tradeoff explicit. Zhao, Yan, and Ni~\cite{zhao2025simple} propose a universal routing strategy that reduces long-range connections by routing information through short data--ancilla loops, at the cost of increased syndrome-extraction depth. This kind of approach is particularly relevant to superconducting processors, where high-fidelity local gates may be easier to engineer than dense long-range coupling.

\subsubsection{\textbf{Code-Circuit Co-Design for Low-Degree Layouts}}
Some works change the code or the measurement circuit itself to fit the device. Morphing circuits~\cite{shaw2025lowering} apply a code-specific parity-check circuit design principle to BB codes, reducing the required qubit connectivity degree from six to five while maintaining competitive circuit-level performance. Directional codes~\cite{geher2025directional} use native iSWAP gates to realize qLDPC syndrome extraction on square-grid and, in some cases, hexagonal-grid connectivity. Their finite-code constructions use periodic boundaries, so planar hardware must still accommodate connections across those boundaries. These works represent an important code--circuit--hardware co-design direction: instead of treating the QEC code and hardware topology as fixed inputs, they reshape the code or extraction circuit so that fault-tolerant syndrome measurement is compatible with low-degree superconducting layouts from the outset.

\subsubsection{\textbf{Multi-Layer and Long-Range-Coupler Layouts}}
A more hardware-forward solution is to add routable physical connectivity to the processor and ask the compiler to place and route the code's connectivity graph under fabrication constraints. Tremblay, Delfosse, and Beverland~\cite{tremblay2022constant} propose thin planar connectivity for qLDPC codes by decomposing Tanner graphs into a small number of planar layers. This avoids uncontrolled edge crossings while preserving bounded-depth stabilizer-measurement circuits. HAL~\cite{mathews2026placing} turns this idea into a placement-and-routing workflow for multi-layer superconducting hardware. HAL starts from the connectivity graph implied by the parity-check matrices. It extracts a large planar subgraph for the qubit tier and places that subgraph on a grid. The remaining edges are routed through higher coupler layers, where fabrication limits such as maximum coupler length and via count decide what can be built. HAL therefore differs from syndrome-extraction schedulers. Its main role is to evaluate whether a chosen code connectivity graph can be physically embedded in a realistic multi-layer superconducting stack, rather than to choose a CNOT ordering.

\subsubsection{\textbf{System-Level Constraints Beyond Connectivity}}
Beyond static connectivity, superconducting QEC implementation also depends on the scalability and stability of the control stack. A large patch needs thousands of control and readout lines that fire in lockstep. At that scale the wiring itself becomes a constraint, and so do calibration drift and crosstalk between neighboring lines. CaliQEC~\cite{fang2024caliscalpelinsitufinegrainedqubit} addresses the calibration aspect of this problem for surface-code computation through in-situ calibration. It uses code deformation to cut the qubits under calibration out of the patch while the rest keeps computing, and picks the calibration schedule from device characterization data. YOUTIAO~\cite{tian2025youtiao} attacks the wiring cost of large superconducting processors. It proposes hybrid multiplexing with dynamic qubit grouping to reduce cryostat-level coaxial wiring complexity while preserving the ability to address qubits required by quantum workloads. Although YOUTIAO is not itself a QEC-code compiler, it is relevant to superconducting QEC realization because scalable error correction requires repeated, highly parallel syndrome-extraction cycles whose control signals must be delivered through a feasible cryogenic wiring stack.

\subsection{Hardware-Aware QEC Realization on Trapped-Ion}

\subsubsection{\textbf{Challenges.}} Trapped‑ion (TI) systems are leading candidates for quantum processors, owing to their long coherence times and high‑fidelity operations. To scale TI hardware, the Quantum Charge‑Coupled Device (QCCD) has emerged as a prominent modular architecture. Unlike superconducting devices with fixed local couplings, QCCD systems support ion shuttling and all‑to‑all connectivity within each trap. These QCCD‑specific primitives must be explicitly scheduled by the compiler to enable efficient implementations of QEC codes. Below, we detail recent advances in QCCD architectural design and the implementation of QEC codes on such hardware.

\subsubsection{\textbf{Surface Code Implementation}}
Prior QCCD architecture and backend-compilation studies established the transport-aware abstraction for segmented ion-trap processors, where logical interactions are mediated by explicit ion transport and chain reordering~\cite{murali2020architecting_iontrap,saki2022muzzle}. Building on this abstraction, TISCC provides a surface-code-oriented lowering flow in which logical patch operations, rounds of error correction, transversal operations, and lattice-surgery primitives are instantiated on a grid-like QCCD architecture and translated into native trapped-ion gate and transport operations with explicit resource estimates~\cite{leblond2023tiscc}. More recent surface-code architecture studies use topology-aware compilation to evaluate QCCD design choices, showing that the hardware layout itself must be co-designed with the target QEC cycle~\cite{jones2026architecting}. Together, these works frame trapped-ion surface-code implementation as a hardware-timed lowering problem, where logical patch scheduling and syndrome extraction must be coordinated with transport-mediated connectivity.

iSwitch/Flexion extends this surface-code implementation perspective to an early fault-tolerant regime in which protection is applied selectively rather than maintaining all program qubits as encoded logical qubits throughout the computation~\cite{yin2025flexion}. Instead of compiling every operation into an always-encoded surface-code execution model, it exploits the high fidelity of trapped-ion single-qubit gates by executing selected one-qubit operations on bare physical qubits, while converting qubits into surface-code logical patches for protected two-qubit interactions. To support this bare-logical hybrid model, iSwitch introduces an in-situ conversion protocol between bare and logical representations, a hybrid instruction set for bare, logical, and boundary regions, and a compiler that allocates logical protection over time under capacity and routing constraints. This reframes surface-code lowering as a dynamic protection problem, where the compiler decides how to implement encoded operations and when encoding overhead is justified by the reliability and interaction structure of the workload.

\subsubsection{\textbf{Color-code implementation.}} 
Lee et al.~\cite{lee2026ion} study a color-code-oriented ion-trap chip architecture that separates horizontal regions for transversal logical gates from vertical regions for non-transversal gates and syndrome extraction. The work provides a dedicated transpiler, scheduler, and error analyzer for two-dimensional color-code execution, with the scheduler inserting swap and shuttling operations to satisfy chip constraints. From a compiler perspective, it complements surface-code QCCD flows by showing that the code family can shape the chip layout, region assignment, motion policy, and placement of periodic QEC rounds.

\subsubsection{\textbf{Stabilizer Code Implementation}}
General stabilizer and CSS codes induce interaction patterns that are not necessarily aligned with surface-code tiles or local lattice-surgery layouts. On QCCD processors, syndrome extraction therefore has to be scheduled around ion movement. Moveless exploits stabilizer-code structure to reduce excess shuttling: because stabilizer checks commute and syndrome ancillas are often interchangeable, the compiler can reorder both the sequence of stabilizer measurements and the gates within each check according to the current physical placement~\cite{khan2025movelessminimizingoverheadqccds}. Its move-ancilla-only policy keeps data ions comparatively stable while using ancilla motion to realize the required interactions, and its round-pairing strategy restores favorable mappings across repeated QEC cycles. Cyclone takes a complementary architecture/compiler co-design approach: it observes that QCCD layouts optimized around surface-code memories may not generalize to non-topological CSS codes, whose stabilizer graphs can require high inter-trap parallelism and nonlocal interaction patterns~\cite{khan2025cyclone}. Cyclone therefore introduces a roadblock-free circular transport architecture and a matching compilation strategy designed to expose parallelism in broad stabilizer workloads. Walking Cat goes further and sketches a complete trapped-ion FTQC blueprint around LDPC codes. It covers the stack from the compiler down to the decoder and the microarchitecture~\cite{tripier2026faulttolerantquantumcomputingtrapped}.

\subsection{Hardware-Aware QEC Realization on Neutral Atom Arrays}

\textbf{Challenges and compiler abstractions.}
Neutral-atom arrays have emerged as a promising substrate for fault-tolerant quantum computation because they combine large qubit arrays, reconfigurable connectivity, and hardware support for spatially separated execution regions. In the context of physical-level QEC implementation, these features define a distinct compilation model in which movable atomic qubits, Rydberg-mediated entangling operations, and dedicated regions must be coordinated by the compiler. Existing neutral-atom compilers provide the backend mechanisms for legalizing these operations at the circuit level~\cite{tan2022qubit_atom,tan2023compiling_atom,tan2024compilation,wang2023q,wang2024atomique,nottingham2024circuit,lin2024reuse,ruan2024powermove,huang2024zap}. For QEC workloads, however, these mechanisms must be lifted from physical-gate routing to QEC-level objects. Recent experiments make this concrete. Reconfigurable logical processors with mid-circuit measurement and loss handling show that the physical layer has to reason about reconfiguration and error information at the same time~\cite{Bluvstein2022Aquantum,bluvstein2024logical,reichardt2024fault,bluvstein2026faulttolerant,chow2024circuit}.

\textbf{Surface-code and logical-processor implementation.}
At the surface-code level, reconfigurable atom arrays can translate physical reconfigurability into improved logical connectivity. Viszlai et al.~\cite{viszlai2023architecture} propose interleaved groups of surface-code logical qubits, enabling efficient transversal CNOT gates within each group and interleaved lattice-surgery mechanisms between groups. This design uses neutral-atom reconfigurability to reduce spacetime overhead and enlarge the logical routing space, with the preferred mechanism depending on circuit scale and routing distance. Complementary proposals such as long-range-enhanced surface codes introduce a limited number of nonlocal parity checks, interpolating between local surface-code structure and higher-rate LDPC-like behavior while remaining compatible with platforms such as neutral atoms in mobile optical tweezers~\cite{hong2024long}. Recent work on transversal logical Clifford gates for rotated surface codes similarly exploits neutral-atom reconfigurability: logical $H$ can be implemented through patch rotation, while logical $S$ can be realized by embedding a fold-transversal operation into the syndrome-extraction cycle~\cite{chen2024transversallogicalcliffordgates}. On neutral atoms the compiler can choose its mechanism. Interleaving and transversal gates exploit atom movement directly. A few nonlocal checks or lattice surgery keep the layout closer to a static grid. Which one wins depends on the reconfiguration hardware and the workload.

\textbf{High-rate and qLDPC-code implementation.}
For qLDPC and bivariate-bicycle-style codes, the central physical-level challenge is implementing nonlocal checks without losing the overhead advantages of the code. qSIEVE~\cite{viszlai2023matching} addresses this problem by co-designing generalized-bicycle and related qLDPC memories with systolic movement in atom arrays. Xu et al.~\cite{xu2023constantoverhead} provide an architecture-level argument that product structure in qLDPC codes can be matched to atom rearrangement to approach constant-overhead fault tolerance in practical regimes. A complementary direction studies high-rate quantum LDPC codes for long-range-connected neutral-atom registers, where limited long-range interactions are used to implement code families that can outperform surface codes in certain circuit-level regimes~\cite{pecorari2025high}. Zhao et al.~\cite{zhao2026ultrahighrate} push toward ultra-high-rate codes for neutral atoms. Their finding is that the code cannot be chosen first and mapped later: its syndrome circuit and the atom rearrangement it needs have to be designed with it. Because erasure-biased noise is especially natural in neutral-atom systems, qLDPC designs under erasure-biased circuit-level models also provide a relevant bridge between physical noise engineering and code-family selection~\cite{sahay2023highthreshold,pecorari2025quantum}.

\textbf{Concatenated codes and logical-operation scheduling.}
Concatenated codes induce a different compiler structure because the implementation must preserve hierarchical code organization while still exploiting neutral-atom parallelism and mobility. ConiQ~\cite{liu2025coniqenablingconcatenatedquantum} addresses this setting through a dedicated intermediate representation for concatenated QEC on neutral atom arrays, exposing many-hypercube structure and code automorphisms during legalization. At the logical-operation level, zoned neutral-atom architectures require compilers to route logical entangling operations and prepare logical arrays under storage, entangling, and readout-zone constraints. Stade et al.~\cite{stade2024abstract} model logical entangling-gate routing on zoned architectures, and their subsequent work formulates logical-array state preparation as a solver-based scheduling problem~\cite{stade2025optimalstatepreparation}. Universal fault-tolerant operation further requires efficient non-Clifford resources. Wang et al.~\cite{wang2024efficient} study how reconfigurable atom arrays can support non-Clifford logical operations through magic-state and concatenation-based mechanisms that exploit nonlocal connectivity, collective mobility, parallel gates, and native multi-controlled operations. These works suggest a compiler stack with four layers. At the top, a QEC-aware logical IR selects code-specific operations. A layout pass then maps logical blocks to zones. Below it, a QEC scheduler interleaves syndrome extraction with the logical operations. At the bottom, a neutral-atom backend legalizes the movement and the gates.

\textbf{Atom Loss-aware runtime support.}
Neutral-atom QEC compilation must also expose information about loss and leakage to the runtime decoder. A major motivation is that several neutral-atom proposals and experiments convert dominant physical error mechanisms into detectable erasures, allowing error locations to become decoder-visible information~\cite{wu2022erasureconversion,sahay2023highthreshold,chow2024circuit,reichardt2024fault}. For surface codes, Kobayashi and Nagayama~\cite{kobayashi2024erasure} study erasure-tolerance schemes based on code deformation and spare-array recovery, addressing the fact that erasures can accumulate over repeated cycles. Perrin et al.~\cite{perrin2025atomloss} introduce loss-detection-unit protocols and adaptive decoding for atom loss, while later work treats correlated atom loss as structured decoding information through loss graphs and dynamically updated loss probabilities~\cite{perrin2026correlated}. Native Rydberg multi-qubit gates may also reduce stabilizer-measurement depth, but their correlated-error behavior must be incorporated into circuit-level fault-tolerance analysis. Pecorari et al.~\cite{pecorari2025lowdepth} preserve circuit-level fault tolerance in unrotated surface-code readout using a specific pairing of three-qubit Rydberg gates. A neutral-atom QEC compiler therefore has more to optimize than depth and gate count. Cycle latency and decoder throughput enter the cost directly. So do erasure information and atom reloading, and so does the way readout circuits interact with correlated hardware errors.

\section{Decoding in the Classical Control Stack}
\label{sec:decoder}

In an FTQC stack, the decoder is the classical inference between repeated syndrome extraction and the runtime actions that maintain the logical state, as illustrated in Figure~\ref{fig:main_figure}. Its input is a time-ordered measurement record together with circuit metadata such as detector definitions, logical observables, and fault probabilities. Its output is a recovery class that the runtime may realize as a frame update, a feedforward decision, or a correction committed at a later boundary.

Section~\ref{sec:decoding_problem} starts from ideal syndrome extraction, then extends the problem to circuit-level noise through the detector error model. Section~\ref{sec:decoding_algorithms} surveys matching, search- and hypergraph-based decoding, Union-Find, belief propagation, and neural decoders together with their implementation costs. Section~\ref{sec:ftqc_decoding_systems} then turns from algorithms to decoding systems. It covers frame tracking and window decoding, and then the side information a decoder can use at runtime, from erasure flags to soft readout.

Decoder capability also shapes the time overhead of logical operations. Conventional distance-$d$ surface-code lattice surgery typically uses $O(d)$ noisy syndrome-extraction rounds per logical parity measurement~\cite{horsman2012surface}. Transversal algorithmic fault tolerance (AFT) instead combines suitable transversal operations with correlated decoding to support universal computation with $O(1)$ extraction rounds per logical operation, given magic-state inputs and feedforward~\cite{zhou2025lowoverhead}. Architecture studies already incorporate AFT into layout synthesis and application-level resource estimates~\cite{zhou2025transversalresources}, but further validation and optimization using actual decoder implementations within practical compilation systems are still needed.

\subsection{From Syndromes to Detector Error Models}
\label{sec:decoding_problem}

QEC relies on the extraction of error syndromes to identify and correct noise without destroying the encoded quantum information. Unlike classical codes where data bits can be read directly, measuring data qubits collapses their state. Therefore, error information is extracted indirectly by measuring the generators of the stabilizer group $\mathcal{S} = \langle S_1, \dots, S_m \rangle$.

\subsubsection{\textbf{Ideal Syndrome Extraction}}

We first consider the \textit{code capacity} noise model, which assumes that errors occur only on the data qubits, while syndrome measurements are perfect and instantaneous. In this model, let $\mathcal{P}_n$ denote the $n$-qubit Pauli group. For a Pauli error $E \in \mathcal{P}_n$, the \textit{error syndrome} $\mathbf{s} \in \mathbb{F}_2^m$ is the vector of measurement outcomes of the stabilizer generators $\{S_i\}$. The $i$-th component $s_i$ corresponds to the eigenvalue of the measured generator, determined by the commutation relation:
\begin{equation}
    s_i(E) = \begin{cases} 0 & \text{if } [E, S_i] = 0 \quad (\text{commute}) \\ 1 & \text{if } \{E, S_i\} = 0 \quad (\text{anticommute}). \end{cases}
\end{equation}
In the symplectic representation used for CSS codes, let $H$ be the binary parity-check matrix of the code, and let $\mathbf{e}$ be the binary vector representation of the error $E$. The syndrome mapping is linear:
\begin{equation}
    \mathbf{s}^T = H \mathbf{e}^T.
\end{equation}

A unique challenge in QEC is that multiple distinct physical errors can be logically equivalent. A stabilizer code is degenerate when distinct low-weight errors differ by a non-trivial stabilizer. In particular, if $S\in\mathcal{S}$ is a stabilizer, then $E$ and $E' = E S$ produce the same syndrome and act identically on the code space up to the stabilizer $S$. They should therefore be treated as the same decoding outcome, even though they are different physical error patterns. This equivalence complicates decoders that attempt to identify a specific physical error, such as belief propagation, because probability mass
can be split across many equivalent representatives of the same logical coset.

Because of degeneracy, decoding aims to identify a recovery operation that restores the logical state rather than the exact physical error that occurred. Maximum-likelihood QEC decoding therefore seeks the most probable error equivalence class (coset) for an observed syndrome $\mathbf{s}$. Let $\mathcal{L}$ be the set of logical operators and $P(E)$ be the physical error probability distribution. The decoder evaluates the probability of each logical coset $\bar{L} = \{ E_{ref} \cdot L \cdot S \mid S \in \mathcal{S} \}$, where $E_{ref}$ is any fixed error consistent with the syndrome $\mathbf{s}$ and $L \in \mathcal{L}$. The probability of a coset is the sum of the probabilities of all errors within it:
\begin{equation}
    P(\bar{L} | \mathbf{s}) \propto \sum_{S \in \mathcal{S}} P(E_{ref} \cdot L \cdot S).
\end{equation}
The decoder outputs a correction operator $R$ corresponding to the logical class with the maximum posterior probability:
\begin{equation}
    \text{Correction } \sim \arg \max_{L \in \mathcal{L}} P(\bar{L} | \mathbf{s}).
\end{equation}

\subsubsection{\textbf{Extending to Circuit-Level Noise}}

The code capacity model provides a theoretical upper bound on performance. In real devices, however, the syndrome extraction circuit itself is composed of noisy physical operations. Faults can occur during state preparation, gates, idle periods, and measurement.

One often starts from a \textit{spacetime} description where nodes represent events in space (qubits) and time (QEC cycles). Define a set of elementary fault mechanisms $\mathcal{F} = \{f_1, \dots, f_{|\mathcal{F}|}\}$ where each fault $f_j$ occurs independently with probability $p_j$. The primary types of circuit faults are:

\begin{itemize}
    \item A state-preparation error is a bit-flip occurring immediately after qubit initialization (reset). It can be modeled as a channel $\mathcal{E}_{\text{prep}}$ acting on the ideal state $\rho = \ket{0}\bra{0}$:
    \begin{equation}
        \mathcal{E}_{\text{prep}}(\rho) = (1-p) \ket{0}\bra{0} + p \ket{1}\bra{1}.
    \end{equation}
    \item An idle error is a Pauli error occurring on a data qubit while it waits for ancillary operations, modeled as a single-qubit depolarizing channel.
    \item A gate error is a Pauli error occurring after a one- or two-qubit unitary gate operation. For a CNOT gate, this is typically modeled as a two-qubit depolarizing channel:
    \begin{equation}
        \mathcal{E}_{\text{CNOT}}(\rho) = (1-p_{g}) U \rho U^\dagger + \frac{p_{g}}{15} \sum_{P \in \{I,X,Y,Z\}^{\otimes 2} \setminus \{II\}} P (U \rho U^\dagger) P^\dagger.
    \end{equation}
    \item A measurement error is a classical flip of the measurement readout bit $m_{\text{ideal}} \in \{0, 1\}$. The observed outcome is:
    \begin{equation}
        m_{\text{observed}} = m_{\text{ideal}} \oplus e_{\text{meas}}, \quad \text{where } P(e_{\text{meas}}=1) = p_{meas}.
    \end{equation}
\end{itemize}

With noisy repeated measurements, a single syndrome measurement is unreliable. A stabilizer might appear violated due to a measurement flip rather than a data qubit error. Therefore, circuit-level decoding uses more compact variables known as \textit{detectors}. A detector $D_k$ is a deterministic parity check on a subset of measurement outcomes $\mathbf{m}$ in the spacetime volume. In the absence of any faults, a detector must have even parity: $D_k(\mathbf{m}_{ideal}) = 0$. For repetitive QEC cycles, a detector typically compares the value of a stabilizer $S$ measured at time $t$ with its value at time $t-1$:
\begin{equation}
    D_{S, t} = m_{S, t} \oplus m_{S, t-1}.
\end{equation}
A value $D_{S, t}=1$ signifies a detector event, indicating that an odd number of faults occurred within the specific spacetime region bounded by measurements $t-1$ and $t$.

This framework converts a noisy circuit into a Detector Error Model (DEM), which lists stochastic fault mechanisms, the detectors they flip, the logical observables they affect, and their probabilities. A graphlike DEM, where each retained mechanism touches at most two detectors, can be decoded directly by matching. General circuit-level DEMs may contain mechanisms that touch more than two detectors or that should be treated as correlated groups. Reducing such a DEM to a graph is an approximation made for the decoder's sake.

A common independent-mechanism DEM can be written as a linear mapping defined by a binary matrix $H_{\text{DEM}} \in \mathbb{F}_2^{N_{det} \times N_{faults}}$. The entry $(i, j)$ of $H_{\text{DEM}}$ is $1$ if the elementary fault $f_j$ triggers detector $D_i$. Given a vector of physical faults $\mathbf{f}$ (representing the occurrence of faults in $\mathcal{F}$), the observed detector syndrome $\mathbf{s}_{\text{det}}$ is:
\begin{equation}
    \mathbf{s}_{\text{det}}^T = H_{\text{DEM}} \cdot \mathbf{f}^T.
\end{equation}
A common approximation to optimal logical decoding is to find the most likely individual fault configuration consistent with the observed detector events. For independent mechanisms, this objective becomes:
\begin{equation}
    \arg \min_{\mathbf{f}} \sum_{j : f_j=1} \log \left( \frac{1-p_j}{p_j} \right) \quad \text{subject to} \quad H_{\text{DEM}} \cdot \mathbf{f}^T = \mathbf{s}_{\text{det}}^T.
\end{equation}

For surface codes after graph-like decomposition, $H_{\text{DEM}}$ often corresponds to the incidence matrix of a matching graph in 3D spacetime. For qLDPC codes there is no single standard. The decoder may work on the code's Tanner graph or on a circuit-level detector graph, and hybrid representations keep the measurement history along with selected correlations.

The temporal parity above describes repeated memory experiments, but it does not by itself define detectors for circuits whose stabilizers or logical observables change during execution, as occurs in lattice surgery and other logical protocols. LightStim~\cite{fang2026lightstimframeworkqecprotocol} addresses this problem by deriving detectors and logical observables automatically during circuit compilation, enabling DEM construction for protocols ranging from quantum memories to distillation and cross-code lattice surgery.

\subsection{Algorithms and Implementation Trade-offs}
\label{sec:decoding_algorithms}

While the general problem of finding the most likely error configuration is NP-hard for stabilizer codes~\cite{iyer2015hardness}, practical quantum error correction relies on specialized algorithms that exploit structure in the syndrome-extraction circuit and the noise model.

An ideal decoding algorithm is expected to be fast, accurate, and scalable. Table~\ref{tab:decoder_comparison} summarizes the trade-offs between mainstream decoding paradigms across key performance metrics. Matching-based algorithms remain the benchmark for graphlike surface-code decoding, but graph maintenance and long-range matching operations create real-time implementation challenges at scale. Union-Find and local-clustering decoders give up some accuracy for nearly local growth and merge rules, which is what makes them easy to put in hardware. BP-based algorithms, usually with post-processing, are central to qLDPC codes but must be adapted to degeneracy and short cycles. Neural-network decoders can learn decoding rules tailored to realistic or calibrated noise distributions, at the price of training data and accelerator constraints. Exact logical maximum-likelihood decoding minimizes the logical failure probability for a fixed code and noise model. Tensor-network methods evaluate or approximate the required coset-probability sums~\cite{bravyi2014maximumlikelihood,chubb2021statistical}. High-accuracy circuit-level implementations often serve as offline references due to the high complexity.

\begin{table}[!ht]
\centering
\caption{Comparison of mainstream decoding algorithms. Accuracy and complexity are indicative and depend on the code family and implementation. $N$: number of error mechanisms or detector events per round.}
\label{tab:decoder_comparison}
\scriptsize
\renewcommand{\decodercontentheight}{31pt}
\hypersetup{citecolor=ratingcite}
\arrayrulecolor{ratingrule}

\setlength{\tabcolsep}{1.5pt}
\renewcommand{\arraystretch}{1.0}
\setlength{\aboverulesep}{1pt}
\setlength{\belowrulesep}{1pt}
\begin{tabularx}{\linewidth}{@{}
  >{\hsize=1.08\hsize\linewidth=\hsize\RaggedRight\arraybackslash}X
  >{\hsize=.92\hsize\linewidth=\hsize\RaggedRight\arraybackslash}X
  >{\hsize=.93\hsize\linewidth=\hsize\RaggedRight\arraybackslash}X
  >{\hsize=1.12\hsize\linewidth=\hsize\RaggedRight\arraybackslash}X
  >{\hsize=1.00\hsize\linewidth=\hsize\RaggedRight\arraybackslash}X
  >{\hsize=.95\hsize\linewidth=\hsize\RaggedRight\arraybackslash}X@{}}
\toprule
\parbox[c][20pt][c]{\linewidth}{\centering\fontsize{7.5}{8.5}\selectfont\bfseries Algorithm}
& \parbox[c][20pt][c]{\linewidth}{\centering\fontsize{7.5}{8.5}\selectfont\bfseries Complexity}
& \parbox[c][20pt][c]{\linewidth}{\centering\fontsize{7.5}{8.5}\selectfont\bfseries Accuracy}
& \parbox[c][20pt][c]{\linewidth}{\centering\fontsize{7.5}{8.5}\selectfont\bfseries Hardware\\Feasibility}
& \parbox[c][20pt][c]{\linewidth}{\centering\fontsize{7.5}{8.5}\selectfont\bfseries Noise\\Adaptivity}
& \parbox[c][20pt][c]{\linewidth}{\centering\fontsize{7.5}{8.5}\selectfont\bfseries Code\\Applicability} \\
\midrule
\noalign{\gdef\decodercontentheight{31pt}}
\decodercell{white}{\textbf{Matching}\\ \cite{higgott2025sparse, wu2023fusion, tian2025enhancing}}
& \decodercell{white}{$\tilde{O}(N)$-$O(\text{poly}(N))$}
& \ratingcell{ratinghighbg}{high}{High}{}
& \ratingcell{ratinghighbg}{high}{High}{graph maintenance and parallelism}
& \ratingcell{ratingmediumbg}{medium}{Medium}{calibrated or reweighted edges}
& \ratingcell{ratingmediumbg}{medium}{Medium}{graphlike codes} \\
\midrule[.25pt]
\decodercell{white}{\textbf{Search /\\ Hypergraph}\\ \cite{beni2025tesseract, wu2025mwpf}}
& \decodercell{white}{Model dependent; exponential in search width}
& \ratingcell{ratinghighbg}{high}{High}{}
& \ratingcell{ratinglowmediumbg}{low}{Low--Medium}{mostly software-oriented}
& \ratingcell{ratinghighbg}{high}{High}{multi-detector faults}
& \ratingcell{ratinghighbg}{high}{High}{codes with useful DEM structure} \\
\midrule[.25pt]
\decodercell{white}{\textbf{Union-Find}\\ \cite{delfosse2021almost, huang2020fault, delfosse2022toward, chan2023actis}}
& \decodercell{white}{$\tilde{O}(N)$}
& \ratingcell{ratingmediumbg}{medium}{Medium}{}
& \ratingcell{ratinghighbg}{high}{High}{naturally parallel}
& \ratingcell{ratingmediumbg}{medium}{Medium}{local weights or side information}
& \ratingcell{ratingmediumbg}{medium}{Medium}{local / cluster-\newline friendly graphs} \\
\midrule[.25pt]
\decodercell{white}{\textbf{BP +\\ Post-processing}\\ \cite{poulin2008iterative, panteleev2021degenerate, hillmann2025localizedstatistics, muller2025improved, wang2025fullyparallelized}}
& \decodercell{white}{$\tilde{O}(N)$-$O(\text{poly}(N))$}
& \ratingcell{ratinghighbg}{high}{High}{}
& \ratingcell{ratingmediumhighbg}{medium}{Medium--High}{local message passing, parallel}
& \ratingcell{ratinghighbg}{high}{High}{bias, degeneracy, correlated priors}
& \ratingcell{ratinghighbg}{high}{High}{qLDPC codes} \\
\midrule[.25pt]
\decodercell{white}{\textbf{Neural \\ Network}\\ \cite{bausch2024learning, lee2025scalableneural, hu2025efficient}}
& \decodercell{white}{$O(\text{poly}(N))$}
& \ratingcell{ratinghighbg}{high}{High}{}
& \ratingcell{ratingmediumbg}{medium}{Medium}{requires optimization for hardware}
& \ratingcell{ratingveryhighbg}{veryhigh}{Very High}{learns realistic noise}
& \ratingcell{ratinghighbg}{high}{High}{training-noise\newline dependent} \\
\midrule[.25pt]
\noalign{\gdef\decodercontentheight{37pt}}
\decodercell{white}{\textbf{Tensor Network /\\ MLE}\\ \cite{bravyi2014maximumlikelihood,chubb2021statistical}}
& \decodercell{white}{Generally intractable, cost depends on bond dimension}
& \ratingcell{ratingveryhighbg}{veryhigh}{Very High}{exact or approximate ML}
& \ratingcell{ratinglowbg}{low}{Low}{contraction and storage costs}
& \ratingcell{ratinghighbg}{high}{High}{explicit noise distribution}
& \ratingcell{ratingmediumbg}{medium}{Medium}{codes with tractable tensor networks} \\
\bottomrule
\end{tabularx}
\vspace{3pt}

{\scriptsize\textbf{Qualitative rating:}\quad
\ratingkey{ratinglowbg}{low}{Low}\quad
\ratingkey{ratingmediumbg}{medium}{Medium}\quad
\ratingkey{ratinghighbg}{high}{High}\quad
\ratingkey{ratingveryhighbg}{veryhigh}{Very High}\par}
\end{table}

\paragraph{Matching} Work on matching decoders has gone in two directions, faster graph processing and richer error models.

On the efficiency front, efforts have focused on overcoming the traditional bottlenecks of the Minimum Weight Perfect Matching (MWPM) problem~\cite{edmonds1965paths}. \textit{Sparse Blossom}~\cite{higgott2025sparse} reduces average complexity from cubic to nearly linear by operating directly on the sparse decoding graph, while \textit{Fusion Blossom}~\cite{wu2023fusion} partitions the graph into temporal chunks for parallel and streaming decoding. For the compiler, this means the detector graph should stay sparse, local, and streamable, since those properties are what these decoders exploit.

To surpass the accuracy limit imposed by the standard assumption of independent errors, hybrid strategies known as belief matching (or BP+MWPM)~\cite{higgott2023beliefmatching, caune2023beliefpartial} have been developed. These methods use BP to estimate the marginal probability $p_i$ of each physical error. The edge weights in the matching graph are then dynamically adjusted to $w_i \approx \ln((1-p_i)/p_i)$, thereby injecting soft information about degeneracy and error likelihoods into the rigid matching framework~\cite{criger2018multi, higgott2025sparse}.
Complementing this, correlated matching~\cite{fowler2013optimal, paler2023pipelined} addresses local circuit-level correlations inside a matching instance, such as Y-errors in surface codes where bit-flip and phase-flip errors are correlated. Approaches like Iterative Reweighting MWPM (IRMWPM)~\cite{tian2025enhancing} treat these correlations by iteratively updating edge weights based on preliminary decoding outcomes, improving logical error rates in biased noise regimes. Benhemou et al.~\cite{benhemou2025minimising} map surface-code syndromes to a color-code decoding problem to exploit correlations in depolarizing noise, improving low-error-rate performance over separate bit-flip and phase-flip decoding. Liu et al.~\cite{liu2026correlatedcolor} extend correlated matching to the $4.8.8$ color code by retaining correlations between restricted lattices, with improvements evaluated under code-capacity and phenomenological noise models.

On the hardware front, matching-derived designs span complete matching engines, compiled lookup decoders, and deliberately approximate variants. \textit{Micro Blossom}~\cite{wu2025micro} implements exact MWPM with CPU-managed primal structures and FPGA-based graph-local dual updates, reporting an average decoding latency of approximately $0.8\,\mu\mathrm{s}$ at $d=13$ and physical error rate $p=0.1\%$. For smaller distances ($d \le 7$), \textit{Astrea}~\cite{vittal2023astrea} enumerates candidate perfect matchings with lookup tables and brute-force parallel search. \textit{LILLIPUT} uses an offline software MWPM decoder, together with the device error model, to populate compressed lookup tables that are queried at runtime on an FPGA~\cite{das2022lilliput}. Thus, its runtime datapath performs table lookup rather than the matching search itself.

Several cryogenic and online designs trade exactness for a more local matching procedure. \textit{NISQ+} implements a distributed SFQ decoder whose algorithm is a greedy 2-approximation to maximum-weight matching~\cite{holmes2020nisqplus}. \textit{QECOOL} similarly uses spike propagation to realize greedy, MWPM-inspired matching on a streaming three-dimensional syndrome lattice, and \textit{QULATIS} extends that online SFQ scheme to the changing boundaries and logical measurements of lattice surgery~\cite{ueno2021qecool,ueno2022qulatis}. Several accelerators are front ends for a matching decoder. \textit{Clique} applies local parity rules to decode common isolated-error patterns on chip and forwards complex signatures to an off-chip MWPM decoder~\cite{ravi2023clique}; \textit{Promatch} instead performs locality-aware greedy prematching to lower syndrome Hamming weight before the residual problem is passed to a real-time MWPM engine such as Astrea-G~\cite{alavisamani2024promatch}. \textit{Pinball} extends this predecoding line to realistic circuit-level noise. A pipelined 4-K cryo-CMOS front end handles common sparse patterns locally and offloads harder syndromes to a room-temperature decoder~\cite{knapen2026pinball}.

The matching family also anchors the practical software workflow used by many compiler and architecture studies. A stabilizer circuit can be compiled into detector definitions, logical observables, and a detector error model using \textit{Stim}~\cite{gidney2021stim}, graphlike detector error models can then be decoded with \textit{PyMatching}~\cite{higgott2022pymatching}.

\paragraph{Search- and Hypergraph-Based Decoding}
\begin{newcontentblock}
A separate line of recent work keeps richer detector-error structure than an ordinary matching graph. \textit{Tesseract} performs an $A^*$-guided most-likely-error search over subsets of detector-error mechanisms for surface, color, and bivariate-bicycle codes~\cite{beni2025tesseract}. \textit{HyperBlossom} formulates decoding as a minimum-weight parity-factor problem on hypergraphs, encompassing MWPM and hypergraph Union-Find as special graph-structured cases while preserving multi-detector fault structure~\cite{wu2025mwpf}. These are currently software-oriented decoders, but they sharpen an important compiler question: which correlated or high-order fault mechanisms should be retained before reducing a circuit-level model to pairwise edges?
\end{newcontentblock}

\paragraph{Union-Find} The Union-Find (UF) algorithm~\cite{delfosse2021almost} has evolved from a naive approximation of MWPM to a mature scheme for scalable decoding architectures. Algorithmic advances, such as Weighted Union-Find~\cite{huang2020fault} and Union-Intersection~\cite{delfosse2022toward} approaches, have reduced the performance gap between Union-Find and matching-based approaches by modeling error probability correlations in the cluster growth algorithm. Actis~\cite{chan2023actis} made Union-Find scalable in hardware by restricting all messages to nearest-neighbor processing nodes.

Hardware implementations exploit the local cluster operations of Union-Find. \textit{AFS} pipelines graph generation, depth-first search, and correction while sharing decoder resources across logical qubits~\cite{das2022afs}. \textit{Helios} maps parallel cluster processing to an FPGA with a hybrid tree-grid topology~\cite{liyanage2023scalable}, whereas \textit{Collision Clustering} implements a UF variant as a compact, low-latency ASIC~\cite{barber2025realtime}. \textit{DECONET} extends the Helios line into a network-integrated multi-FPGA system, connecting decoder nodes to support lattice-surgery decoding graphs and exposing inter-board communication as a scaling constraint~\cite{liyanage2025deconet}. The \textit{Coset Ensemble Decoder} improves accuracy by aggregating candidates from multiple coset-consistent forests and scales its FPGA design through temporal resource reuse, banked memory, and hierarchical identifier mapping~\cite{liang2026coset}. \begin{newcontentblock}The \textit{Local Clustering Decoder} (LCD) adds leakage adaptivity by pregrowing flagged edges in its local growth-and-merge pipeline~\cite{ziad2025localclustering}.\end{newcontentblock}

\paragraph{Belief Propagation} The resurgence of Belief Propagation (BP) is driven by the adoption of qLDPC codes~\cite{kuo2022exploiting}, whose short cycles and degeneracy can trap standard BP decoders~\cite{poulin2008iterative}. The baseline practical recipe has therefore been BP followed by a post-processing step, Ordered Statistics Decoding (OSD)~\cite{roffe2020qldpclandscape,panteleev2021degenerate}. OSD improves accuracy but introduces matrix operations and search steps that are difficult to place in a real-time decoder.

\begin{newcontentblock}
Recent BP-based decoders can be organized by how they reduce this BP-OSD bottleneck. Reliability-guided inversion methods localize or simplify the linear algebra: Localized Statistics Decoding (LSD) solves local unreliable regions with on-the-fly elimination~\cite{hillmann2025localizedstatistics}, while BP+OTF removes variables until the residual Tanner graph becomes tree-like enough for BP to converge~\cite{demarti2024almostlinear}. Clustering and search methods keep BP as the first pass but confine expensive work to ambiguous regions: Ambiguity Clustering partitions the residual problem into independent clusters~\cite{wolanski2024ambiguity}, BP Guided Decimation freezes confident variables~\cite{yao2024belief}, and Beam Search explores a bounded set of unreliable candidates~\cite{ye2025beam}. Message-passing variants and ensembles instead modify the BP dynamics: Relay-BP~\cite{muller2025improved} builds on memory-style BP~\cite{chen2025improved} updates for stabilizing oscillatory messages, Restart Belief reruns BP from branch-and-bound-inspired restarts~\cite{valentini2025restart}, BP-SF speculates on oscillating syndrome bits~\cite{wang2025fullyparallelized}, AutDEC runs BP over automorphism-transformed syndromes in parallel~\cite{koutsioumpas2025automorphism}, and GARI rewires correlated detector error models before normalized min-sum decoding~\cite{maan2026correlatedqldpc}. The same BP-plus-local-postprocessing pattern is also changing the outlook for color codes: VibeLSD combines schedule diversity with LSD and reports color-code performance comparable to surface-code baselines under practical decoding assumptions~\cite{koutsioumpas2025vibe}.
\end{newcontentblock}

Hardware realizations of BP are proving compact and fast. Recent FPGA implementations of Relay-BP~\cite{maurer2025real} demonstrate iteration times as low as 24 ns using reduced-precision fixed-point arithmetic, while co-designed accelerators such as Vegapunk's sparse decoding unit~\cite{zhou2025vegapunk} target high-throughput qLDPC workloads without BP-OSD's Gaussian-elimination bottleneck~\cite{valls2021syndrome}. \begin{newcontentblock}The newer qLDPC decoders above also make implementation choices more explicit: LSD and BP+OTF expose locality and sparsification as hardware parameters, AC exposes cluster-size tails, ensemble methods expose parallel decoder replication, and GARI has moved into FPGA architecture with reported sub-microsecond per-round latency for bivariate-bicycle-code instances while reusing resources across multiple decoder cores~\cite{bascones2026gari}.\end{newcontentblock}

\paragraph{Neural-Network Based}
Neural decoders learn complex error correlations from syndrome data, enabling high decoding accuracy under realistic noise. Transformer-based models such as \textit{AlphaQubit}~\cite{bausch2024learning} learn non-local error correlations but face quadratic scaling in the syndrome volume. \textit{AlphaQubit 2} extends this approach to streaming memory decoding, with accuracy benchmarks reaching surface-code distance 23 and color-code distance 27, while its real-time variant achieves sub-microsecond processing per cycle at distances 11 and 9, respectively~\cite{senior2025alphaqubit2}. Zhang et al.~\cite{zhang2026selfcoordinating} train self-coordinating parallel-window decoders, reporting sub-microsecond amortized processing per round up to distance 25 on a TPU v6e and accuracy benchmarks on Zuchongzhi 3.2 data up to distance 7. Graph-based models such as \textit{GraphQEC}~\cite{hu2025efficient} provide code-agnostic message passing. Recent neural algorithms refine this direction: Mamba-based QEC decoders target lower-complexity temporal modeling for real-time surface-code streams~\cite{lee2025scalableneural}, Sparse Mamba processes only active detection events~\cite{sayedsalehi2026sparsemamba}, and SAQ and DiffQEC explore constraint-aware transformer decoding and generative posterior inference~\cite{zenati2025saq,xu2026diffqec}. Beyond memory decoding, Bonilla Ataides et al.~\cite{bonillaataides2026neuralalgorithms} use logical circuit structure to learn gate-induced correlations, support loss information, and predict algorithm-specific observables, exposing a direct compiler-decoder interface.

In deployed systems, learned modules usually sit in front of a classical decoder. AI pre-decoders apply local corrections before sending a sparser residual syndrome to a global decoder: NVIDIA's GPU implementation couples the surface-code pre-decoder to PyMatching~\cite{chamberland2026fastpredecoders}, while its color-code extension composes with Chromobius and supports spacelike and timelike local corrections~\cite{olle2026fastpredecoderscolor}. The accompanying open-source training and deployment recipes include pretrained models and ONNX/TensorRT export for integration with CUDA-Q QEC~\cite{nvidia2026isingdecoding}. An FPGA neural decoder has also demonstrated distance-3 closed-loop surface-code feedback with 124 ns neural-decoding latency~\cite{yang2026realtimefpga}. Deployment studies now report quantization, pruning, and FPGA resource usage alongside logical error rate~\cite{yan2026rethink}.

\subsection{Real-Time Decoder Integration}
\label{sec:ftqc_decoding_systems}
A complementary perspective surveys the throughput, hardware, and system-integration challenges of real-time QEC decoding~\cite{battistel2023realtimedecoding}. Here we connect those requirements to the compiler and runtime boundary. In a fault-tolerant quantum device, the decoder is a runtime service: it must accept a continuous measurement record and side information from the control stack, return frame updates before dependent operations, and operate under throughput constraints. At the software-system level, the open-source \textsc{deq} prototype makes this boundary explicit by lowering declarative QEC programs into detector information for dynamic logical instructions and exposing a stable interface to pluggable decoder backends~\cite{microsoft2026deq}. CUDA-Q QEC provides a complementary GPU-oriented software layer for code definitions, detector-error-model generation, and pluggable CPU/GPU decoder implementations~\cite{nvidia2026cudaqx,nvidia2026cudaq}. NVQLink defines the underlying low-latency connection between quantum-system controllers and heterogeneous HPC resources, so that GPU decoding can overlap quantum execution~\cite{caldwell2025nvqlink,caldwell2026cudaqrealtime}. THQLink connects an FPGA emulating the quantum controller to CPU-based decoding through the TH-Express interconnect~\cite{lao2026thqlink}. Its surface-code memory benchmarks demonstrate parallel-window decoding with correction feedback to the FPGA. A recent superconducting experiment that integrates FPGA decoding into the control stack provides a direct demonstration of this closed-loop requirement~\cite{caune2026realtime}. This subsection therefore focuses on three system interfaces: how decoded corrections are represented as frames, how a long syndrome history is partitioned for real-time processing, and how hardware-specific information changes the messages seen by the decoder.

\subsubsection{\textbf{Pauli and Clifford Frames}}
\label{sec:pauli_frame}

A naive QEC control loop would run a block of computation, collect several rounds of stabilizer measurements, decode them, and then apply physical recovery operations. That ordering is too slow for a fault-tolerant machine for two reasons:

\begin{enumerate}
    \item The quantum processor must remain idle, waiting for the classical decoder to complete, during which time the qubits are susceptible to decoherence.
    \item The explicit application of recovery operations can be time consuming and can itself be a source of errors.
\end{enumerate}

The Pauli frame~\cite{knill2005quantum, divincenzo2007effective, riesebos2017pauli} avoids most immediate recoveries by tracking accumulated Pauli errors in software. This works because the syndrome-measurement circuits used in many QEC codes are Clifford circuits, and logical Clifford operations also preserve the Pauli group under conjugation. Formally, the Clifford group $\Clifford_2$ normalizes the Pauli group $\Pauli$ (which corresponds to $\Clifford_1$ in the Clifford hierarchy~\cite{gottesman1999quantum}). Thus, for any Clifford operator $U \in \Clifford_2$ and any Pauli operator $P \in \Pauli$, the conjugated operator $U P U^{\dagger}$ is another Pauli operator $P' \in \Pauli$. When a Clifford gate $U$ is applied to a state $|\psi\rangle$ carrying a Pauli error $P$, the error can be commuted through the gate:
$$
U (P |\psi\rangle) = (U P U^{\dagger}) (U |\psi\rangle) = P' (U |\psi\rangle).
$$
The machine can continue with the physical state uncorrected while the controller updates a classical frame entry from $P$ to $P'$. A single-qubit Pauli frame has only four states ($\{I, X, Z, Y \equiv iXZ\}$), so two classical bits per qubit are enough for this part of the bookkeeping. Quantum execution and classical tracking can then proceed concurrently.

The propagation rules through common Clifford gates are deterministic and can be pre-computed. Table~\ref{tab:pauli_frame_propagation} shows how an initial Pauli error on a data qubit changes after each gate.

\begin{table}[!ht]
\centering
\caption{Pauli frame update rules for common Clifford gates. The table shows the resulting Pauli error(s) after a gate is applied to a qubit with an initial error. For CNOT, `c` denotes the control qubit and `t` denotes the target qubit.}
\label{tab:pauli_frame_propagation}
\small
\begin{tabular}{lccc}
\hline
\textbf{Gate} & \textbf{Initial Error} & \textbf{Final Error(s)} & \textbf{Location} \\ \hline
Hadamard (H)  & $X$ & $Z$ & Same qubit \\
              & $Z$ & $X$ & Same qubit \\ \hline
Phase (S)     & $X$ & $XZ$ & Same qubit \\
              & $Z$ & $Z$ & Same qubit \\ \hline
CNOT          & $X_c$ & $X_c X_t$ & Control \& Target \\
              & $Z_c$ & $Z_c$ & Control \\
              & $X_t$ & $X_t$ & Target \\
              & $Z_t$ & $Z_c Z_t$ & Control \& Target \\ \hline
CZ            & $X_c$ & $X_c Z_t$ & Control \& Target \\
              & $Z_c$ & $Z_c$ & Control \\
              & $X_t$ & $Z_c X_t$ & Control \& Target \\
              & $Z_t$ & $Z_t$ & Target \\ \hline
\end{tabular}
\end{table}

Frame tracking also changes how measurement outcomes are interpreted. When a physical qubit with a recorded frame $F_q$ is measured in a basis defined by an operator $M$, the measurement is effectively performed on the transformed operator $F_q M F_q^{\dagger}$. The classical bit returned by the hardware is then flipped to obtain the correct logical measurement outcome.

An ordinary Pauli frame has a hard boundary at non-Clifford gates. Pending Pauli errors propagate through Clifford operations, but after a $T$-type operation they no longer remain Pauli errors under conjugation. The controller must therefore update the frame before the non-Clifford gadget reaches the point where that frame matters: in $T$ teleportation, for example, the pending frame must be folded into the choice and interpretation of the conditional Clifford correction before that correction is applied. Auto-corrected non-Clifford gates reduce this pressure by moving the correction to an additional ancilla or correction qubit, so the late classical decision becomes a measurement or interpretation of that qubit rather than an immediate Clifford correction on the data~\cite{litinski2019game,gidney2019autoccz}. The blocking point is then the correction ancilla, or a later operation that consumes its outcome, not necessarily the non-Clifford data-block operation itself~\cite{ryananderson2021realtime,acharya2024quantum,bluvstein2026faulttolerant}. Table~\ref{tab:pauli_frame_workflow} summarizes the physical actions and corresponding frame updates for common operations under this model.

\begin{table}[!ht]
\centering
\caption{Summary of operations within the Pauli frame paradigm.}
\label{tab:pauli_frame_workflow}
\small
\begin{tabular}{p{3.5cm}p{4.7cm}p{4.3cm}}
\hline
\textbf{Operations} & \textbf{Physical Action} & \textbf{Pauli Frame Update} \\ \hline
Z/X-basis Initialization & Prepare the qubit in the $|0\rangle$/$|+\rangle$ state. & The frame is reset to identity. \newline $F' \leftarrow I$ \\ \hline
Pauli Operation ($P$) & Do nothing. & The Pauli operator is absorbed into the frame. \newline $F' \leftarrow P \cdot F$ \\ \hline
Clifford Operation ($C$) & Apply the gate to the qubit. & The frame is conjugated with the Clifford gate. \newline $F' \leftarrow C F C^{\dagger}$ \\ \hline
Z/X-basis Measurement ($M$)& Perform corresponding measurement to get outcome. If $\{M, F\}=0$, the outcome is flipped. &The frame is reset to identity. \newline $F' \leftarrow I$ \\ \hline
Non-Clifford Operation & Apply, commute, or reinterpret frame-dependent corrections according to the operation protocol. & The runtime must expose any required feedforward bit before the protocol deadline. \\ \hline
\end{tabular}
\end{table}

The frame can also be enlarged from Pauli errors to Clifford-valued errors~\cite{chamberland2018fault}. Certain non-Clifford gates transform Pauli errors into Clifford operators, for instance, conjugating a Pauli operator $P$ by a $T$ gate can produce a Clifford operator $C\in \Clifford_2$. Tracking such errors in software postpones physical state correction until a later operation would move the tracked error outside the Clifford group, often at another non-Clifford gate. Clifford corrections in addition to Pauli corrections can also improve thresholds under certain correlated errors~\cite{chamberland2017hard}.

\subsubsection{\textbf{Window Decoding}}
\label{sec:window_decoding}

\begin{newcontentblock}
Frame tracking lets many corrections wait, but commit points remain. Some logical measurements, basis choices, and non-Clifford feedforward operations need a decoded value before the next dependent operation. Window decoding manages this constraint by finalizing a local part of the spacetime syndrome history while carrying boundary information forward.
\end{newcontentblock}

Large-scale FTQC produces a continuous spacetime stream whose volume grows with code distance, circuit depth, and the number of logical blocks involved in an operation. Running a superlinear decoder on the full accumulated history is incompatible with real-time feedforward, and multi-qubit logical operations such as lattice surgery raise a second question: how should the shared syndrome volume be split without losing correlations across block boundaries? Window decoding addresses both issues by committing a local part of the syndrome history while carrying unresolved boundary information forward.

Inspired by classical communication systems, window decoding partitions the check nodes into overlapping spacetime regions. Each window is decoded locally, while a buffer region between adjacent windows preserves syndrome consistency.

\paragraph{Buffer-region consistency.}
Let the current window contain a commit region and an adjacent buffer region. The decoder uses syndrome data from both regions to infer a candidate correction, but commits only the part assigned to the commit region. The buffer provides additional context that reduces the effect of the artificial window boundary on committed corrections. Its size depends on the code and decoding strategy. \(d\) rounds is a common choice when data and measurement error rates are comparable~\cite{skoric2023parallel}.

When the candidate correction crosses the commit boundary, truncating it produces artificial defects at that boundary. These defects are combined modulo two with the observed defects in the uncommitted region and passed to the next window. The next decoder therefore receives a syndrome updated to account for the committed correction, rather than reconstructing missing syndrome data from the unknown endpoints of the actual error.

The same buffer condition has been generalized beyond a single time axis. Modular decoding decomposes lattice-surgery circuits into logical-block subtasks and requires sufficient separation between committed corrections and unavailable data~\cite{bombin2023modular}. Spatially parallel decoding instead partitions large lattice-surgery patches into overlapping hardware-compatible regions~\cite{lin2025spatially}. Recent adaptive approaches make the buffer data-dependent. ADaPT starts from smaller windows and expands or redecodes when decoder confidence near the boundary is low~\cite{oberoi2026adapt}, while the spatiotemporal complementary gap provides boundary-specific soft information for adapting the buffer size directly~\cite{mishima2026adaptivewindow}.

\paragraph{Window-scheduling strategies.}

Representative strategies arrange these windows differently. Sliding window decoding processes windows sequentially, so window \(n\) waits until window \(n-1\) has been committed~\cite{dennis2002topological}. Parallel window decoding runs non-adjacent windows concurrently when they share no direct syndrome dependency~\cite{skoric2023parallel,tan2023scalable}. Speculative window decoding goes further by pre-processing the first layers of the next window to predict inter-window dependencies, allowing adjacent windows to start nearly simultaneously~\cite{viszlai2025swiper}.
\begin{newcontentblock}
Adaptive and deadline-aware methods sit at the runtime scheduler, where window priority and buffer size can be adjusted according to feedforward urgency and boundary uncertainty~\cite{oberoi2026adapt,chen2026triage}. At the coarser resource-allocation level, elastic decoding virtualizes a shared pool of hardware decoders across logical qubits and schedules decoding jobs to reduce dedicated decoder provisioning~\cite{maurya2026elastic}.
\end{newcontentblock}

\subsubsection{\textbf{Noise- and Side-Information-Aware Decoding}}
\label{sec:noise_side_info_decoding}

The circuit-level detector error model introduced in Section~\ref{sec:decoding_problem} is a useful abstraction. Yet many FTQC devices expose information that is more specific than an averaged Pauli fault rate. A decoding system may receive erasure flags, leakage indicators, analog readout likelihoods, calibration-dependent edge weights, or correlations induced by the current schedule. Treating these facts as decoder inputs can improve the decoding accuracy.

\paragraph{Loss-, erasure-, leakage-, and crosstalk-aware decoding.}
Detectable loss and erasure differ from unknown Pauli faults because the decoder receives the affected location, and sometimes the time. Early threshold analyses handled known loss locations by modifying the decoding lattice~\cite{stace2009thresholds}. More recent hardware proposals use erasure conversion to expose dominant faults to the decoder, including alkaline-earth neutral atoms and superconducting erasure qubits~\cite{wu2022erasureconversion,kubica2023erasure}. This interface has now been demonstrated with metastable neutral atoms through $[[4,2,2]]$ logical encoding with mid-circuit erasure measurements and logical teleportation with erasure-conditioned ancilla selection~\cite{zhang2026logicalqubits}. Neutral-atom biased-erasure models and loss-detection units make this a shot-dependent interface, since the decoder may need to update weights, effective checks, or the detector error model for each run~\cite{sahay2023highthreshold,perrin2025atomloss}. Leakage has a similar interface but different physics. A leaked qubit can keep interacting with neighbors and generate time-correlated faults until it returns to the computational subspace~\cite{fowler2013leakage}. Adaptive hardware decoders can use heralded leakage to pregrow clusters or update the decoding graph in real time~\cite{ziad2025localclustering}. Crosstalk is more compiler dependent, since its rate depends on parallel gates and placement, requiring schedule-aware noise metadata rather than only static code geometry~\cite{huang2020crosstalk}.

\paragraph{Soft-information decoding.}
The standard syndrome stream records thresholded bits, but measurements often produce richer observations before thresholding, such as continuous readout values, photon counts, or outcome likelihoods. Soft-information decoders preserve this evidence as log-likelihood ratios or dynamic edge weights. Pattison et al. show that both MWPM and Union-Find can use soft measurement information and outperform hard decoders that see only binary outcomes~\cite{pattison2021soft}. The same idea appears in bosonic-code concatenations, where surface-GKP matching weights are computed dynamically from analog GKP correction information over the syndrome history~\cite{noh2022surfacegkp}. Learned decoders can also consume such inputs. AlphaQubit's scaling experiments use analog I/Q-derived probabilities and leakage information as soft decoder inputs~\cite{bausch2024learning}.

\paragraph{Correlated-error decoding.}
Here the correlations come from the logical operation being executed, and they span more than one code block. Neutral-atom processors provide a concrete example because reconfigurable arrays support transversal operations between logical qubits~\cite{bluvstein2024logical}. During a transversal entangling gate, faults and stabilizer-measurement errors can propagate across code blocks, so independent per-block decoding loses useful information. Cain et al. formalize correlated decoding as joint inference over the participating logical qubits~\cite{cain2024correlated}, reducing the noisy syndrome-extraction rounds needed between transversal Clifford gates from $O(d)$ to $O(1)$. A later protocol decodes propagated logical-operator products, yielding an MWPM-compatible surface-code implementation with smaller decoding instances~\cite{cain2025fastcorrelated}.

\section{Conclusion and Future Direction}
\label{sec:conclusion}

In this survey, we examined quantum error correction through the lens of compilation, focusing on how an abstract logical algorithm is turned into fault-tolerant operations that a real device can run. We first laid the groundwork, covering the main QEC code families, fault-tolerant gate sets, and the connectivity and noise constraints of leading hardware platforms. We then organized the field into three layers where compilation decisions are made, and reviewed the techniques and trade-offs at each.

At the logical level, Clifford operations such as the CNOT are realized through lattice surgery. Here the compiler places logical patches, routes the ancilla regions that carry joint measurements, and schedules those measurements to minimize space-time volume on both surface-code and qLDPC architectures. Universality also demands non-Clifford resources, so we reviewed magic-state distillation and cultivation, code switching, and the schedulers that keep these resources flowing. At the physical level, we turned to syndrome-extraction scheduling and to hardware-aware realization on superconducting circuits, trapped ions, and neutral-atom arrays. On these platforms, movement, limited connectivity, and timing constraints reshape the mapping and scheduling problem. Finally, in the control stack, we surveyed detector error models and the decoders that run on them. We then looked at frame tracking and window decoding, which keep decoding inside the control loop.

A common thread connects these layers. Most current work optimizes one layer in isolation, yet the dominant costs of fault-tolerant computation appear at the seams between them. A layout that eases routing can strain the decoder, and a hardware-friendly schedule can waste magic states. Closing these gaps calls for compilation that reasons across the whole stack at once, which motivates the directions discussed below.

\subsection{Future Directions}

The next stage of FTQC compiler design should address the bottlenecks between layers of the QEC stack. Logical-operation compilers, syndrome-extraction schedulers, hardware-aware mappers, and real-time decoders are often evaluated separately. In a fault-tolerant quantum computer, however, their costs interact through the same actual execution. A schedule that reduces logical depth may increase decoder pressure, while a hardware-friendly extraction circuit may consume routing area or magic-state resources that the logical compiler assumed to be available. Future compilers therefore need to optimize the execution of fault-tolerant programs as a coupled systems problem.

\paragraph{From component-level demonstrations to algorithm-level FTQC}
Recent experiments and systems papers have demonstrated important pieces of the FTQC stack in different settings, from logical memories and gates to syndrome extraction and real-time decoder feedback~\cite{ryananderson2021realtime,acharya2024quantum,bluvstein2024logical}. The compiler challenge is now to compose these components into algorithm-level executions. A compiled program needs more than a logical gate list or a lattice-surgery layout. It also needs to specify the measured logical observables and required code distances. Detector definitions, frame-update semantics, and feed-forward deadlines should remain visible because they determine whether the execution is fault tolerant. Execution-oriented compiler frameworks are early steps in this direction~\cite{Watkins2024highperformance,trochatos2025trace,zhu2025ecmas+}. A general FTQC compiler still needs an IR that carries correctness assumptions from logical synthesis through syndrome extraction and decoding.

\paragraph{Managing logical resource flow.}
Surface-code compilation has often optimized spacetime volume, routing distance, or logical depth. These metrics remain useful, while large computations are also limited by the flow of protected resource states. Non-Clifford support introduces time-dependent resources and conversion procedures. Distillation and cultivation produce magic states with finite latency and possible rejection events. Code switching and early fault-tolerant rotations add their own conversion and resource-state scheduling constraints~\cite{litinski2019magic,gidney2024magic,claes2025cultivating,weilandt2025minimizing,stein2024architecture,akahoshi2024partially,ismail2026transversal}. Existing schedulers show that factory placement and pooling matter in practice. Retry policies and partial preparation can also change the cost of an algorithm~\cite{ding2018magic,holmes2019resource,hirano2024magicpool}. Future compilers should therefore treat resource states as live objects in the logical schedule. They need to model where each state is prepared, how long it is stored, when it is consumed, and how its production affects decoder load.

\paragraph{Turning qLDPC advantages into executable compiler stacks.}
Many qLDPC code families are attractive because they can reduce spatial overhead relative to surface-code patches. Their compiler stack, however, is less mature. Recent code-surgery and qLDPC proposals offer several routes to logical operations. One is adapter-based surgery. Another is a homological or automorphism-based construction. A third exploits high-rate transversal structure and batches many measurements together~\cite{cowtan2024ssip,cross2024improved,poirson2025engineering,he2025extractors,yoder2025tour,swaroop2026universal,cowtan2026parallel,baspin2025fast,yuan2026parsimonious,xu2025batched,koh2026entangling,he2026logical}. The open problem is to turn these constructions into an automated path from a code description to executable logical operations. Such a path begins with operation synthesis and syndrome-extraction circuit generation. It then has to include hardware embedding and a compatible decoder. Favorable code parameters can be offset once routing, measurement scheduling, and decoding overheads are included.

\paragraph{Synthesizing hardware-realizable QEC cycles.}
Physical-level QEC implementation has become part of the compiler problem. Syndrome-extraction circuits shape hook errors and detector locality. They also determine decoder load and timing. On superconducting devices the compiler must respect a sparse coupling graph and its calibration constraints. Wiring and multilayer layouts add further placement restrictions~\cite{kishony2026surfacecodeoffthehookdiagonal,viszlai2026prophunt,strikis2026high,yin2024qeccsynth,mathews2026placing,fang2024caliscalpelinsitufinegrainedqubit}. Trapped-ion and neutral-atom systems constrain the schedule differently. Transport and rearrangement decide where a check can be measured. Zone control and atom reloading decide when, and erasure handling adds its own timing~\cite{tan2024compilation,huang2024zap,reichardt2024fault,bluvstein2026faulttolerant}. A mature physical-level compiler should synthesize QEC cycles under these platform constraints, then close the loop through detector-error-model generation and decoding. A schedule that is hardware-friendly at the gate level still has to preserve effective distance and produce a syndrome stream that can be decoded within the available classical latency.

\paragraph{Making compilation decoder-aware by construction.}
The decoder defines the real-time boundary of an FTQC system. Matching-oriented decoders such as sparse matching, Fusion Blossom, and micro-blossom emphasize locality and hardware implementation. Other decoder families expose different trade-offs: union-find variants emphasize simple growth rules, BP-family qLDPC decoders emphasize iterative message passing, and neural decoders learn correlations at the cost of training and model-specific deployment constraints~\cite{higgott2025sparse,wu2023fusion,wu2025micro,barber2025realtime,hillmann2025localizedstatistics,muller2025improved,bascones2026gari,bausch2024learning,yang2026realtimefpga}. The compiler also has to preserve side information that changes the decoding instance. Erasure and leakage flags change the graph seen by the decoder. Analog readout changes its weights. Multi-block logical gates and crosstalk add correlations that depend on which operation is running~\cite{kubica2023erasure,chow2024circuit,pattison2021soft,cain2024correlated,cain2025fastcorrelated}. A decoder-aware compiler should therefore generate schedules and detector models together. It should also expose which frame updates are urgent, so that window decoding and decoder-capacity allocation can be scheduled together with feed-forward operations~\cite{viszlai2025swiper,chen2026triage,maurya2026elastic}.

\paragraph{Toward end-to-end evaluation.}
Future FTQC compiler work should move beyond reporting a single resource metric. Patch count and spacetime volume describe layout pressure. Logical error rate describes reliability. Factory throughput and decoder latency answer yet other questions, and each number usually comes from its own set of assumptions. Stronger evidence would come from compiling benchmarks through the whole stack under a shared model. That model should cover logical-operation compilation, resource-state supply, and QEC-cycle generation. It should then carry the result through hardware mapping, detector-error-model construction, and an implemented decoder. This style of evaluation would make it easier to compare code families, hardware platforms, and scheduling choices on the same baseline~\cite{beverland2022assessing,kang2025quits,bravyi2024high,xu2023constantoverhead,zhao2026ultrahighrate,acharya2024quantum,ryananderson2021realtime}. \textit{decoder-bench} is an early step toward this goal: it provides a shared Stim-based framework for comparing decoder accuracy and latency across various QEC codes and lattice-surgery workloads~\cite{maurya2025decoderbench}. Such end-to-end evaluation would also make negative results more useful, because a failed compilation could identify which interface is the current bottleneck.

\section*{Acknowledgment}

We would like to thank Zi-Wen Liu, Xiang Fang, Chen Zhao for useful suggestions.
This work was supported by the National Key R\&D Program of China (Grant No.~2024YFB4504004), the National Natural Science Foundation of China (Grant. No.~92576114, 12447107), the Guangdong Provincial Quantum Science Strategic Initiative (Grant No.~GDZX2403008, GDZX2503001), and the Guangdong Provincial Key Lab of Integrated Communication, Sensing and Computation for Ubiquitous Internet of Things (Grant No.~2023B1212010007).

\bibliographystyle{ACM-Reference-Format}
\bibliography{ref}

@article{bravyi2005universal,
  title={Universal quantum computation with ideal Clifford gates and noisy ancillas},
  author={Bravyi, Sergey and Kitaev, Alexei},
  journal={Physical Review A—Atomic, Molecular, and Optical Physics},
  volume={71},
  number={2},
  pages={022316},
  year={2005},
  publisher={APS}
}

@article{guemard2026good,
  title={Good quantum codes with addressable and parallelizable non-Clifford gates},
  author={Gu{\'e}mard, Virgile},
  year={2026}
}

@article{he2025asymptotically,
  title={Asymptotically good quantum codes with addressable and transversal non-Clifford gates},
  author={He, Zhiyang and Vaikuntanathan, Vinod and Wills, Adam and Zhang, Rachel Yun},
  journal={arXiv preprint arXiv:2507.05392},
  year={2025}
}

@article{bhardwaj2026high,
  title={High-rate qLDPC processors},
  author={Bhardwaj, Aditya and Ma, Muzhou and Meister, Nadine and King, Robbie and Bluvstein, Dolev and Preskill, John and Cain, Madelyn and Xu, Qian and Huang, Hsin-Yuan},
  journal={arXiv preprint arXiv:2607.28795},
  year={2026}
}

@article{li2026transversal,
  title={Transversal non-Clifford gates on almost-good quantum LDPC and quantum locally testable codes},
  author={Li, Yiming and Li, Zimu and Liu, Zi-Wen},
  journal={arXiv preprint arXiv:2604.01874},
  year={2026}
}

@inproceedings{golowich2025asymptotically,
  title={Asymptotically good quantum codes with transversal non-Clifford gates},
  author={Golowich, Louis and Guruswami, Venkatesan},
  booktitle={Proceedings of the 57th Annual ACM Symposium on Theory of Computing},
  pages={707--717},
  year={2025}
}

@article{breuckmann2026cups,
  title={Cups and Gates I: Cohomology Invariants and Logical Quantum Operations: NP Breuckmann, M. Davydova, N. Tantivasadakarn},
  author={Breuckmann, Nikolas P and Davydova, Margarita and Eberhardt, Jens N and Tantivasadakarn, Nathanan},
  journal={Communications in Mathematical Physics},
  volume={407},
  number={5},
  pages={86},
  year={2026},
  publisher={Springer}
}

@article{bombin2015gauge,
  title={Gauge color codes: optimal transversal gates and gauge fixing in topological stabilizer codes},
  author={Bomb{\'\i}n, H{\'e}ctor},
  journal={New Journal of Physics},
  volume={17},
  number={8},
  pages={083002},
  year={2015},
  publisher={IOP Publishing}
}

@article{anderson2014fault,
  title={Fault-tolerant conversion between the steane and reed-muller quantum codes},
  author={Anderson, Jonas T and Duclos-Cianci, Guillaume and Poulin, David},
  journal={Physical review letters},
  volume={113},
  number={8},
  pages={080501},
  year={2014},
  publisher={APS}
}

@article{paetznick2013universal,
   title={Universal Fault-Tolerant Quantum Computation with Only Transversal Gates and Error Correction},
   volume={111},
   ISSN={1079-7114},
   url={http://dx.doi.org/10.1103/PhysRevLett.111.090505},
   DOI={10.1103/physrevlett.111.090505},
   number={9},
   journal={Physical Review Letters},
   publisher={American Physical Society (APS)},
   author={Paetznick, Adam and Reichardt, Ben W.},
   year={2013},
   month=Aug }

@misc{lu2026intrinsiclocalitydimensionquantum,
      title={Intrinsic locality dimension of quantum codes}, 
      author={Yimin Lu and Esther Xiaozhen Fu and Zi-Wen Liu},
      year={2026},
      eprint={2605.31441},
      archivePrefix={arXiv},
      primaryClass={quant-ph},
      url={https://arxiv.org/abs/2605.31441}, 
}

@misc{microsoft2026deq,
  author={{Microsoft}},
  title={{deq}: Dynamic and Generic {QEC} Decoding System},
  year={2026},
  howpublished={\url{https://github.com/microsoft/qdk-ec/tree/main/deq}},
  note={Software repository. Accessed: 2026-08-06},
}

@article{bravyi2013classification,
   title={Classification of Topologically Protected Gates for Local Stabilizer Codes},
   volume={110},
   ISSN={1079-7114},
   url={http://dx.doi.org/10.1103/PhysRevLett.110.170503},
   DOI={10.1103/physrevlett.110.170503},
   number={17},
   journal={Physical Review Letters},
   publisher={American Physical Society (APS)},
   author={Bravyi, Sergey and König, Robert},
   year={2013},
   month=Apr }

@misc{wang2026trapping11000,
  title={Trapping 11,000 Atoms in a Tweezer Array Generated by a Single Metasurface},
  author={Wang, Yuqing and Zhang, Zhongchi and Zhang, Tao and Liao, Yuxuan and Wang, Hanteng and Tian, Ye and Ji, Binjie and Wu, Yujia and Ma, Luming and Qing, Chen and Li, Chengshu and Zhang, Wei and Huang, Yidong and Zhang, Wenjun and Feng, Xue and Chen, Wenlan and Zhai, Hui},
  year={2026},
  eprint={2606.02715},
  archivePrefix={arXiv},
  primaryClass={quant-ph},
  url={https://arxiv.org/abs/2606.02715},
}

@misc{atomcomputing2026toric,
  title={Quantum Error Correction with the Toric Code},
  author={{Atom Computing and Collaborators}},
  year={2026},
  eprint={2606.04079},
  archivePrefix={arXiv},
  primaryClass={quant-ph},
  url={https://arxiv.org/abs/2606.04079},
}

@misc{kasai2026orthogonality,
  title={Breaking the Orthogonality Barrier in Quantum {LDPC} Codes},
  author={Kasai, Kenta},
  year={2026},
  eprint={2601.08824},
  archivePrefix={arXiv},
  primaryClass={quant-ph},
  url={https://arxiv.org/abs/2601.08824},
}

@misc{okada2026highgirth,
  title={High-Girth Regular Quantum {LDPC} Codes from Affine-Coset Structures},
  author={Okada, Koki and Kasai, Kenta},
  year={2026},
  eprint={2604.20838},
  archivePrefix={arXiv},
  primaryClass={quant-ph},
  url={https://arxiv.org/abs/2604.20838},
}

@misc{okada2026twobranch,
  title={A Two-Branch Finite-Field Construction for Regular {CSS LDPC} Bases},
  author={Okada, Koki and Kasai, Kenta},
  year={2026},
  eprint={2605.23894},
  archivePrefix={arXiv},
  primaryClass={quant-ph},
  url={https://arxiv.org/abs/2605.23894},
}

@misc{okada2026ratetwothirds,
  title={Rate-2/3 Girth-8 (3,18)-Regular Quantum {LDPC} Codes from Two-Branch Finite-Field Bases and {CPM} Lifts},
  author={Okada, Koki and Kasai, Kenta},
  year={2026},
  eprint={2606.27130},
  archivePrefix={arXiv},
  primaryClass={quant-ph},
  url={https://arxiv.org/abs/2606.27130},
}

@misc{okada2026pairpartition,
  title={Pair-Partition Constructions for {CPM}-Based Quantum {LDPC} Codes},
  author={Okada, Koki and Kasai, Kenta},
  year={2026},
  eprint={2607.14091},
  archivePrefix={arXiv},
  primaryClass={quant-ph},
  url={https://arxiv.org/abs/2607.14091},
}

@article{zhang2026logicalqubits,
  title={Logical Qubits with Erasure Conversion Using Metastable Neutral Atoms},
  author={Zhang, Bichen and Liu, Genyue and Bornet, Guillaume and Horvath, Sebastian P. and Peng, Pai and Ma, Shuo and Huang, Shilin and Puri, Shruti and Thompson, Jeff D.},
  journal={Nature Physics},
  volume={22},
  number={6},
  pages={910--916},
  year={2026},
  doi={10.1038/s41567-026-03309-0},
}

@misc{nvidia2026cudaqx,
  author={{NVIDIA Corporation, CUDA-QX Development Team}},
  title={{CUDA-QX}},
  year={2026},
  howpublished={\url{https://github.com/NVIDIA/cudaqx}},
  note={Software repository, version 0.7.0. Accessed: 2026-08-06},
}

@misc{nvidia2026cudaq,
  author={{The CUDA-Q Development Team}},
  title={{CUDA-Q}},
  year={2026},
  howpublished={\url{https://github.com/NVIDIA/cuda-quantum}},
  note={Software repository, version 0.15.0. Accessed: 2026-08-06},
}

@misc{olle2026fastpredecoderscolor,
  title={Fast and Accurate {AI}-Based Pre-Decoders for Color Codes},
  author={Olle, Jan and Chamberland, Christopher and Li, Muyuan and Baratta, Igor},
  year={2026},
  eprint={2607.10058},
  archivePrefix={arXiv},
  primaryClass={quant-ph},
  url={https://arxiv.org/abs/2607.10058},
}

@misc{nvidia2026isingdecoding,
  author={{NVIDIA Corporation}},
  title={{Ising-Decoding}: A Set of Training Recipes for {AI} Quantum Error Correction Decoders},
  year={2026},
  howpublished={\url{https://github.com/NVIDIA/Ising-Decoding}},
  note={Software repository, version 0.1.2. Accessed: 2026-08-06},
}

@misc{caldwell2025nvqlink,
  title={Platform Architecture for Tight Coupling of High-Performance Computing with Quantum Processors},
  author={Caldwell, Shane A. and Khazraee, Moein and Agostini, Elena and Lassiter, Tom and Simpson, Corey and Kahalon, Omri and Kanuri, Mrudula and Kim, Jin-Sung and Stanwyck, Sam and Li, Muyuan and Olle, Jan and Chamberland, Christopher and Howe, Ben and Schmitt, Bruno and Lietz, Justin G. and McCaskey, Alex and Ye, Jun and Li, Ang and Magann, Alicia B. and Ostrove, Corey I. and Rudinger, Kenneth and Blume-Kohout, Robin and Young, Kevin and Miller, Nathan E. and Xu, Yilun and Huang, Gang and Siddiqi, Irfan and Lange, John and Zimmer, Christopher and Humble, Travis},
  year={2025},
  eprint={2510.25213},
  archivePrefix={arXiv},
  primaryClass={quant-ph},
  url={https://arxiv.org/abs/2510.25213},
}

@misc{caldwell2026cudaqrealtime,
  title={Introducing {cudaq-realtime} for Programming the Logical {QPU}},
  author={Caldwell, Shane and Ketcham, Chuck and Nguyen, Thien and Schmitt, Bruno and Agostini, Elena and Simpson, Corey and Bonde, Jeffrey and Lassiter, Tom and Khazraee, Moein and Howe, Ben and Heim, Bettina},
  year={2026},
  month={mar},
  howpublished={NVIDIA Quantum Technical Blog},
  url={https://nvidia.github.io/cuda-quantum/blogs/blog/2026/03/16/launching-cudaq-realtime/},
  note={Accessed: 2026-08-06},
}

@misc{li2025noncliffordfusiontgateoptimization,
      title={{Non-Clifford Fusion}: {$T$}-Gate Optimization for Quantum Simulation},
      author={Yingheng Li and Xulong Tang and Paul Hovland and Ji Liu},
      year={2025},
      eprint={2510.13573},
      archivePrefix={arXiv},
      primaryClass={quant-ph},
      url={https://arxiv.org/abs/2510.13573},
      note={Accepted to the 59th IEEE/ACM International Symposium on Microarchitecture (MICRO '26)},
}

@misc{fang2026lightstimframeworkqecprotocol,
      title={{LightStim}: A Framework for {QEC} Protocol Evaluation and Prototyping with Automated {DEM} Construction},
      author={Xiang Fang and Ming Wang and Yue Wu and Sharanya Prabhu and Dean Tullsen and Narasinga Rao Miniskar and Frank Mueller and Travis Humble and Yufei Ding},
      year={2026},
      eprint={2604.21472},
      archivePrefix={arXiv},
      primaryClass={quant-ph},
      url={https://arxiv.org/abs/2604.21472},
}

@misc{ismail2026fastparallelhighratestar,
      title={Fast and Parallel High-Rate {STAR} Architecture for Megaquop Quantum Simulation},
      author={Refaat Ismail and Milan Kornjača and Hong-Ye Hu and Nishad Maskara and Sheng-Tao Wang and Hengyun Zhou and Chen Zhao},
      year={2026},
      eprint={2606.25011},
      archivePrefix={arXiv},
      primaryClass={quant-ph},
      url={https://arxiv.org/abs/2606.25011},
}

@article{ismail2026transversal,
  title={Transversal architecture for megaquop-scale quantum simulation with neutral atoms},
  author={Ismail, Refaat and Chen, I-Chi and Zhao, Chen and Weiss, Ronen and Liu, Fangli and Zhou, Hengyun and Wang, Sheng-Tao and Sornborger, Andrew and Kornja{\v{c}}a, Milan},
  journal={PRX Quantum},
  volume={7},
  number={2},
  pages={020343},
  year={2026},
  publisher={American Physical Society},
  doi={10.1103/j2fw-ccmy},
  eprint={2509.18294},
  archivePrefix={arXiv},
  primaryClass={quant-ph},
  url={https://arxiv.org/abs/2509.18294},
}

@article{Bravyi_2005,
   title={Universal quantum computation with ideal Clifford gates and noisy ancillas},
   volume={71},
   ISSN={1094-1622},
   url={http://dx.doi.org/10.1103/PhysRevA.71.022316},
   DOI={10.1103/physreva.71.022316},
   number={2},
   journal={Physical Review A},
   publisher={American Physical Society (APS)},
   author={Bravyi, Sergey and Kitaev, Alexei},
   year={2005},
   month=Feb }

@article{dasu2026computing,
  title={Computing with many encoded logical qubits beyond break-even},
  author={Dasu, Shival and DeCross, Matthew and Guo, Andrew Y and Lavasani, Ali and Behrends, Jan and Benhemou, Asmae and Chen, Yi-Hsiang and Mayer, Karl and Self, Chris N and Simsek, Selwyn and others},
  journal={arXiv preprint arXiv:2602.22211},
  year={2026}
}

@article{rines2025demonstration,
  title={Demonstration of a Logical Architecture Uniting Motion and In-Place Entanglement: Shor's Algorithm, Constant-Depth CNOT Ladder, and Many-Hypercube Code},
  author={Rines, Rich and Hall, Benjamin and Teo, Mariesa H and Viszlai, Joshua and Cole, Daniel C and Mason, David and Barker, Cameron and Bedalov, Matt J and Blakely, Matt and Bothwell, Tobias and others},
  journal={arXiv preprint arXiv:2509.13247},
  year={2025}
}

@article{tham2026breakeven,
  title={Breakeven demonstration of quantum low-density parity-check codes},
  author={Tham, Edwin and Goldman, Michael L and Debnath, Shantanu and Patel, Ashay N and Saraladevi, Jyothi and Nguyen, Jason and Nielsen, Erik and Pisenti, Neal and Wright, Kenneth and Gamble, John and others},
  journal={arXiv preprint arXiv:2606.06455},
  year={2026}
}

@article{reichardt2024demonstration,
  title={Demonstration of quantum computation and error correction with a tesseract code},
  author={Reichardt, Ben W and Aasen, David and Chao, Rui and Chernoguzov, Alex and van Dam, Wim and Gaebler, John P and Gresh, Dan and Lucchetti, Dominic and Mills, Michael and Moses, Steven A and others},
  journal={arXiv preprint arXiv:2409.04628},
  year={2024}
}

@article{xu2025batched,
  title={Batched high-rate logical operations for quantum LDPC codes},
  author={Xu, Qian and Zhou, Hengyun and Bluvstein, Dolev and Cain, Madelyn and Kalinowski, Marcin and Preskill, John and Lukin, Mikhail D and Maskara, Nishad},
  journal={arXiv preprint arXiv:2510.06159},
  year={2025}
}

@article{yuan2026parsimonious,
  title={Parsimonious Quantum Low-Density Parity-Check Code Surgery},
  author={Yuan, Andrew C and Cowtan, Alexander and He, Zhiyang and Lin, Ting-Chun and Williamson, Dominic J},
  journal={arXiv preprint arXiv:2603.05082},
  year={2026}
}

@article{baspin2025fast,
  title={Fast surgery for quantum LDPC codes},
  author={Baspin, Nou{\'e}dyn and Berent, Lucas and Cohen, Lawrence Z},
  journal={arXiv preprint arXiv:2510.04521},
  year={2025}
}

@article{cowtan2026parallel,
  title={Parallel logical measurements via quantum code surgery},
  author={Cowtan, Alexander and He, Zhiyang and Williamson, Dominic J and Yoder, Theodore J},
  journal={PRX Quantum},
  volume={7},
  number={2},
  pages={020325},
  year={2026},
  publisher={APS},
  doi={10.1103/gj8x-n5gg},
}

@article{swaroop2026universal,
  title={Universal adapters between quantum low-density parity check codes},
  author={Swaroop, Esha and Jochym-O'Connor, Tomas and Yoder, Theodore J},
  journal={PRX Quantum},
  volume={7},
  number={1},
  pages={010324},
  year={2026},
  publisher={APS},
  doi={10.1103/1g44-jp62},
}

@article{ide2025fault,
  title={Fault-tolerant logical measurements via homological measurement},
  author={Ide, Benjamin and Gowda, Manoj G and Nadkarni, Priya J and Dauphinais, Guillaume},
  journal={Physical Review X},
  volume={15},
  number={2},
  pages={021088},
  year={2025},
  publisher={APS},
  doi={10.1103/physrevx.15.021088},
}

@article{williamson2026low,
  title={Low-overhead fault-tolerant quantum computation by gauging logical operators},
  author={Williamson, Dominic J and Yoder, Theodore J},
  journal={Nature Physics},
  pages={1--6},
  year={2026},
  publisher={Nature Publishing Group UK London},
  doi={10.1038/s41567-026-03220-8},
}

@article{zhang2025time,
  title={Time-efficient logical operations on quantum low-density parity check codes},
  author={Zhang, Guo and Li, Ying},
  journal={Physical Review Letters},
  volume={134},
  number={7},
  pages={070602},
  year={2025},
  publisher={APS},
  doi={10.1103/physrevlett.134.070602},
}

@article{cross2024improved,
  title={Improved QLDPC surgery: Logical measurements and bridging codes},
  author={Cross, Andrew W and He, Zhiyang and Rall, Patrick J and Yoder, Theodore J},
  journal={arXiv preprint arXiv:2407.18393},
  year={2024}
}

@article{cohen2022low,
  title={Low-overhead fault-tolerant quantum computing using long-range connectivity},
  author={Cohen, Lawrence Z and Kim, Isaac H and Bartlett, Stephen D and Brown, Benjamin J},
  journal={Science Advances},
  volume={8},
  number={20},
  pages={eabn1717},
  year={2022},
  publisher={American Association for the Advancement of Science},
  doi={10.1126/sciadv.abn1717},
}

@article{liu2026assessing,
  title={Assessing System Capabilities and Bottlenecks of an Early Fault-Tolerant Bicycle Architecture},
  author={Liu, Kun and Foxman, Ben and Anselmetti, Gian-Luca R and Ding, Yongshan},
  journal={arXiv preprint arXiv:2604.20013},
  year={2026},
}

@article{he2025extractors,
  title={Extractors: {QLDPC} Architectures for Efficient {Pauli}-Based Computation},
  author={He, Zhiyang and Cowtan, Alexander and Williamson, Dominic J and Yoder, Theodore J},
  journal={arXiv preprint arXiv:2503.10390},
  year={2025},
}

@article{zhou2026genecs,
  title={GeneCS: Synthesizing Resource-Efficient Code Surgery for Arbitrary Quantum Stabilizer Codes},
  author={Zhou, Junyu and Javadi-Abhari, Ali and Li, Gushu},
  journal={arXiv preprint arXiv:2605.21746},
  year={2026}
}

@article{poirson2025engineering,
  title={Engineering CSS surgery: compiling any CNOT in any code},
  author={Poirson, Cl{\'e}ment and Roffe, Joschka and Booth, Robert I},
  journal={arXiv preprint arXiv:2505.01370},
  year={2025}
}

@article{cowtan2024ssip,
  title={SSIP: automated surgery with quantum LDPC codes},
  author={Cowtan, Alexander},
  journal={arXiv preprint arXiv:2407.09423},
  year={2024}
}

@article{hofmeyr2026puremagicdynamicschedulerlattice,
      title={{PureMagic}: A Dynamic Scheduler for Lattice Surgery}, 
      author={Steven Hofmeyr and Mathias Weiden and Justin Kalloor and John Kubiatowicz and Costin Iancu},
      journal={arXiv preprint arXiv:2512.06484},
      year={2026}
}

@article{zhou2026topols,
  title={TopoLS: Lattice Surgery Compilation via Topological Program Transformations},
  author={Zhou, Junyu and Liu, Yuhao and Decker, Ethan and Kalloor, Justin and Weiden, Mathias and Chen, Kean and Iancu, Costin and Li, Gushu},
  journal={arXiv preprint arXiv:2601.23109},
  year={2026}
}

@article{zhu2025ecmas+,
  title={Ecmas+: Efficient Circuit Mapping and Scheduling for Surface Code Encoded Circuit on Quantum Cloud Platform},
  author={Zhu, Mingzheng and Fu, Hao and Song, Haishan and Wu, Jun and Zhang, Chi and Xie, Wei and Li, Xiangyang},
  journal={ACM Transactions on Architecture and Code Optimization},
  volume={22},
  number={3},
  pages={1--25},
  year={2025},
  publisher={ACM New York, NY},
  doi={10.1145/3760783},
}

@inproceedings{zhu2024ecmas,
  title={Ecmas: Efficient circuit mapping and scheduling for surface code},
  author={Zhu, Mingzheng and Fu, Hao and Wu, Jun and Zhang, Chi and Xie, Wei and Li, Xiang-Yang},
  booktitle={Proceedings of the IEEE/ACM International Symposium on Code Generation and Optimization (CGO '24)},
  pages={158--169},
  year={2024},
  organization={IEEE},
  doi={10.1109/cgo57630.2024.10444874},
}

@article{silva2024multi,
  title={Multi-qubit lattice surgery scheduling},
  author={Silva, Allyson and Zhang, Xiangyi and Webb, Zak and Kramer, Mia and Yang, Chan Woo and Liu, Xiao and Lemieux, Jessica and Chen, Ka-Wai and Scherer, Artur and Ronagh, Pooya},
  journal={arXiv preprint arXiv:2405.17688},
  year={2024}
}

@article{liao2026design,
  title={Design automation and space-time reduction for surface-code logical operations using a SAT-based EDA kernel compatible with general encodings},
  author={Liao, Wang and Tokami, Rei and Suzuki, Yasunari},
  journal={arXiv preprint arXiv:2604.12560},
  year={2026}
}

@article{tremblay2022constant,
  title={Constant-overhead quantum error correction with thin planar connectivity},
  author={Tremblay, Maxime A and Delfosse, Nicolas and Beverland, Michael E},
  journal={Physical Review Letters},
  volume={129},
  number={5},
  pages={050504},
  year={2022},
  publisher={APS},
  doi={10.1103/physrevlett.129.050504},
}

@misc{ibm2021heavyhex,
  author = {{IBM Quantum}},
  title = {The IBM Quantum Heavy Hex Lattice},
  year = {2021},
  howpublished = {\url{https://www.ibm.com/quantum/blog/heavy-hex-lattice}},
  note = {Accessed 2026-06-23}
}

@misc{google2024willowSpec,
  author = {{Google Quantum AI}},
  title = {Willow Spec Sheet},
  year = {2024},
  howpublished = {\url{https://quantumai.google/static/site-assets/downloads/willow-spec-sheet.pdf}},
  note = {Accessed 2026-06-23}
}

@misc{aws2024ankaa2,
  author = {{Amazon Web Services}},
  title = {Amazon Braket Launches the Rigetti Ankaa-2 Superconducting Device},
  year = {2024},
  howpublished = {\url{https://aws.amazon.com/blogs/quantum-computing/amazon-braket-launches-the-rigetti-ankaa-2-superconducting-device-2/}},
  note = {Accessed 2026-06-23}
}

@article{geher2025directional,
  title={Directional codes: a new family of quantum LDPC codes on hexagonal-and square-grid connectivity hardware},
  author={Geh{\'e}r, Gy{\"o}rgy P and Byfield, David and Ruban, Archibald},
  journal={arXiv preprint arXiv:2507.19430},
  year={2025}
}

@article{shaw2025lowering,
  title={Lowering connectivity requirements for bivariate bicycle codes using morphing circuits},
  author={Shaw, Mackenzie H and Terhal, Barbara M},
  journal={Physical Review Letters},
  volume={134},
  number={9},
  pages={090602},
  year={2025},
  publisher={APS},
  doi={10.1103/physrevlett.134.090602},
}

@article{zhao2025simple,
  title={A simple universal routing strategy for reducing the connectivity requirements of quantum LDPC codes},
  author={Zhao, Guangqi and Yan, Fei and Ni, Xiaotong},
  journal={arXiv preprint arXiv:2509.00850},
  year={2025}
}

@article{mcewen2023relaxing,
  title={Relaxing hardware requirements for surface code circuits using time-dynamics},
  author={McEwen, Matt and Bacon, Dave and Gidney, Craig},
  journal={Quantum},
  volume={7},
  pages={1172},
  year={2023},
  publisher={Verein zur F{\"o}rderung des Open Access Publizierens in den Quantenwissenschaften},
  doi={10.22331/q-2023-11-07-1172},
}

@inproceedings{vittal2024flag,
  title={Flag-proxy networks: Overcoming the architectural, scheduling and decoding obstacles of quantum ldpc codes},
  author={Vittal, Suhas and Javadi-Abhari, Ali and Cross, Andrew W and Bishop, Lev S and Qureshi, Moinuddin},
  booktitle={Proceedings of the IEEE/ACM International Symposium on Microarchitecture (MICRO '24)},
  pages={718--734},
  year={2024},
  organization={IEEE},
  doi={10.1109/micro61859.2024.00059},
}

@article{sato2025scheduling,
  title={Scheduling of syndrome measurements with a few ancillary qubits},
  author={Sato, Shintaro and Suzuki, Yasunari},
  journal={arXiv preprint arXiv:2508.07913},
  year={2025}
}

@article{strikis2026high,
  title={High-performance syndrome extraction circuits for quantum codes},
  author={Strikis, Armands and Browne, Dan E and Beverland, Michael E},
  journal={arXiv preprint arXiv:2603.05481},
  year={2026}
}

@inproceedings{tan2026syndrome,
author = {Tan, Daniel Bochen and Bonilla Ataides, J. Pablo and Menon, Varun and Koh, Jin Ming and Diaconu, Andrei C. and Lukin, Mikhail D.},
title = {Syndrome Extraction Circuits with Near-Optimal Depths for Practical Quantum Error Correcting Code Families},
year = {2026},
publisher = {Association for Computing Machinery},
address = {New York, NY, USA},
booktitle={Proceedings of the ACM/IEEE Design Automation Conference (DAC '26)},
location = {Long Beach, CA, USA},
numpages = {7},
url = {https://63dac.conference-program.com/presentation/?id=RESEARCH144&sess=sess134}
}

@inproceedings{viszlai2026prophunt,
  title={{PropHunt}: Automated optimization of quantum syndrome measurement circuits},
  author={Viszlai, Joshua and Maurya, Satvik and Tannu, Swamit and Martonosi, Margaret and Chong, Frederic T},
  booktitle={Proceedings of the ACM International Conference on Architectural Support for Programming Languages and Operating Systems (ASPLOS '26)},
  pages={1476--1491},
  year={2026},
  doi={10.1145/3779212.3790205},
}

@inproceedings{zhang2026optimal,
  title={Optimal Compilation of Syndrome Extraction Circuits for General Quantum LDPC Codes},
  author={Zhang, Kai and Gao, Dingchao and Yang, Zhaohui and Zhou, Runshi and Liu, Fangming and Ji, Zhengfeng and Chen, Jianxin},
  booktitle={Proceedings of the Design, Automation \& Test in Europe Conference (DATE '26)},
  pages={1--7},
  year={2026},
  organization={IEEE},
  doi={10.23919/date69613.2026.11539585},
}

@article{menon2026magic,
  title={Magic Tricycles: Efficient Magic-State Generation with Finite Block-Length Quantum LDPC Codes},
  author={Menon, Varun and Bonilla Ataides, J Pablo and Mehta, Rohan and Gu, Andi and Tan, Daniel Bochen and Lukin, Mikhail D},
  journal={Physical Review X},
  volume={16},
  number={2},
  pages={021014},
  year={2026},
  publisher={APS},
  doi={10.1103/ghhp-cytl},
}

@article{kang2025quits,
  title={QUITS: A modular Qldpc code circUIT Simulator},
  author={Kang, Mingyu and Lin, Yingjia and Yao, Hanwen and G{\"o}kduman, Mert and Meinking, Arianna and Brown, Kenneth R},
  journal={Quantum},
  volume={9},
  pages={1931},
  year={2025},
  publisher={Verein zur F{\"o}rderung des Open Access Publizierens in den Quantenwissenschaften},
  doi={10.22331/q-2025-12-05-1931},
}

@article{zhao2022realization,
  title={Realization of an error-correcting surface code with superconducting qubits},
  author={Zhao, Youwei and Ye, Yangsen and Huang, He-Liang and Zhang, Yiming and Wu, Dachao and Guan, Huijie and Zhu, Qingling and Wei, Zuolin and He, Tan and Cao, Sirui and others},
  journal={Physical Review Letters},
  volume={129},
  number={3},
  pages={030501},
  year={2022},
  publisher={APS},
  doi={10.1103/physrevlett.129.030501},
}

@inproceedings{liu2026alphasyndrome,
  title={{AlphaSyndrome}: Tackling the syndrome measurement circuit scheduling problem for {QEC} codes},
  author={Liu, Yuhao and Ping, Shuohao and Zhou, Junyu and Decker, Ethan and Kalloor, Justin and Weiden, Mathias and Chen, Kean and Shi, Yunong and Javadi-Abhari, Ali and Iancu, Costin and others},
  booktitle={Proceedings of the ACM International Conference on Architectural Support for Programming Languages and Operating Systems (ASPLOS '26)},
  pages={77--93},
  year={2026},
  doi={10.1145/3779212.3790123},
}

@article{manes2025distance,
  title={Distance-preserving stabilizer measurements in hypergraph product codes},
  author={Manes, Argyris Giannisis and Claes, Jahan},
  journal={Quantum},
  volume={9},
  pages={1618},
  year={2025},
  publisher={Verein zur F{\"o}rderung des Open Access Publizierens in den Quantenwissenschaften},
  doi={10.22331/q-2025-01-30-1618},
}

@article{geher2024tangling,
  title={Tangling schedules eases hardware connectivity requirements for quantum error correction},
  author={Geh{\'e}r, Gy{\"o}rgy P and Crawford, Ophelia and Campbell, Earl T},
  journal={PRX Quantum},
  volume={5},
  number={1},
  pages={010348},
  year={2024},
  publisher={APS},
  doi={10.1103/prxquantum.5.010348},
}

@article{kishony2026surfacecodeoffthehookdiagonal,
      title={Surface code off-the-hook: diagonal syndrome-extraction scheduling}, 
      author={Gilad Kishony and Austin Fowler},
      journal={arXiv preprint arXiv:2602.09099},
      year={2026}
}

@article{mathews2026placing,
  title={Placing and routing quantum LDPC codes in multilayer superconducting hardware},
  author={Mathews, Melvin and Pahl, Lukas and Pahl, David and Addala, Vaishnavi L and Tang, Catherine and Oliver, William D and Grover, Jeffrey A},
  journal={npj Quantum Information},
  year={2026},
  publisher={Nature Publishing Group UK London},
  doi={10.1038/s41534-026-01243-w},
}

@article{zhu2026o3ls,
  title={O3LS: Optimizing Lattice Surgery via Automatic Layout Searching and Loose Scheduling},
  author={Zhu, Chenghong and Wu, Xian and Chen, Jiahan and He, Keming and Wu, Junjie and Wang, Xin and Lao, Lingling},
  journal={arXiv preprint arXiv:2604.15099},
  year={2026}
}

@article{perrin2026correlated,
  title={Correlated Atom Loss as a Resource for Quantum Error Correction},
  author={Perrin, Hugo and Roger, Gatien and Pupillo, Guido},
  journal={arXiv preprint arXiv:2603.24237},
  year={2026}
}

@article{pecorari2025lowdepth,
  title={Low-depth quantum error correction via three-qubit gates in Rydberg atom arrays},
  author={Pecorari, Laura and Jandura, Sven and Pupillo, Guido},
  journal={arXiv preprint arXiv:2507.06096},
  year={2025}
}

@article{kobayashi2024erasure,
  title={Erasure-tolerance scheme for the surface codes on neutral atom quantum computers},
  author={Kobayashi, Fumiyoshi and Nagayama, Shota},
  journal={IEEE Transactions on Quantum Engineering},
  year={2025},
  publisher={IEEE},
  doi={10.1109/tqe.2025.3627918},
}

@article{pecorari2025quantum,
  title={Quantum low-density parity-check codes for erasure-biased atomic quantum processors},
  author={Pecorari, Laura and Pupillo, Guido},
  journal={Physical Review A},
  volume={112},
  number={5},
  pages={052417},
  year={2025},
  publisher={APS},
  doi={10.1103/mgkt-ctv8},
}

@inproceedings{stade2025optimalstatepreparation,
  title={Optimal state preparation for logical arrays on zoned neutral atom quantum computers},
  author={Stade, Yannick and Schmid, Ludwig and Burgholzer, Lukas and Wille, Robert},
  booktitle={Proceedings of the Design, Automation \& Test in Europe Conference (DATE '25)},
  pages={1--7},
  year={2025},
  organization={IEEE},
  doi={10.23919/date64628.2025.10993241},
}

@article{zhao2026ultrahighrate,
      title={Towards Ultra-High-Rate Quantum Error Correction with Reconfigurable Atom Arrays}, 
      author={Chen Zhao and Casey Duckering and Andi Gu and Nishad Maskara and Hengyun Zhou},
      journal={arXiv preprint arXiv:2604.16209},
      year={2026}
}

@article{pecorari2025high,
  title={High-rate quantum LDPC codes for long-range-connected neutral atom registers},
  author={Pecorari, Laura and Jandura, Sven and Brennen, Gavin K and Pupillo, Guido},
  journal={Nature Communications},
  volume={16},
  number={1},
  pages={1111},
  year={2025},
  publisher={Nature Publishing Group UK London},
  doi={10.1038/s41467-025-56255-5},
}

@article{chen2024transversallogicalcliffordgates,
      title={Transversal Logical Clifford gates on rotated surface codes with reconfigurable neutral atom arrays}, 
      author={Zi-Han Chen and Ming-Cheng Chen and Chao-Yang Lu and Jian-Wei Pan},
      journal={arXiv preprint arXiv:2412.01391},
      year={2024}
}

@article{hong2024long,
  title={Long-range-enhanced surface codes},
  author={Hong, Yifan and Marinelli, Matteo and Kaufman, Adam M and Lucas, Andrew},
  journal={Physical Review A},
  volume={110},
  number={2},
  pages={022607},
  year={2024},
  publisher={APS},
  doi={10.1103/physreva.110.022607},
}

@article{chow2024circuit,
  title={Circuit-based leakage-to-erasure conversion in a neutral-atom quantum processor},
  author={Chow, Matthew NH and Buchemmavari, Vikas and Omanakuttan, Sivaprasad and Little, Bethany J and Pandey, Saurabh and Deutsch, Ivan H and Jau, Yuan-Yu},
  journal={PRX Quantum},
  volume={5},
  number={4},
  pages={040343},
  year={2024},
  publisher={APS},
  doi={10.1103/prxquantum.5.040343},
}

@article{reichardt2024fault,
  title={Fault-tolerant quantum computation with a neutral atom processor},
  author={Reichardt, Ben W and Paetznick, Adam and Aasen, David and Basov, Ivan and Bello-Rivas, Juan M and Bonderson, Parsa and Chao, Rui and van Dam, Wim and Hastings, Matthew B and Mishmash, Ryan V and others},
  journal={arXiv preprint arXiv:2411.11822},
  year={2024}
}

@inproceedings{nottingham2024circuit,
  title={Circuit decompositions and scheduling for neutral atom devices with limited local addressability},
  author={Nottingham, Natalia and Perlin, Michael A and Shah, Dhirpal and White, Ryan and Bernien, Hannes and Chong, Frederic T and Baker, Jonathan M},
  booktitle={Proceedings of the IEEE International Conference on Quantum Computing and Engineering (QCE '24)},
  volume={1},
  pages={854--865},
  year={2024},
  organization={IEEE},
  doi={10.1109/qce60285.2024.00105},
}

@article{tripier2026faulttolerantquantumcomputingtrapped,
      title={Fault-Tolerant Quantum Computing with Trapped Ions: The Walking Cat Architecture}, 
      author={Felix Tripier and Woo Chang Chung and Jacob Young and Safwan Alam and Bryce Bjork and Aharon Brodutch and Finn Lasse Buessen and Nolan J. Coble and Thomas Dellaert and Dmitri Maslov and Martin Roetteler and Edwin Tham and Mark Webster and Min Ye and John Gamble and Andrii Maksymov and J. P. Marceaux and Nicolas Delfosse},
      journal={arXiv preprint arXiv:2604.19481},
      year={2026}
}

@article{lee2026ion,
  title={Ion-trap chip architecture optimized for the implementation of quantum error-correcting codes},
  author={Lee, Jeonghoon and Jeon, Hyeongjun and Kim, Taehyun},
  journal={Physical Review A},
  volume={113},
  number={3},
  pages={032432},
  year={2026},
  publisher={APS},
  doi={10.1103/tk99-76gb},
}

@inproceedings{saki2022muzzle,
  title={Muzzle the shuttle: Efficient compilation for multi-trap trapped-ion quantum computers},
  author={Saki, Abdullah Ash and Topaloglu, Rasit Onur and Ghosh, Swaroop},
  booktitle={Proceedings of the Design, Automation \& Test in Europe Conference (DATE '22)},
  pages={322--327},
  year={2022},
  organization={IEEE},
  doi={10.23919/date54114.2022.9774619},
}

@inproceedings{jones2026architecting,
  title={Architecting scalable trapped ion quantum computers using surface codes},
  author={Jones, Scott and Murali, Prakash},
  booktitle={Proceedings of the ACM International Conference on Architectural Support for Programming Languages and Operating Systems (ASPLOS '26)},
  pages={175--190},
  year={2026},
  doi={10.1145/3779212.3790128},
}

@article{Barenco_1995,
   title={A universal two-bit gate for quantum computation},
   volume={449},
   ISSN={2053-9177},
doi={10.1098/rspa.1995.0066},
   number={1937},
   journal={Proceedings of the Royal Society of London. Series A: Mathematical and Physical Sciences},
   publisher={The Royal Society},
   author={Barenco, Adriano},
   year={1995},
   month={June}, pages={679--683} }

@article{wu2023enablingfullstackquantumcomputing,
      title={Enabling Full-Stack Quantum Computing with Changeable Error-Corrected Qubits}, 
      author={Anbang Wu and Keyi Yin and Andrew W. Cross and Ang Li and Yufei Ding},
      journal={arXiv preprint arXiv:2305.07072},
      year={2023}
}

@inproceedings{tan2022qubit_atom,
  title={Qubit Mapping for Reconfigurable Atom Arrays},
  author={Tan, Bochen and Bluvstein, Dolev and Lukin, Mikhail D and Cong, Jason},
  booktitle={Proceedings of the IEEE/ACM International Conference on Computer-Aided Design (ICCAD '22)},
  pages={1--9},
  year={2022},
  doi={10.1145/3508352.3549331},
}

@inproceedings{ruan2024powermove,
  title={{PowerMove}: Optimizing Compilation for Neutral Atom Quantum Computers with Zoned Architecture},
  author={Ruan, Jixuan and Fang, Xiang and Zhang, Hezi and Li, Ang and Humble, Travis and Ding, Yufei},
  booktitle={Proceedings of the ACM International Conference on Architectural Support for Programming Languages and Operating Systems (ASPLOS '25)},
  pages={163--178},
  year={2025},
  doi={10.1145/3676642.3736128},
}

@article{lin2024reuse,
  title={Reuse-aware compilation for zoned quantum architectures based on neutral atoms},
  author={Lin, Wan-Hsuan and Tan, Daniel Bochen and Cong, Jason},
  journal={arXiv preprint arXiv:2411.11784},
  year={2024}
}

@article{tan2023compiling_atom,
  title={Compiling Quantum Circuits for Dynamically Field-Programmable Neutral Atoms Array Processors},
  author={Tan, Daniel Bochen and Bluvstein, Dolev and Lukin, Mikhail D and Cong, Jason},
  journal={Quantum},
  volume={8},
  pages={1281},
  year={2024},
  doi={10.22331/q-2024-03-14-1281},
}

@article{huang2024zap,
  title={{ZAP}: Zoned Architecture and Performant Compiler for Field Programmable Atom Array},
  author={Huang, Chen and Zhao, Xi and Xu, Hongze and Zhuang, Weifeng and Hu, Meng-Jun and Liu, Dong E and Wang, Jingbo},
  journal={IEEE Transactions on Quantum Engineering},
  volume={7},
  pages={3103619--3103619},
  year={2026},
  doi={10.1109/tqe.2026.3696707},
}

@article{wang2023q,
  title={Q-Pilot: field programmable quantum array compilation with flying ancillas},
  author={Wang, Hanrui and Tan, Bochen and Liu, Pengyu and Liu, Yilian and Gu, Jiaqi and Cong, Jason and Han, Song},
  journal={arXiv preprint arXiv:2311.16190},
  year={2023}
}

@inproceedings{wang2024atomique,
  title={Atomique: A quantum compiler for reconfigurable neutral atom arrays},
  author={Wang, Hanrui and Liu, Pengyu and Tan, Daniel Bochen and Liu, Yilian and Gu, Jiaqi and Pan, David Z and Cong, Jason and Acar, Umut A and Han, Song},
  booktitle={Proceedings of the ACM/IEEE International Symposium on Computer Architecture (ISCA '24)},
  pages={293--309},
  year={2024},
  organization={IEEE},
  doi={10.1109/isca59077.2024.00030},
}

@article{tan2024compilation,
  title={Compilation for Dynamically Field-Programmable Qubit Arrays with Efficient and Provably Near-Optimal Scheduling},
  author={Tan, Daniel Bochen and Lin, Wan-Hsuan and Cong, Jason},
  journal={arXiv preprint arXiv:2405.15095},
  year={2024}
}

@inproceedings{baker2021exploiting_atom,
  title={Exploiting long-distance interactions and tolerating atom loss in neutral atom quantum architectures},
  author={Baker, Jonathan M and Litteken, Andrew and Duckering, Casey and Hoffmann, Henry and Bernien, Hannes and Chong, Frederic T},
  booktitle={Proceedings of the ACM/IEEE International Symposium on Computer Architecture (ISCA '21)},
  pages={818--831},
  year={2021},
  organization={IEEE},
  doi={10.1109/isca52012.2021.00069},
}

@article{zhu2025s,
  title={S-SYNC: Shuttle and Swap Co-Optimization in Quantum Charge-Coupled Devices},
  author={Zhu, Chenghong and Wu, Xian and Wang, Jingbo and Wang, Xin},
  journal={arXiv preprint arXiv:2505.01316},
  year={2025}
}

@article{ruan2025trapsimdsimdawarecompileroptimization,
      title={TrapSIMD: SIMD-Aware Compiler Optimization for 2D Trapped-Ion Quantum Machines}, 
      author={Jixuan Ruan and Hezi Zhang and Xiang Fang and Ang Li and Wesley C. Campbell and Eric Hudson and David Haye and Hartmut Haeffner and Travis Humble and Jens Palsberg and Yufei Ding},
      journal={arXiv preprint arXiv:2504.17886},
      year={2025}
}

@inproceedings{schoenberger2024using,
  title={Using Boolean satisfiability for exact shuttling in trapped-ion quantum computers},
  author={Schoenberger, Daniel and Hillmich, Stefan and Brandl, Matthias and Wille, Robert},
  booktitle={Proceedings of the Asia and South Pacific Design Automation Conference (ASP-DAC '24)},
  pages={127--133},
  year={2024},
  organization={IEEE},
  doi={10.1109/asp-dac58780.2024.10473902},
}

@article{schoenberger2024shuttling,
  title={Shuttling for Scalable Trapped-Ion Quantum Computers},
  author={Schoenberger, Daniel and Hillmich, Stefan and Brandl, Matthias and Wille, Robert},
  journal={IEEE Transactions on Computer-Aided Design of Integrated Circuits and Systems},
  volume={44},
  number={6},
  pages={2144--2155},
  year={2025},
  doi={10.1109/tcad.2024.3513262},
}

@inproceedings{murali2020architecting_iontrap,
  title={Architecting noisy intermediate-scale trapped ion quantum computers},
  author={Murali, Prakash and Debroy, Dripto M and Brown, Kenneth R and Martonosi, Margaret},
  booktitle={Proceedings of the ACM/IEEE International Symposium on Computer Architecture (ISCA '20)},
  pages={529--542},
  year={2020},
  organization={IEEE},
  doi={10.1109/isca45697.2020.00051},
}

@article{kaushal2020shuttling,
  title={Shuttling-based trapped-ion quantum information processing},
  author={Kaushal, Vidyut and Lekitsch, Bjoern and Stahl, A and Hilder, J and Pijn, Daniel and Schmiegelow, C and Bermudez, Alejandro and M{\"u}ller, M and Schmidt-Kaler, Ferdinand and Poschinger, U},
  journal={AVS Quantum Science},
  volume={2},
  number={1},
  year={2020},
  publisher={AIP Publishing},
  doi={10.1116/1.5126186},
}

@article{pino2021demonstration,
  title={Demonstration of the trapped-ion quantum CCD computer architecture},
  author={Pino, Juan M and Dreiling, Jennifer M and Figgatt, Caroline and Gaebler, John P and Moses, Steven A and Allman, MS and Baldwin, CH and Foss-Feig, Michael and Hayes, David and Mayer, Karl and others},
  journal={Nature},
  volume={592},
  number={7853},
  pages={209--213},
  year={2021},
  publisher={Nature Publishing Group UK London},
  doi={10.1038/s41586-021-03318-4},
}

@inproceedings{wu2025muss,
  title={MUSS-TI: Multi-level Shuttle Scheduling for Large-Scale Entanglement Module Linked Trapped-Ion},
  author={Wu, Xian and Zhu, Chenghong and Wang, Jingbo and Wang, Xin},
  booktitle={Proceedings of the IEEE/ACM International Symposium on Microarchitecture (MICRO '25)},
  pages={749--763},
  year={2025},
  doi={10.1145/3725843.3756129},
}

@inproceedings{tian2025youtiao,
  title={YOUTIAO: Hybrid Multiplexing with Dynamic Qubit Grouping for Low-cost and Scalable Quantum Wiring},
  author={Tian, Wuwei and Lu, Liqiang and Tan, Siwei and Li, Shiyu and Li, Hengyi and Chu, Tianyao and Zhang, Xuhong and Chen, Mingshuai and Yin, Jianwei},
  booktitle={Proceedings of the IEEE/ACM International Symposium on Microarchitecture (MICRO '25)},
  pages={595--608},
  year={2025},
  doi={10.1145/3725843.3756061},
}

@article{weilandt2025minimizing,
  title={Minimizing the Number of Code Switching Operations in Fault-Tolerant Quantum Circuits},
  author={Weilandt, Erik and Peham, Tom and Wille, Robert},
  journal={arXiv preprint arXiv:2512.04170},
  year={2025},
}

@misc{latticesurgery_compiler_github,
  author       = {{latticesurgery-com}},
  title        = {{Lattice Surgery Compiler}},
  howpublished = {\url{https://github.com/latticesurgery-com/lattice-surgery-compiler}},
  note         = {Compile logical quantum circuits to lattice surgery operations on a surface code lattice. Accessed: 2025-12-28},
  year         = {2025}
}

@article{leblond2023realistic,
  title={Realistic Cost to Execute Practical Quantum Circuits using Direct Clifford+ T Lattice Surgery Compilation},
  author={LeBlond, Tyler and Dean, Christopher and Watkins, George and Bennink, Ryan},
  journal={ACM Transactions on Quantum Computing},
  year={2023},
  publisher={ACM New York, NY},
  doi={10.1145/3689826},
}

@inproceedings{shor1994algorithms,
  title={Algorithms for quantum computation: discrete logarithms and factoring},
  author={Shor, Peter W.},
  booktitle={Proceedings of the IEEE Symposium on Foundations of Computer Science (FOCS '94)},
  pages={124--134},
  year={1994},
  organization={IEEE},
  doi={10.1109/sfcs.1994.365700}
}

@inproceedings{grover1996fast,
  title={A fast quantum mechanical algorithm for database search},
  author={Grover, Lov K.},
  booktitle={Proceedings of the ACM Symposium on Theory of Computing (STOC '96)},
  pages={212--219},
  year={1996},
  doi={10.1145/237814.237866}
}

@article{aspuru2005simulated,
  title={Simulated quantum computation of molecular energies},
  author={Aspuru-Guzik, Alan and Dutoi, Anthony D. and Love, Peter J. and Head-Gordon, Martin},
  journal={Science},
  volume={309},
  number={5741},
  pages={1704--1707},
  year={2005},
  publisher={American Association for the Advancement of Science},
  doi={10.1126/science.1113479}
}

@article{biamonte2017quantum,
  title={Quantum machine learning},
  author={Biamonte, Jacob and Wittek, Peter and Pancotti, Nicola and Rebentrost, Patrick and Wiebe, Nathan and Lloyd, Seth},
  journal={Nature},
  volume={549},
  number={7671},
  pages={195--202},
  year={2017},
  publisher={Nature Publishing Group},
  doi={10.1038/nature23474}
}

@article{arute2019quantum,
  title={Quantum supremacy using a programmable superconducting processor},
  author={Arute, Frank and Arya, Kunal and Babbush, Ryan and Bacon, Dave and Bardin, Joseph C. and Barends, Rami and Biswas, Rupak and Boixo, Sergio and Brandao, Fernando G. S. L. and Buell, David A. and others},
  journal={Nature},
  volume={574},
  number={7779},
  pages={505--510},
  year={2019},
  publisher={Nature Publishing Group},
  doi={10.1038/s41586-019-1666-5}
}

@article{zhong2020quantum,
  title={Quantum computational advantage using photons},
  author={Zhong, Han-Sen and Wang, Hui and Deng, Yu-Hao and Chen, Ming-Cheng and Peng, Li-Chao and Luo, Yi-Han and Qin, Jian and Wu, Dian and Ding, Xing and Hu, Yi and others},
  journal={Science},
  volume={370},
  number={6523},
  pages={1460--1463},
  year={2020},
  publisher={American Association for the Advancement of Science},
  doi={10.1126/science.abe8770}
}

@article{liu2025certified,
  title={Certified randomness using a trapped-ion quantum processor},
  author={Liu, Minzhao and Shaydulin, Ruslan and Niroula, Pradeep and DeCross, Matthew and Hung, Shih-Han and Kon, Wen Yu and Cervero-Mart{\'i}n, Enrique and Chakraborty, Kaushik and Amer, Omar and Aaronson, Scott and others},
  journal={Nature},
  volume={640},
  number={8058},
  pages={343--348},
  year={2025},
  doi={10.1038/s41586-025-08737-1}
}

@article{bruzewicz2019trapped,
  title={Trapped-ion quantum computing: Progress and challenges},
  author={Bruzewicz, Colin D. and Chiaverini, John and McConnell, Robert and Sage, Jeremy M.},
  journal={Applied Physics Reviews},
  volume={6},
  number={2},
  pages={021314},
  year={2019},
  publisher={AIP Publishing},
  doi={10.1063/1.5088164}
}

@article{henriet2020quantum,
  title={Quantum computing with neutral atoms},
  author={Henriet, Loic and Beguin, Lucas and Signoles, Adrien and Lahaye, Thierry and Browaeys, Antoine and Reymond, Georges-Olivier and Jurczak, Christophe},
  journal={Quantum},
  volume={4},
  pages={327},
  year={2020},
  publisher={Verein zur F{\"o}rderung des Open Access Publizierens in den Quantenwissenschaften},
  doi={10.22331/q-2020-09-21-327}
}

@article{kjaergaard2020superconducting,
  title={Superconducting qubits: Current state of play},
  author={Kjaergaard, Morten and Schwartz, Mollie E. and Braumuller, Jochen and Krantz, Philip and Wang, Joel I.-J. and Gustavsson, Simon and Oliver, William D.},
  journal={Annual Review of Condensed Matter Physics},
  volume={11},
  pages={369--395},
  year={2020},
  publisher={Annual Reviews},
  doi={10.1146/annurev-conmatphys-031119-050605}
}

@article{castelvecchi2023ibm,
  title={IBM releases first-ever 1,000-qubit quantum chip},
  author={Castelvecchi, Davide},
  journal={Nature},
  volume={624},
  number={7991},
  pages={238--238},
  year={2023},
  publisher={Nature},
  doi={10.1038/d41586-023-03854-1},
}

@article{ransford2025helios98qubittrappedionquantum,
      title={A 98-qubit trapped-ion quantum computer with all-to-all connectivity}, 
      author={Anthony Ransford and M. S. Allman and Jake Arkinstall and J. P. Campora III and Samuel F. Cooper and Robert D. Delaney and Joan M. Dreiling and Brian Estey and Caroline Figgatt and Alex Hall and Ali A. Husain and Akhil Isanaka and Colin J. Kennedy and Nikhil Kotibhaskar and Ivaylo S. Madjarov and Karl Mayer and Alistair R. Milne and Annie J. Park and Adam P. Reed and Riley Ancona and Molly P. Andersen and Pablo Andres-Martinez and Will Angenent and Liz Argueta and Benjamin Arkin and Leonardo Ascarrunz and William Baker and Corey Barnes and John Bartolotta and Jordan Berg and Ryan Besand and Bryce Bjork and Matt Blain and Paul Blanchard and Robin Blume-Kohout and Matt Bohn and Agustin Borgna and Daniel Y. Botamanenko and Robert Boutelle and Natalie Brown and Grant T. Buckingham and Nathaniel Q. Burdick and William Cody Burton and Varis Carey and Christopher J. Carron and Joe Chambers and John Children and Victor E. Colussi and Steven Crepinsek and Andrew Cureton and Joe Davies and Daniel Davis and Matthew DeCross and David Deen and Conor Delaney and Davide DelVento and B. J. DeSalvo and Jason Dominy and Ross Duncan and Vanya Eccles and Alec Edgington and Neal Erickson and Stephen Erickson and Christopher T. Ertsgaard and Bruce Evans and Tyler Evans and Maya I. Fabrikant and Andrew Fischer and Cameron Foltz and Michael Foss-Feig and David Francois and Brad Freyberg and Charles Gao and Robert Garay and Jane Garvin and David M. Gaudiosi and Christopher N. Gilbreth and Josh Giles and Erin Glynn and Jeff Graves and Azure Hansen and David Hayes and Lukas Heidemann and Bob Higashi and Tyler Hilbun and Jordan Hines and Ariana Hlavaty and Kyle Hoffman and Ian M. Hoffman and Craig Holliman and Isobel Hooper and Bob Horning and James Hostetter and Daniel Hothem and Jack Houlton and Jared Hout and Ross Hutson and Ryan T. Jacobs and Trent Jacobs and Melf Johannsen and Jacob Johansen and Loren Jones and Sydney Julian and Ryan Jung and Aidan Keay and Todd Klein and Mark Koch and Ryo Kondo and Chang Kong and Asa Kosto and Alan Lawrence and David Liefer and Michelle Lollie and Dominic Lucchetti and Nathan K. Lysne and Christian Lytle and Callum MacPherson and Andrew Malm and Spencer Mather and Brian Mathewson and Daniel Maxwell and Lauren McCaffrey and Hannah McDougall and Robin Mendoza and Michael Mills and Richard Morrison and Louis Narmour and Nhung Nguyen and Lora Nugent and Scott Olson and Daniel Ouellette and Jeremy Parks and Zach Peters and Jessie Petricka and Juan M. Pino and Frank Polito and Matthias Preidl and Gabriel Price and Timothy Proctor and McKinley Pugh and Noah Ratcliff and Daisy Raymondson and Peter Rhodes and Conrad Roman and Craig Roy and Ciaran Ryan-Anderson and Fernando Betanzo Sanchez and George Sangiolo and Tatiana Sawadski and Andrew Schaffer and Peter Schow and Jon Sedlacek and Henry Semenenko and Peter Shevchuk and Susan Shore and Peter Siegfried and Kartik Singhal and Seyon Sivarajah and Thomas Skripka and Lucas Sletten and Ben Spaun and R. Tucker Sprenkle and Paul Stoufer and Mariel Tader and Stephen F. Taylor and Travis H. Thompson and Raanan Tobey and Anh Tran and Tam Tran and Grahame Vittorini and Curtis Volin and Jim Walker and Sam White and Douglas Wilson and Quinn Wolf and Chester Wringe and Kevin Young and Jian Zheng and Kristen Zuraski and Charles H. Baldwin and Alex Chernoguzov and John P. Gaebler and Steven J. Sanders and Brian Neyenhuis and Russell Stutz and Justin G. Bohnet},
      journal={Nature},
      year={2026},
doi={10.1038/s41586-026-10676-4}, 
}

@article{Manetsch_2025,
   title={A tweezer array with 6,100 highly coherent atomic qubits},
   volume={647},
   ISSN={1476-4687},
DOI={10.1038/s41586-025-09641-4},
   number={8088},
   journal={Nature},
   publisher={Springer Science and Business Media LLC},
   author={Manetsch, Hannah J. and Nomura, Gyohei and Bataille, Elie and Lv, Xudong and Leung, Kon H. and Endres, Manuel},
   year={2025},
   month=sep, pages={60–67} }

@article{beverland2022assessing,
  title={Assessing requirements to scale to practical quantum advantage},
  author={Beverland, Michael E and Murali, Prakash and Troyer, Matthias and Svore, Krysta M and Hoefler, Torsten and Kliuchnikov, Vadym and Low, Guang Hao and Soeken, Mathias and Sundaram, Aarthi and Vaschillo, Alexander},
  journal={arXiv preprint arXiv:2211.07629},
  year={2022}
}

@article{khan2025cyclone,
  title={Cyclone: Designing Efficient and Highly Parallel QCCD Architectural Codesigns for Fault Tolerant Quantum Memory},
  author={Khan, Sahil and Anand, Abhinav and Brown, Kenneth R and Baker, Jonathan M},
  journal={arXiv preprint arXiv:2511.15910},
  year={2025}
}

@article{ross2016optimalancillafreecliffordtapproximation,
      title={Optimal ancilla-free Clifford+{T} approximation of {Z}-rotations}, 
      author={Neil J. Ross and Peter Selinger},
      journal={Quantum Information and Computation},
      volume={16},
      number={11--12},
      pages={901--953},
      year={2016},
doi={10.26421/qic16.11-12-1},
}

@article{dawson2005solovaykitaevalgorithm,
      title={The Solovay-Kitaev algorithm}, 
      author={Christopher M. Dawson and Michael A. Nielsen},
      journal={Quantum Information and Computation},
      volume={6},
      number={1},
      pages={81--95},
      year={2006},
doi={10.26421/qic6.1-6},
}

@inproceedings{sethi2025rescq,
  title={{RESCQ}: Realtime Scheduling for Continuous Angle Quantum Error Correction Architectures},
  author={Sethi, Sayam and Baker, Jonathan Mark},
  booktitle={Proceedings of the ACM International Conference on Architectural Support for Programming Languages and Operating Systems (ASPLOS '25)},
  pages={1028--1043},
  year={2025},
  doi={10.1145/3676641.3716018},
}

@article{akahoshi2025compilation,
  title={Compilation of trotter-based time evolution for partially fault-tolerant quantum computing architecture},
  author={Akahoshi, Yutaro and Toshio, Riki and Fujisaki, Jun and Oshima, Hirotaka and Sato, Shintaro and Fujii, Keisuke},
  journal={PRX Quantum},
  volume={6},
  number={4},
  pages={040319},
  year={2025},
  publisher={APS},
  doi={10.1103/93zr-1ykb},
}

@inproceedings{dangwal2025variational,
  title={Variational quantum algorithms in the era of early fault tolerance},
  author={Dangwal, Siddharth and Vittal, Suhas and Seifert, Lennart Maximilian and Chong, Frederic T and Ravi, Gokul Subramanian},
  booktitle={Proceedings of the ACM/IEEE International Symposium on Computer Architecture (ISCA '25)},
  pages={1417--1431},
  year={2025},
  doi={10.1145/3695053.3731112},
}

@article{akahoshi2024partially,
  title={Partially Fault-Tolerant Quantum Computing Architecture with Error-Corrected {Clifford} Gates and Space-Time Efficient Analog Rotations},
  author={Akahoshi, Yutaro and Maruyama, Kazunori and Oshima, Hirotaka and Sato, Shintaro and Fujii, Keisuke},
  journal={PRX Quantum},
  volume={5},
  number={1},
  pages={010337},
  year={2024},
  publisher={APS},
  doi={10.1103/prxquantum.5.010337},
}

@article{zhou2025louvre,
  title={{Louvre}: Relaxing Hardware Requirements of Quantum {LDPC} Codes by Routing with Expanded Quantum Instruction Set},
  author={Zhou, Runshi and Zhang, Fang and Zhao, Hui-Hai and Wu, Feng and Kong, Linghang and Chen, Jianxin},
  journal={arXiv preprint arXiv:2508.20858},
  year={2025},
}

@article{trochatos2025trace,
  title={Trace-Based Reconstruction of Quantum Circuit Dataflow in Surface Codes},
  author={Trochatos, Theodoros and Kang, Christopher and Wang, Andrew and Chong, Frederic T and Szefer, Jakub},
  journal={arXiv preprint arXiv:2508.14533},
  year={2025}
}

@article{wang2025tableau,
  title={Tableau-Based Framework for Efficient Logical Quantum Compilation},
  author={Wang, Meng and Liu, Chenxu and Garner, Sean and Stein, Samuel and Ding, Yufei and Nair, Prashant J and Li, Ang},
  journal={arXiv preprint arXiv:2509.02721},
  year={2025},
}

@article{herr2017optimization,
  title={Optimization of lattice surgery is NP-hard},
  author={Herr, Daniel and Nori, Franco and Devitt, Simon J},
  journal={Npj quantum information},
  volume={3},
  number={1},
  pages={35},
  year={2017},
  publisher={Nature Publishing Group UK London},
  doi={10.1038/s41534-017-0035-1},
}

@article{zhou2025lowoverhead,
  title={Low-overhead transversal fault tolerance for universal quantum computation},
  author={Zhou, Hengyun and Zhao, Chen and Cain, Madelyn and Bluvstein, Dolev and Maskara, Nishad and Duckering, Casey and Hu, Hong-Ye and Wang, Sheng-Tao and Kubica, Aleksander and Lukin, Mikhail D.},
  journal={Nature},
  volume={646},
  number={8084},
  pages={303--308},
  year={2025},
  doi={10.1038/s41586-025-09543-5}
}

@article{benhemou2025minimising,
  title={Minimising surface-code failures using a color-code decoder},
  author={Benhemou, Asmae and Sahay, Kaavya and Lao, Lingling and Brown, Benjamin J.},
  journal={Quantum},
  volume={9},
  pages={1632},
  year={2025},
  doi={10.22331/q-2025-02-17-1632}
}

@article{senior2025alphaqubit2,
  title={A scalable and real-time neural decoder for topological quantum codes},
  author={Senior, Andrew W. and Edlich, Thomas and Heras, Francisco J. H. and Zhang, Lei M. and Higgott, Oscar and Spencer, James S. and others},
  journal={arXiv preprint arXiv:2512.07737},
  year={2025},
  doi={10.48550/arXiv.2512.07737},
  note={Version 2, revised March 2026}
}

@article{zhang2026selfcoordinating,
  title={Learning to Decode in Parallel: Self-Coordinating Neural Network for Real-Time Quantum Error Correction},
  author={Zhang, Kai and Yi, Zhengzhong and Guo, Shaojun and Kong, Linghang and Wang, Situ and Zhan, Xiaoyu and He, Tan and Lin, Weiping and Jiang, Tao and Gao, Dongxin and Zhang, Yiming and Liu, Fangming and Zhang, Fang and Ji, Zhengfeng and Chen, Fusheng and Chen, Jianxin},
  journal={arXiv preprint arXiv:2601.09921},
  year={2026},
  doi={10.48550/arXiv.2601.09921}
}

@article{bonillaataides2026neuralalgorithms,
  title={Neural Decoders for Universal Quantum Algorithms},
  author={Bonilla Ataides, J. Pablo and Gu, Andi and Yelin, Susanne F. and Lukin, Mikhail D.},
  journal={PRX Intelligence},
  volume={1},
  number={1},
  pages={013008},
  year={2026},
  doi={10.1103/vjn1-mbxl}
}

@article{liu2026correlatedcolor,
  title={The correlated matching decoder for the 4.8.8 color code},
  author={Liu, Yantong and Wu, Junjie and Lao, Lingling},
  journal={Science China Information Sciences},
  volume={69},
  number={8},
  pages={180508},
  year={2026},
  doi={10.1007/s11432-026-5027-y}
}

@article{lao2026thqlink,
  title={Real-time decoding of quantum error correction codes using high-performance computing},
  author={Lao, Lingling and Wang, Qiang and Liu, Yuanqi and Liu, Yantong and Wang, Haowen and Chen, Yitao and Zhao, Yankang and Wu, Zhenwei and Zhang, Wei and Dong, Yong and Liu, Yingwen and Lai, Mingche and Wu, Junjie},
  journal={arXiv preprint arXiv:2608.03948},
  year={2026},
  doi={10.48550/arXiv.2608.03948}
}

@inproceedings{zhou2025transversalresources,
  title={Resource Analysis of Low-Overhead Transversal Architectures for Reconfigurable Atom Arrays},
  author={Zhou, Hengyun and Duckering, Casey and Zhao, Chen and Bluvstein, Dolev and Cain, Madelyn and Kubica, Aleksander and Wang, Sheng-Tao and Lukin, Mikhail D.},
  booktitle={Proceedings of the 52nd Annual International Symposium on Computer Architecture},
  pages={1432--1448},
  publisher={ACM},
  year={2025},
  doi={10.1145/3695053.3731039}
}

@article{molavi2025dependency,
  title={Dependency-aware compilation for surface code quantum architectures},
  author={Molavi, Abtin and Xu, Amanda and Tannu, Swamit and Albarghouthi, Aws},
  journal={Proceedings of the ACM on Programming Languages},
  volume={9},
  number={OOPSLA1},
  pages={57--84},
  year={2025},
  month={apr},
  articleno={82},
  numpages={28},
  publisher={ACM New York, NY, USA},
  doi={10.1145/3720416},
}

@inproceedings{khan2025movelessminimizingoverheadqccds,
      title={Moveless: Minimizing {QEC} Overhead on {QCCDs} via Versatile Execution and Low Excess Shuttling}, 
      author={Sahil Khan and Suhas Vittal and Kenneth Brown and Jonathan Baker},
      booktitle={Proceedings of the IEEE International Conference on Quantum Computing and Engineering (QCE '25)},
      pages={637--648},
      year={2025},
doi={10.1109/qce65121.2025.00075},
}

@article{wang2024efficient,
  title={Efficient fault-tolerant implementations of non-Clifford gates with reconfigurable atom arrays},
  author={Wang, Yifei and Wang, Yixu and Chen, Yu-An and Zhang, Wenjun and Zhang, Tao and Hu, Jiazhong and Chen, Wenlan and Gu, Yingfei and Liu, Zi-Wen},
  journal={npj Quantum Information},
  volume={10},
  number={1},
  pages={136},
  year={2024},
  publisher={Nature Publishing Group UK London},
  doi={10.1038/s41534-024-00945-3},
}

@inproceedings{liu2025coniqenablingconcatenatedquantum,
      title={{ConiQ}: Enabling Concatenated Quantum Error Correction on Neutral Atom Arrays}, 
      author={Pengyu Liu and Mingkuan Xu and Hengyun Zhou and Hanrui Wang and Umut A. Acar and Yunong Shi},
      booktitle={Proceedings of the IEEE International Conference on Quantum Computing and Engineering (QCE '25)},
      pages={615--626},
      year={2025},
doi={10.1109/qce65121.2025.00073},
}

@article{zhu2025quantumcompilerdesignqubit,
      title={Quantum Compiler Design for Qubit Mapping and Routing: A Cross-Architectural Survey of Superconducting, Trapped-Ion, and Neutral Atom Systems}, 
      author={Chenghong Zhu and Xian Wu and Zhaohui Yang and Jingbo Wang and Anbang Wu and Shenggen Zheng and Xin Wang},
      journal={arXiv preprint arXiv:2505.16891},
      year={2025}
}

@inproceedings{stade2024abstract,
  title={An abstract model and efficient routing for logical entangling gates on zoned neutral atom architectures},
  author={Stade, Yannick and Schmid, Ludwig and Burgholzer, Lukas and Wille, Robert},
  booktitle={Proceedings of the IEEE International Conference on Quantum Computing and Engineering (QCE '24)},
  volume={1},
  pages={784--795},
  year={2024},
  organization={IEEE},
  doi={10.1109/qce60285.2024.00098},
}

@book{nielsen2010quantum,
  title={Quantum Computation and Quantum Information},
  author={Nielsen, Michael A and Chuang, Isaac L},
  year={2010},
  publisher={Cambridge University Press},
  edition={10th Anniversary},
  doi={10.1017/cbo9780511976667},
}

@article{eastin2009restrictions,
  title={Restrictions on transversal encoded quantum gate sets},
  author={Eastin, Bryan and Knill, Emanuel},
  journal={Physical Review Letters},
  volume={102},
  number={11},
  pages={110502},
  year={2009},
  publisher={APS},
  doi={10.1103/physrevlett.102.110502},
}

@article{horsman2012surface,
  title={Surface code quantum computing by lattice surgery},
  author={Horsman, Dominic and Fowler, Austin G and Devitt, Simon and Van Meter, Rodney},
  journal={New Journal of Physics},
  volume={14},
  number={12},
  pages={123011},
  year={2012},
  publisher={IOP Publishing},
  doi={10.1088/1367-2630/14/12/123011},
}

@article{litinski2019game,
  title={A game of surface codes: Large-scale quantum computing with lattice surgery},
  author={Litinski, Daniel},
  journal={Quantum},
  volume={3},
  pages={128},
  year={2019},
  publisher={Verein zur F{\"o}rderung des Open Access Publizierens in den Quantenwissenschaften},
  doi={10.22331/q-2019-03-05-128},
}

@article{lao2018mapping,
  title={Mapping of lattice surgery-based quantum circuits on surface code architectures},
  author={Lao, Lingling and van Wee, Bas and Ashraf, Imran and Van Someren, J and Khammassi, Nader and Bertels, Koen and Almudever, Carmen G},
  journal={Quantum Science and Technology},
  volume={4},
  number={1},
  pages={015005},
  year={2018},
  publisher={IOP Publishing},
  doi={10.1088/2058-9565/aadd1a},
}

@article{fowler2018low,
  title={Low overhead quantum computation using lattice surgery},
  author={Fowler, Austin G and Gidney, Craig},
  journal={arXiv preprint arXiv:1808.06709},
  year={2018}
}

@article{fowler2012surface,
  title={Surface codes: Towards practical large-scale quantum computation},
  author={Fowler, Austin G and Mariantoni, Matteo and Martinis, John M and Cleland, Andrew N},
  journal={Physical Review A—Atomic, Molecular, and Optical Physics},
  volume={86},
  number={3},
  pages={032324},
  year={2012},
  publisher={APS},
  doi={10.1103/physreva.86.032324},
}

@inproceedings{leblond2023tiscc,
  title={Tiscc: A surface code compiler and resource estimator for trapped-ion processors},
  author={LeBlond, Tyler and Bennink, Ryan S and Lietz, Justin G and Seck, Christopher M},
  booktitle={Proceedings of the IEEE/ACM International Conference for High Performance Computing, Networking, Storage and Analysis Workshops (SC Workshops '23)},
  pages={1426--1435},
  year={2023},
  doi={10.1145/3624062.3624214},
}

@article{landahl2011fault,
  title={Fault-tolerant quantum computing with color codes},
  author={Landahl, Andrew J and Anderson, Jonas T and Rice, Patrick R},
  journal={arXiv preprint arXiv:1108.5738},
  year={2011}
}

@article{viszlai2023matching,
  title={{qSIEVE}: Efficient {qLDPC} Memory via Systolic Movement in Atom Arrays},
  author={Viszlai, Joshua and Yang, Willers and Lin, Sophia Fuhui and Liu, Junyu and Nottingham, Natalia and Baker, Jonathan M and Chong, Frederic T},
  journal={ACM Transactions on Quantum Computing},
  volume={7},
  number={2},
  articleno={9},
  pages={1--26},
  numpages={26},
  year={2026},
  publisher={ACM},
  doi={10.1145/3779066},
}

@article{acharya2024quantum,
  title={Quantum error correction below the surface code threshold},
  author={Acharya, Rajeev and Abanin, Dmitry A and Aghababaie-Beni, Laleh and Aleiner, Igor and Andersen, Trond I and Ansmann, Markus and Arute, Frank and Arya, Kunal and Asfaw, Abraham and Astrakhantsev, Nikita and others},
  journal={Nature},
  volume={638},
  number={8052},
  pages={920--926},
  year={2025},
  doi={10.1038/s41586-024-08449-y}
}

@article{bluvstein2024logical,
  title={Logical quantum processor based on reconfigurable atom arrays},
  author={Bluvstein, Dolev and Evered, Simon J and Geim, Alexandra A and Li, Sophie H and Zhou, Hengyun and Manovitz, Tom and Ebadi, Sepehr and Cain, Madelyn and Kalinowski, Marcin and Hangleiter, Dominik and others},
  journal={Nature},
  volume={626},
  number={7997},
  pages={58--65},
  year={2024},
  publisher={Nature Publishing Group UK London},
  doi={10.1038/s41586-023-06927-3}
}

@article{cain2024correlated,
  title={Correlated decoding of logical algorithms with transversal gates},
  author={Cain, Madelyn and Zhao, Chen and Zhou, Hengyun and Meister, Nadine and Bonilla Ataides, J. Pablo and Jaffe, Arthur and Bluvstein, Dolev and Lukin, Mikhail D.},
  journal={Physical Review Letters},
  volume={133},
  number={24},
  pages={240602},
  year={2024},
  publisher={APS},
  doi={10.1103/physrevlett.133.240602}
}

@article{cain2025fastcorrelated,
  title={Fast correlated decoding of transversal logical algorithms},
  author={Cain, Madelyn and Bluvstein, Dolev and Zhao, Chen and Gu, Shouzhen and Maskara, Nishad and Kalinowski, Marcin and Geim, Alexandra A. and Kubica, Aleksander and Lukin, Mikhail D. and Zhou, Hengyun},
  journal={arXiv preprint arXiv:2505.13587},
  year={2025},
}

@article{viszlai2023architecture,
  author={Viszlai, Joshua and Lin, Sophia Fuhui and Dangwal, Siddharth and Baker, Jonathan M. and Chong, Frederic T.},
  title={An Architecture for Improved Surface Code Connectivity in Neutral Atoms},
  journal={arXiv preprint arXiv:2309.13507},
  year={2023},
}

@inproceedings{Tan2024SATScalpel,
  author       = {Daniel Bochen Tan and Murphy Yuezhen Niu and Craig Gidney},
  title        = {{A SAT Scalpel for Lattice Surgery: Representation and Synthesis of Subroutines for Surface-Code Fault-Tolerant Quantum Computing}},
  booktitle={Proceedings of the ACM/IEEE International Symposium on Computer Architecture (ISCA '24)},
  pages        = {325--339},
  year         = {2024},
  doi          = {10.1109/isca59077.2024.00032},
}

@inproceedings{Liu2023SubstrateScheduler,
  title = {A Substrate Scheduler for Compiling Arbitrary Fault-Tolerant Graph States},
  booktitle={Proceedings of the IEEE International Conference on Quantum Computing and Engineering (QCE '23)},
  author = {Liu, Sitong and Benchasattabuse, Naphan and Morgan, Darcy QC and Hajdu{\v s}ek, Michal and Devitt, Simon J. and Van Meter, Rodney},
  year = {2023},
  month = sep,
  pages = {870--880},
  publisher = {IEEE},
  doi = {10.1109/qce57702.2023.00101},
}

@article{Watkins2024highperformance,
  title = {A {{High Performance Compiler}} for {{Very Large Scale Surface Code Computations}}},
  author = {Watkins, George and Nguyen, Hoang Minh and Watkins, Keelan and Pearce, Steven and Lau, Hoi-Kwan and Paler, Alexandru},
  year = {2024},
  month = may,
  journal = {Quantum},
  volume = {8},
  pages = {1354},
  publisher = {Verein zur F{\"o}rderung des Open Access Publizierens in den Quantenwissenschaften},
  issn = {2521-327X},
  doi = {10.22331/q-2024-05-22-1354},
}

@article{terhal2015quantum,
  title={Quantum error correction for quantum memories},
  author={Terhal, Barbara M},
  journal={Reviews of Modern Physics},
  volume={87},
  number={2},
  pages={307--346},
  year={2015},
  month={apr},
  publisher={APS},
  doi={10.1103/revmodphys.87.307},
}

@article{beverland2022SurfaceCodeCompilation,
  title = {Surface Code Compilation via Edge-Disjoint Paths},
  author = {Beverland, Michael and Kliuchnikov, Vadym and Schoute, Eddie},
  year = {2022},
  month = may,
  journal = {PRX Quantum},
  volume = {3},
  number = {2},
  publisher = {American Physical Society (APS)},
  issn = {2691-3399},
  doi = {10.1103/prxquantum.3.020342},
}

@article{hamada2024efficient,
  title = {Efficient and High-Performance Routing of Lattice-Surgery Paths on Three-Dimensional Lattice},
  author = {Hamada, Kou and Suzuki, Yasunari and Tokunaga, Yuuki},
  journal={arXiv preprint arXiv:2401.15829},
  year = {2024}
}

@article{hirano2025localityaware,
  title = {Locality-Aware {{Pauli-based}} Computation for Local Magic State Preparation},
  author = {Hirano, Yutaka and Fujii, Keisuke},
  journal={arXiv preprint arXiv:2504.12091},
  year = {2025}
}

@article{ueno2024highperformancescalable,
  title = {High-Performance and Scalable Fault-Tolerant Quantum Computation with Lattice Surgery on a {2.5D} Architecture},
  author = {Ueno, Yosuke and Saito, Taku and Tanimoto, Teruo and Suzuki, Yasunari and Tabuchi, Yutaka and Tamate, Shuhei and Nakamura, Hiroshi},
  journal={arXiv preprint arXiv:2411.17519},
  year = {2024}
}

@article{wang2024optimizingftqcprogramsqec,
  title = {Transpiler-Architecture Co-Design to Curb {Clifford} Costs in Fault-Tolerant Quantum Computing},
  author = {Wang, Meng and Liu, Chenxu and Stein, Samuel and Ding, Yufei and Das, Poulami and Nair, Prashant J. and Li, Ang},
  journal={arXiv preprint arXiv:2412.15434},
  year = {2024}
}

@article{chatterjee2025qspellbook,
  title = {The Q-Spellbook: {{Crafting}} Surface Code Layouts and Magic State Protocols for Large-Scale Quantum Computing},
  author = {Chatterjee, Avimita and Ghosh, Archisman and Ghosh, Swaroop},
  journal={arXiv preprint arXiv:2502.11253},
  year = {2025}
}

@article{kan2025sparo,
  title = {{{SPARO}}: {{Surface-code}} Pauli-Based Architectural Resource Optimization for Fault-Tolerant Quantum Computing},
  author = {Kan, Shuwen and Du, Zefan and Liu, Chenxu and Wang, Meng and Ding, Yufei and Li, Ang and Mao, Ying and Stein, Samuel},
  journal={arXiv preprint arXiv:2504.21854},
  year = {2025}
}

@article{herzog2025latticesurgerycompilation,
  title = {Lattice Surgery Compilation beyond the Surface Code},
  author = {Herzog, Laura S. and Berent, Lucas and Kubica, Aleksander and Wille, Robert},
  journal={arXiv preprint arXiv:2504.10591},
  year = {2025}
}

@article{stein2024architecture,
  title = {Architectures for Heterogeneous Quantum Error Correction Codes},
  author = {Stein, Samuel and Xu, Shifan and Cross, Andrew W. and Yoder, Theodore J. and {Javadi-Abhari}, Ali and Liu, Chenxu and Liu, Kun and Zhou, Zeyuan and Guinn, Charles and Ding, Yufei and Ding, Yongshan and Li, Ang},
  journal={arXiv preprint arXiv:2411.03202},
  year = {2024}
}

@article{xu2023constantoverhead,
  title = {Constant-Overhead Fault-Tolerant Quantum Computation with Reconfigurable Atom Arrays},
  author = {Xu, Qian and Ataides, J. Pablo Bonilla and Pattison, Christopher A. and Raveendran, Nithin and Bluvstein, Dolev and Wurtz, Jonathan and Vasic, Bane and Lukin, Mikhail D. and Jiang, Liang and Zhou, Hengyun},
  journal = {Nature Physics},
  volume = {20},
  number = {7},
  pages = {1084--1090},
  year = {2024},
doi = {10.1038/s41567-024-02479-z},
}

@inproceedings{yin2025flexion,
  title={{iSwitch}: {QEC} on Demand via In-Situ Encoding of Bare Qubits for Ion Trap Architectures},
  author={Yin, Keyi and Fang, Xiang and Chen, Zhuo and Li, Ang and Hayes, David and Kaur, Eneet and Nejabati, Reza and Haeffner, Hartmut and Campbell, Wes and Hudson, Eric and Palsberg, Jens and Humble, Travis and Ding, Yufei},
  booktitle={Proceedings of the ACM International Conference on Architectural Support for Programming Languages and Operating Systems (ASPLOS '26)},
  pages={1007--1021},
  year={2026},
  publisher={Association for Computing Machinery},
  doi={10.1145/3779212.3790177},
}

@article{lao2020fault,
  title={Fault-tolerant quantum error correction on near-term quantum processors using flag and bridge qubits},
  author={Lao, Lingling and Almudever, Carmen G},
  journal={Physical Review A},
  volume={101},
  number={3},
  pages={032333},
  year={2020},
  publisher={APS},
  doi={10.1103/physreva.101.032333},
}

@article{Bluvstein2022Aquantum,
   title={A quantum processor based on coherent transport of entangled atom arrays},
   volume={604},
   ISSN={1476-4687},
DOI={10.1038/s41586-022-04592-6},
   number={7906},
   journal={Nature},
   publisher={Springer Science and Business Media LLC},
   author={Bluvstein, Dolev and Levine, Harry and Semeghini, Giulia and Wang, Tout T. and Ebadi, Sepehr and Kalinowski, Marcin and Keesling, Alexander and Maskara, Nishad and Pichler, Hannes and Greiner, Markus and Vuletić, Vladan and Lukin, Mikhail D.},
   year={2022},
   month=apr, pages={451–456} }

@inproceedings{Wu2022synthesis,
  title = {A Synthesis Framework for Stitching Surface Code with Superconducting Quantum Devices},
  booktitle={Proceedings of the ACM/IEEE International Symposium on Computer Architecture (ISCA '22)},
  author = {Wu, Anbang and Li, Gushu and Zhang, Hezi and Guerreschi, Gian Giacomo and Ding, Yufei and Xie, Yuan},
  year = {2022},
  pages = {337--350},
  publisher = {Association for Computing Machinery},
  address = {New York, NY, USA},
  doi = {10.1145/3470496.3527381},
}

@article{yin2024qeccsynth,
  title = {{{QECC-synth}}: A Layout Synthesizer for Quantum Error Correction Codes on Sparse Hardware Architectures},
  author = {Yin, Keyi and Zhang, Hezi and Fang, Xiang and Shi, Yunong and Humble, Travis and Li, Ang and Ding, Yufei},
  journal={arXiv preprint arXiv:2308.06428},
  year = {2024}
}

@article{fang2024caliscalpelinsitufinegrainedqubit,
  title = {{{CaliScalpel}}: {{In-Situ}} and Fine-Grained Qubit Calibration Integrated with Surface Code Quantum Error Correction},
  author = {Fang, Xiang and Yin, Keyi and Zhu, Yuchen and Ruan, Jixuan and Tullsen, Dean and Liang, Zhiding and Sornborger, Andrew and Li, Ang and Humble, Travis and Ding, Yufei and Shi, Yunong},
  journal={arXiv preprint arXiv:2412.02036},
  year = {2024}
}

@misc{haner2026computing256bitellipticcurve,
      title={Computing 256-bit elliptic curve discrete logarithms in 26 days on a fault-tolerant trapped-ion quantum computer with 20,000 qubits}, 
      author={Thomas Haner and Felix Tripier and Jacob Young and Michael Naehrig and Andrii Maksymov and Safwan Alam and Dmitri Maslov and Matthew Parrott and Yvette de Sereville and Jordan Sullivan and Mark Webster and Nicolas Delfosse and John Gamble and Martin Roetteler},
      year={2026},
      eprint={2609.05625},
      archivePrefix={arXiv},
      primaryClass={quant-ph},
      url={https://arxiv.org/abs/2609.05625}, 
}

@article{bravyi2024high,
  title={High-threshold and low-overhead fault-tolerant quantum memory},
  author={Bravyi, Sergey and Cross, Andrew W and Gambetta, Jay M and Maslov, Dmitri and Rall, Patrick and Yoder, Theodore J},
  journal={Nature},
  volume={627},
  number={8005},
  pages={778--782},
  year={2024},
  publisher={Nature Publishing Group UK London},
  doi={10.1038/s41586-024-07107-7}
}

@article{yoder2025tour,
  title={Tour de gross: A modular quantum computer based on bivariate bicycle codes},
  author={Yoder, Theodore J and Schoute, Eddie and Rall, Patrick and Pritchett, Emily and Gambetta, Jay M and Cross, Andrew W and Carroll, Malcolm and Beverland, Michael E},
  journal={arXiv preprint arXiv:2506.03094},
  year={2025},
}

@article{breuckmann2021quantum,
  title={Quantum Low-Density Parity-Check Codes},
  author={Breuckmann, Nikolas P and Eberhardt, Jens Niklas},
  journal={PRX Quantum},
  volume={2},
  number={4},
  pages={040101},
  year={2021},
  publisher={APS},
  doi={10.1103/prxquantum.2.040101},
}

@article{kovalev2013quantum,
  title={Quantum Kronecker sum-product low-density parity-check codes with finite rate},
  author={Kovalev, Alexey A and Pryadko, Leonid P},
  journal={Physical Review A},
  volume={88},
  number={1},
  pages={012311},
  year={2013},
  publisher={APS},
  doi={10.1103/physreva.88.012311},
}

@article{eberhardt2024logical,
  title={Logical operators and fold-transversal gates of bivariate bicycle codes},
  author={Eberhardt, Jens Niklas and Steffan, Vincent},
  journal={IEEE Transactions on Information Theory},
  year={2024},
  publisher={IEEE},
  doi={10.1109/tit.2024.3521638},
}

@article{gottesman2013fault,
  title={Fault-tolerant quantum computation with constant overhead},
  author={Gottesman, Daniel},
  journal={Quantum Information and Computation},
  volume={14},
  number={15--16},
  pages={1338--1371},
  year={2014},
  eprint={1310.2984},
  archivePrefix={arXiv},
  primaryClass={quant-ph},
  url={https://arxiv.org/abs/1310.2984}
}

@article{bombin2006topological,
  title={Topological quantum distillation},
  author={Bombin, Hector and Martin-Delgado, Miguel Angel},
  journal={Physical review letters},
  volume={97},
  number={18},
  pages={180501},
  year={2006},
  publisher={APS},
  doi={10.1103/physrevlett.97.180501},
}

@article{koh2026entangling,
      title={Entangling logical qubits without physical operations}, 
      author={Jin Ming Koh and Anqi Gong and Andrei C. Diaconu and Daniel Bochen Tan and Alexandra A. Geim and Michael J. Gullans and Norman Y. Yao and Mikhail D. Lukin and Shayan Majidy},
      year={2026},
      eprint={2601.20927},
      archivePrefix={arXiv},
      primaryClass={quant-ph},
      url={https://arxiv.org/abs/2601.20927}, 
}

@article{he2026logical,
      title={Logical Entangling with Phantom Codes in Hypergraph Products}, 
      author={Keming He and Ziao Tang and Zetong Li and Ge Bai and Xin Wang},
      year={2026},
      eprint={2607.12948},
      archivePrefix={arXiv},
      primaryClass={quant-ph},
      url={https://arxiv.org/abs/2607.12948}, 
}

@article{knill2005quantum,
  title={Quantum computing with realistically noisy devices},
  author={Knill, Emanuel},
  journal={Nature},
  volume={434},
  number={7029},
  pages={39--44},
  year={2005},
  month={mar},
  publisher={Nature Publishing Group UK London},
  doi={10.1038/nature03350},
}

@article{divincenzo2007effective,
  title={Effective fault-tolerant quantum computation with slow measurements},
  author={DiVincenzo, David P and Aliferis, Panos},
  journal={Physical review letters},
  volume={98},
  number={2},
  pages={020501},
  year={2007},
  publisher={APS},
  doi={10.1103/physrevlett.98.020501},
}

@inproceedings{riesebos2017pauli,
  title={Pauli frames for quantum computer architectures},
  author={Riesebos, Leon and Fu, Xiang and Varsamopoulos, Savvas and Almudever, Carmen G and Bertels, Koen},
  booktitle={Proceedings of the ACM/IEEE Design Automation Conference (DAC '17)},
  pages={1--6},
  year={2017},
  doi={10.1145/3061639.3062300},
}

@article{gottesman1999quantum,
  title={Quantum Teleportation Is a Universal Computational Primitive},
  author={Gottesman, Daniel and Chuang, Isaac L},
  journal={arXiv preprint quant-ph/9908010},
  year={1999},
}

@article{chamberland2018fault,
  title={Fault-tolerant quantum computing in the Pauli or Clifford frame with slow error diagnostics},
  author={Chamberland, Christopher and Iyer, Pavithran and Poulin, David},
  journal={Quantum},
  volume={2},
  pages={43},
  year={2018},
  publisher={Verein zur F{\"o}rderung des Open Access Publizierens in den Quantenwissenschaften},
  doi={10.22331/q-2018-01-04-43},
}

@article{chamberland2017hard,
  title={Hard decoding algorithm for optimizing thresholds under general markovian noise},
  author={Chamberland, Christopher and Wallman, Joel and Beale, Stefanie and Laflamme, Raymond},
  journal={Physical Review A},
  volume={95},
  number={4},
  pages={042332},
  year={2017},
  publisher={APS},
  doi={10.1103/physreva.95.042332},
}

@article{ryananderson2021realtime,
  title={Realization of Real-Time Fault-Tolerant Quantum Error Correction},
  author={Ryan-Anderson, C. and Bohnet, J. G. and Lee, K. and Gresh, D. and Hankin, A. and Gaebler, J. P. and Francois, D. and Chernoguzov, A. and Lucchetti, D. and Brown, N. C. and Gatterman, T. M. and Halit, S. K. and Gilmore, K. and Gerber, J. and Neyenhuis, B. and Hayes, D. and Stutz, R. P.},
  journal={Physical Review X},
  volume={11},
  number={4},
  pages={041058},
  year={2021},
  publisher={APS},
  doi={10.1103/physrevx.11.041058}
}

@article{dennis2002topological,
  title={Topological quantum memory},
  author={Dennis, Eric and Kitaev, Alexei and Landahl, Andrew and Preskill, John},
  journal={Journal of Mathematical Physics},
  volume={43},
  number={9},
  pages={4452--4505},
  year={2002},
  publisher={American Institute of Physics},
  doi={10.1063/1.1499754},
}

@article{tan2023scalable,
  title={Scalable surface-code decoders with parallelization in time},
  author={Tan, Xinyu and Zhang, Fang and Chao, Rui and Shi, Yaoyun and Chen, Jianxin},
  journal={PRX Quantum},
  volume={4},
  number={4},
  pages={040344},
  year={2023},
  publisher={APS},
  doi={10.1103/prxquantum.4.040344},
}

@article{skoric2023parallel,
  title={Parallel window decoding enables scalable fault tolerant quantum computation},
  author={Skoric, Luka and Browne, Dan E and Barnes, Kenton M and Gillespie, Neil I and Campbell, Earl T},
  journal={Nature Communications},
  volume={14},
  number={1},
  pages={7040},
  year={2023},
  publisher={Nature Publishing Group UK London},
  doi={10.1038/s41467-023-42482-1},
}

@article{bombin2023modular,
  title={Modular decoding: parallelizable real-time decoding for quantum computers},
  author={Bomb{\'i}n, H{\'e}ctor and Dawson, Chris and Liu, Ye-Hua and Nickerson, Naomi and Pastawski, Fernando and Roberts, Sam},
  journal={arXiv preprint arXiv:2303.04846},
  year={2023},
}

@article{lin2025spatially,
  title={Spatially parallel decoding for multi-qubit lattice surgery},
  author={Lin, Sophia Fuhui and Peterson, Eric C. and Sankar, Krishanu and Sivarajah, Prasahnt},
  journal={Quantum Science and Technology},
  year={2025},
  doi={10.1088/2058-9565/adc6b6}
}

@inproceedings{viszlai2025swiper,
  title={SWIPER: Minimizing Fault-Tolerant Quantum Program Latency via Speculative Window Decoding},
  author={Viszlai, Joshua and Chadwick, Jason D and Joshi, Sarang and Ravi, Gokul Subramanian and Li, Yanjing and Chong, Frederic T},
  booktitle={Proceedings of the ACM/IEEE International Symposium on Computer Architecture (ISCA '25)},
  pages={1386--1401},
  year={2025},
  doi={10.1145/3695053.3731022}
}

@article{oberoi2026adapt,
  title={{ADaPT}: Adaptive-window Decoding for Practical fault-Tolerance},
  author={Oberoi, Tina and Viszlai, Joshua and Chong, Frederic T.},
  journal={arXiv preprint arXiv:2605.01149},
  year={2026},
}

@article{chen2026triage,
  title={Triage: An Adaptive Parallel Window Decoding Scheduler for Real-time Fault-Tolerant Quantum Computation},
  author={Chen, Jiahan and Zhu, Chenghong and Bai, Ge and Wang, Xin},
  journal={arXiv preprint arXiv:2605.04459},
  year={2026}
}

@inproceedings{maurya2026elastic,
  title={A Case for Elastic Quantum Error Correction Decoders},
  author={Maurya, Satvik and Molavi, Abtin and Albarghouthi, Aws and Tannu, Swamit},
  booktitle={Proceedings of the 21st European Conference on Computer Systems (EuroSys '26)},
  pages={514--531},
  year={2026},
  publisher={Association for Computing Machinery},
  doi={10.1145/3767295.3803584},
}

@article{iyer2015hardness,
  title={Hardness of Decoding Quantum Stabilizer Codes},
  author={Iyer, Pavithran and Poulin, David},
  journal={IEEE Transactions on Information Theory},
  volume={61},
  number={9},
  pages={5209--5223},
  year={2015},
  publisher={IEEE},
  doi={10.1109/tit.2015.2422294},
}

@article{edmonds1965paths,
  title={Paths, Trees, and Flowers},
  author={Edmonds, Jack},
  journal={Canadian Journal of Mathematics},
  volume={17},
  pages={449--467},
  year={1965},
  publisher={Cambridge University Press},
  doi={10.4153/cjm-1965-045-4},
}

@article{higgott2025sparse,
  title={Sparse blossom: correcting a million errors per core second with minimum-weight matching},
  author={Higgott, Oscar and Gidney, Craig},
  journal={Quantum},
  volume={9},
  pages={1600},
  year={2025},
  publisher={Verein zur F{\"o}rderung des Open Access Publizierens in den Quantenwissenschaften},
  doi={10.22331/q-2025-01-20-1600},
}

@article{higgott2022pymatching,
  title={{PyMatching}: A Python package for decoding quantum codes with minimum-weight perfect matching},
  author={Higgott, Oscar},
  journal={ACM Transactions on Quantum Computing},
  volume={3},
  number={3},
  pages={1--16},
  year={2022},
doi={10.1145/3505637},
}

@article{gidney2021stim,
  title={{Stim}: a fast stabilizer circuit simulator},
  author={Gidney, Craig},
  journal={Quantum},
  volume={5},
  pages={497},
  year={2021},
  doi={10.22331/q-2021-07-06-497},
}

@article{stace2009thresholds,
  title={Thresholds for topological codes in the presence of loss},
  author={Stace, Thomas M. and Barrett, Sean D. and Doherty, Andrew C.},
  journal={Physical Review Letters},
  volume={102},
  number={20},
  pages={200501},
  year={2009},
  publisher={APS},
  doi={10.1103/physrevlett.102.200501}
}

@article{wu2022erasureconversion,
  title={Erasure conversion for fault-tolerant quantum computing in alkaline earth {Rydberg} atom arrays},
  author={Wu, Yue and Kolkowitz, Shimon and Puri, Shruti and Thompson, Jeff D.},
  journal={Nature Communications},
  volume={13},
  number={1},
  pages={4657},
  year={2022},
  publisher={Nature Publishing Group UK London},
  doi={10.1038/s41467-022-32094-6}
}

@article{kubica2023erasure,
  title={Erasure qubits: Overcoming the {$T_1$} limit in superconducting circuits},
  author={Kubica, Aleksander and Haim, Arbel and Vaknin, Yotam and Levine, Harry and Brandao, Fernando and Retzker, Alex},
  journal={Physical Review X},
  volume={13},
  number={4},
  pages={041022},
  year={2023},
  publisher={APS},
  doi={10.1103/physrevx.13.041022}
}

@article{sahay2023highthreshold,
  title={High-threshold codes for neutral-atom qubits with biased erasure errors},
  author={Sahay, Kaavya and Jin, Junlan and Claes, Jahan and Thompson, Jeff D. and Puri, Shruti},
  journal={Physical Review X},
  volume={13},
  number={4},
  pages={041013},
  year={2023},
  publisher={APS},
  doi={10.1103/physrevx.13.041013}
}

@article{perrin2025atomloss,
  title={Quantum error correction resilient against atom loss},
  author={Perrin, Hugo and Jandura, Sven and Pupillo, Guido},
  journal={Quantum},
  volume={9},
  pages={1884},
  year={2025},
  doi={10.22331/q-2025-10-13-1884}
}

@article{bluvstein2026faulttolerant,
  title={A fault-tolerant neutral-atom architecture for universal quantum computation},
  author={Bluvstein, Dolev and Geim, Alexandra A. and Li, Sophie H. and Evered, Simon J. and Bonilla Ataides, J. Pablo and Baranes, Gefen and Gu, Andi and Manovitz, Tom and Xu, Muqing and Kalinowski, Marcin and Majidy, Shayan and Kokail, Christian and Maskara, Nishad and Trapp, Elias C. and Stewart, Luke M. and Hollerith, Simon and Zhou, Hengyun and Gullans, Michael J. and Yelin, Susanne F. and Greiner, Markus and Vuletic, Vladan and Cain, Madelyn and Lukin, Mikhail D.},
  journal={Nature},
  volume={649},
  pages={39--46},
  year={2026},
  publisher={Nature Publishing Group UK London},
  doi={10.1038/s41586-025-09848-5}
}

@article{fowler2013leakage,
  title={Coping with qubit leakage in topological codes},
  author={Fowler, Austin G.},
  journal={Physical Review A},
  volume={88},
  number={4},
  pages={042308},
  year={2013},
  publisher={APS},
  doi={10.1103/physreva.88.042308}
}

@article{ziad2025localclustering,
  title={Local clustering decoder as a fast and adaptive hardware decoder for the surface code},
  author={Ziad, Abbas B. and Zalawadiya, Ankit and Topal, Canberk and Camps, Joan and Geh{\'e}r, Gy{\"o}rgy P. and Stafford, Matthew P. and Turner, Mark L.},
  journal={Nature Communications},
  volume={16},
  number={1},
  pages={11048},
  year={2025},
  publisher={Nature Publishing Group UK London},
  doi={10.1038/s41467-025-66773-x}
}

@article{beni2025tesseract,
  title={Tesseract: A Search-Based Decoder for Quantum Error Correction},
  author={Aghababaie Beni, Laleh and Higgott, Oscar and Shutty, Noah},
  journal={arXiv preprint arXiv:2503.10988},
  year={2025}
}

@article{wu2025mwpf,
  title={Minimum-Weight Parity Factor Decoder for Quantum Error Correction},
  author={Wu, Yue and Li, Binghong and Chang, Kathleen and Puri, Shruti and Zhong, Lin},
  journal={arXiv preprint arXiv:2508.04969},
  year={2025}
}

@article{ye2025beam,
  title={Beam search decoder for quantum {LDPC} codes},
  author={Ye, Min and Wecker, Dave and Delfosse, Nicolas},
  journal={arXiv preprint arXiv:2512.07057},
  year={2025},
}

@article{valentini2025restart,
  title={Restart Belief: A General Quantum {LDPC} Decoder},
  author={Valentini, Lorenzo and Forlivesi, Diego and Talarico, Andrea and Chiani, Marco},
  journal={IEEE Communications Letters},
  volume={30},
  pages={1185--1189},
  year={2026},
  doi={10.1109/lcomm.2026.3666352},
}

@inproceedings{wang2025fullyparallelized,
  title={Fully Parallelized {BP} Decoding for Quantum {LDPC} Codes Can Outperform {BP-OSD}},
  author={Wang, Ming and Li, Ang and Mueller, Frank},
  booktitle={Proceedings of the IEEE International Symposium on High-Performance Computer Architecture (HPCA '26)},
  pages={1--14},
  year={2026},
  publisher={IEEE},
  doi={10.1109/HPCA68181.2026.11408621},
  eprint={2507.00254},
  archivePrefix={arXiv},
  primaryClass={quant-ph},
  url={https://arxiv.org/abs/2507.00254},
}

@article{maan2026correlatedqldpc,
  title={Decoding correlated errors in quantum {LDPC} codes},
  author={Maan, Arshpreet Singh and Garcia Herrero, Francisco Miguel and Paler, Alexandru and Savin, Valentin},
  journal={Nature Communications},
  volume={17},
  pages={3965},
  year={2026},
  doi={10.1038/s41467-026-70556-3}
}

@article{bascones2026gari,
  title={A Scalable {FPGA} Architecture for Real-Time Decoding of Quantum {LDPC} Codes Using {GARI}},
  author={B{\'a}scones, Daniel and Maan, Arshpreet Singh and Savin, Valentin and Garcia-Herrero, Francisco},
  journal={arXiv preprint arXiv:2605.01035},
  year={2026}
}

@article{chamberland2026fastpredecoders,
  title={Fast and accurate {AI}-based pre-decoders for surface codes},
  author={Chamberland, Christopher and Olle, Jan and Li, Muyuan and Thornton, Scott and Baratta, Igor},
  journal={arXiv preprint arXiv:2604.12841},
  year={2026}
}

@article{lee2025scalableneural,
  title={Scalable Neural Decoders for Practical Real-Time Quantum Error Correction},
  author={Lee, Changwon and Hur, Tak and Park, Daniel K.},
  journal={arXiv preprint arXiv:2510.22724},
  year={2025}
}

@article{sayedsalehi2026sparsemamba,
  title={Sparse {Mamba} Decoder for Quantum Error Correction: Efficient Defect-Centric Processing of Surface Code Syndromes},
  author={Sayedsalehi, Samira and Bagherzadeh, Nader and Shcherbakov, Maxim and Gaudiot, Jean-Luc},
  journal={arXiv preprint arXiv:2605.17156},
  year={2026}
}

@article{yang2026realtimefpga,
  title={Real-time Surface-Code Error Correction Using an {FPGA}-based Neural-Network Decoder},
  author={Yang, Xiaohan and Sun, Xuandong and Wu, Zhiyi and Zhang, Jiawei and Jiang, Ji and Linpeng, Xiayu and Zhou, Yuxuan and Chu, Ji and Niu, Jingjing and Zhong, Youpeng and Liu, Song and Yu, Dapeng},
  journal={arXiv preprint arXiv:2605.04892},
  year={2026}
}

@article{zenati2025saq,
  title={{SAQ}: Stabilizer-Aware Quantum Error Correction Decoder},
  author={Zenati, David and Nachmani, Eliya},
  journal={arXiv preprint arXiv:2512.08914},
  year={2025}
}

@article{xu2026diffqec,
  title={{DiffQEC}: A versatile diffusion model for quantum error correction},
  author={Xu, Tianyi and Liu, Qinglong and Wang, Maolin and Zhang, Fei and Zhao, Zhe and Wang, Yang and Wei, Ye},
  journal={arXiv preprint arXiv:2604.24640},
  year={2026},
}

@article{yan2026rethink,
  title={Rethink the Role of Neural Decoders in Quantum Error Correction},
  author={Yan, Ge and Li, Shanchuan and Du, Yuxuan},
  journal={arXiv preprint arXiv:2605.12046},
  year={2026}
}

@article{huang2020crosstalk,
  title={Alibaba Cloud Quantum Development Platform: Surface code simulations with crosstalk},
  author={Huang, Cupjin and Ni, Xiaotong and Zhang, Fang and Newman, Michael and Ding, Dawei and Gao, Xun and Wang, Tenghui and Zhao, Hui-Hai and Wu, Feng and Zhang, Gengyan and Deng, Chunqing and Ku, Hsiang-Sheng and Chen, Jianxin and Shi, Yaoyun},
  journal={arXiv preprint arXiv:2002.08918},
  year={2020}
}

@article{pattison2021soft,
  title={Improved quantum error correction using soft information},
  author={Pattison, Christopher A. and Beverland, Michael E. and da Silva, Marcus P. and Delfosse, Nicolas},
  journal={arXiv preprint arXiv:2107.13589},
  year={2021},
}

@article{noh2022surfacegkp,
  title={Low overhead fault-tolerant quantum error correction with the surface-{GKP} code},
  author={Noh, Kyungjoo and Chamberland, Christopher and Brandao, Fernando G. S. L.},
  journal={PRX Quantum},
  volume={3},
  number={1},
  pages={010315},
  year={2022},
  publisher={APS},
  doi={10.1103/prxquantum.3.010315}
}

@article{criger2018multi,
  title={Multi-path summation for decoding 2{D} topological codes},
  author={Criger, Ben and Ashraf, Imran},
  journal={Quantum},
  volume={2},
  pages={102},
  year={2018},
  doi={10.22331/q-2018-10-19-102}
}

@inproceedings{wu2023fusion,
  title={Fusion Blossom: Fast {MWPM} Decoders for {QEC}},
  author={Wu, Yue and Zhong, Lin},
  booktitle={Proceedings of the IEEE International Conference on Quantum Computing and Engineering (QCE '23)},
  volume={1},
  pages={928--938},
  year={2023},
  organization={IEEE},
  doi={10.1109/qce57702.2023.00107},
}

@article{tian2025enhancing,
  title={Enhancing Fault-Tolerant Surface Code Decoding with Iterative Lattice Reweighting},
  author={Tian, Yi and Zheng, Y and Wang, Xiaoting and Lai, Ching-Yi},
  journal={arXiv preprint arXiv:2509.06756},
  year={2025}
}

@article{higgott2023beliefmatching,
  title={Improved decoding of circuit noise and fragile boundaries of tailored surface codes},
  author={Higgott, Oscar and Bohdanowicz, Thomas C and Kubica, Aleksander and Flammia, Steven T and Campbell, Earl T},
  journal={Physical Review X},
  volume={13},
  number={3},
  pages={031007},
  year={2023},
  publisher={APS},
  doi={10.1103/physrevx.13.031007},
}

@article{fowler2013optimal,
  title={Optimal complexity correction of correlated errors in the surface code},
  author={Fowler, Austin G},
  journal={arXiv preprint arXiv:1310.0863},
  year={2013},
}

@article{paler2023pipelined,
  title={Pipelined correlated minimum weight perfect matching of the surface code},
  author={Paler, Alexandru and Fowler, Austin G},
  journal={Quantum},
  volume={7},
  pages={1205},
  year={2023},
  publisher={Verein zur F{\"o}rderung des Open Access Publizierens in den Quantenwissenschaften},
  doi={10.22331/q-2023-12-12-1205},
}

@inproceedings{zhou2025vegapunk,
  title={Vegapunk: Accurate and Fast Decoding for Quantum LDPC Codes with Online Hierarchical Algorithm and Sparse Accelerator},
  author={Zhou, Kaiwen and Lu, Liqiang and Xiang, Debin and Tao, Chenning and Wu, Anbang and Leng, Jingwen and Liu, Fangxin and Chen, Mingshuai and Yin, Jianwei},
  booktitle={Proceedings of the IEEE/ACM International Symposium on Microarchitecture (MICRO '25)},
  pages={719--732},
  year={2025},
  doi={10.1145/3725843.3756084},
}

@article{caune2023beliefpartial,
  title={Belief propagation as a partial decoder},
  author={Caune, Laura and Reid, Brendan and Camps, Joan and Campbell, Earl},
  journal={arXiv preprint arXiv:2306.17142},
  year={2023}
}

@inproceedings{wu2025micro,
  title={Micro Blossom: Accelerated Minimum-Weight Perfect Matching Decoding for Quantum Error Correction},
  author={Wu, Yue and Liyanage, Namitha and Zhong, Lin},
  booktitle={Proceedings of the ACM International Conference on Architectural Support for Programming Languages and Operating Systems (ASPLOS '25)},
  pages={639--654},
  year={2025},
  doi={10.1145/3676641.3716005},
}

@inproceedings{vittal2023astrea,
  title={{Astrea}: Accurate Quantum Error-Decoding via Practical Minimum-Weight Perfect-Matching},
  author={Vittal, Suhas and Das, Poulami and Qureshi, Moinuddin},
  booktitle={Proceedings of the ACM/IEEE International Symposium on Computer Architecture (ISCA '23)},
  pages={1--16},
  year={2023},
  doi={10.1145/3579371.3589037},
}

@inproceedings{holmes2020nisqplus,
  title={{NISQ+}: Boosting Quantum Computing Power by Approximating Quantum Error Correction},
  author={Holmes, Adam and Jokar, Mohammad Reza and Pasandi, Ghasem and Ding, Yongshan and Pedram, Massoud and Chong, Frederic T.},
  booktitle={Proceedings of the ACM/IEEE International Symposium on Computer Architecture (ISCA '20)},
  pages={556--569},
  year={2020},
  publisher={IEEE},
  address={Piscataway, NJ, USA},
  doi={10.1109/ISCA45697.2020.00053},
}

@inproceedings{das2022lilliput,
  title={{LILLIPUT}: A Lightweight Low-Latency Lookup-Table Decoder for Near-Term Quantum Error Correction},
  author={Das, Poulami and Locharla, Aditya and Jones, Cody},
  booktitle={Proceedings of the ACM International Conference on Architectural Support for Programming Languages and Operating Systems (ASPLOS '22)},
  pages={541--553},
  year={2022},
  publisher={Association for Computing Machinery},
  address={New York, NY, USA},
  doi={10.1145/3503222.3507707},
}

@inproceedings{das2022afs,
  title={{AFS}: Accurate, Fast, and Scalable Error-Decoding for Fault-Tolerant Quantum Computers},
  author={Das, Poulami and Pattison, Christopher A. and Manne, Srilatha and Carmean, Douglas M. and Svore, Krysta M. and Qureshi, Moinuddin and Delfosse, Nicolas},
  booktitle={Proceedings of the IEEE International Symposium on High-Performance Computer Architecture (HPCA '22)},
  pages={259--273},
  year={2022},
  publisher={IEEE},
  address={Piscataway, NJ, USA},
  doi={10.1109/HPCA53966.2022.00027},
}

@inproceedings{ueno2021qecool,
  title={{QECOOL}: On-Line Quantum Error Correction with a Superconducting Decoder for Surface Code},
  author={Ueno, Yosuke and Kondo, Masaaki and Tanaka, Masamitsu and Suzuki, Yasunari and Tabuchi, Yutaka},
  booktitle={Proceedings of the ACM/IEEE Design Automation Conference (DAC '21)},
  pages={451--456},
  year={2021},
  publisher={IEEE},
  address={Piscataway, NJ, USA},
  doi={10.1109/DAC18074.2021.9586326},
}

@inproceedings{ueno2022qulatis,
  title={{QULATIS}: A Quantum Error Correction Methodology toward Lattice Surgery},
  author={Ueno, Yosuke and Kondo, Masaaki and Tanaka, Masamitsu and Suzuki, Yasunari and Tabuchi, Yutaka},
  booktitle={Proceedings of the IEEE International Symposium on High-Performance Computer Architecture (HPCA '22)},
  pages={274--287},
  year={2022},
  publisher={IEEE},
  address={Piscataway, NJ, USA},
  doi={10.1109/HPCA53966.2022.00028},
}

@inproceedings{ravi2023clique,
  title={Better Than Worst-Case Decoding for Quantum Error Correction},
  author={Ravi, Gokul Subramanian and Baker, Jonathan M. and Fayyazi, Arash and Lin, Sophia Fuhui and Javadi-Abhari, Ali and Pedram, Massoud and Chong, Frederic T.},
  booktitle={Proceedings of the ACM International Conference on Architectural Support for Programming Languages and Operating Systems (ASPLOS '23)},
  pages={88--102},
  year={2023},
  publisher={Association for Computing Machinery},
  address={New York, NY, USA},
  doi={10.1145/3575693.3575733},
}

@inproceedings{alavisamani2024promatch,
  title={{Promatch}: Extending the Reach of Real-Time Quantum Error Correction with Adaptive Predecoding},
  author={Alavisamani, Narges and Vittal, Suhas and Ayanzadeh, Ramin and Das, Poulami and Qureshi, Moinuddin},
  booktitle={Proceedings of the ACM International Conference on Architectural Support for Programming Languages and Operating Systems (ASPLOS '24)},
  pages={818--833},
  year={2024},
  publisher={Association for Computing Machinery},
  address={New York, NY, USA},
  doi={10.1145/3620666.3651339},
}

@inproceedings{knapen2026pinball,
  title={{Pinball}: A Cryogenic Predecoder for Quantum Error Correction Decoding Under Circuit-Level Noise},
  author={Knapen, Alexander and Tao, Guanchen and Mack, Jacob and Bruno, Tomas and Saligane, Mehdi and Sylvester, Dennis and Zhang, Qirui and Ravi, Gokul Subramanian},
  booktitle={Proceedings of the IEEE International Symposium on High-Performance Computer Architecture (HPCA '26)},
  pages={1--17},
  year={2026},
  publisher={IEEE},
  doi={10.1109/HPCA68181.2026.11408464},
}

@inproceedings{liang2026coset,
  title={Coset Ensemble Decoder for Quantum Error Correction with Algorithm--Hardware Co-Design},
  author={Liang, Shuang and Xu, Jubo and Bassanino, Giulio and Wang, Qianzhou and Zhou, Yidong and Lu, Yuncheng and Mo, Zhiwen and Kelly, Paul H. J. and Yuan, Bo and Luk, Wayne and Fan, Hongxiang},
  booktitle={Proceedings of the ACM/IEEE International Symposium on Computer Architecture (ISCA '26)},
  year={2026},
  note={To appear},
  eprint={2606.11076},
  archivePrefix={arXiv},
  primaryClass={cs.AR},
  url={https://arxiv.org/abs/2606.11076},
}

@article{bravyi2014maximumlikelihood,
  title={Efficient Algorithms for Maximum Likelihood Decoding in the Surface Code},
  author={Bravyi, Sergey and Suchara, Martin and Vargo, Alexander},
  journal={Physical Review A},
  volume={90},
  number={3},
  pages={032326},
  year={2014},
  doi={10.1103/PhysRevA.90.032326},
}

@article{chubb2021statistical,
  title={Statistical Mechanical Models for Quantum Codes with Correlated Noise},
  author={Chubb, Christopher T. and Flammia, Steven T.},
  journal={Annales de l'Institut Henri Poincar\'{e} D},
  volume={8},
  number={2},
  pages={269--321},
  year={2021},
  doi={10.4171/AIHPD/105},
}

@article{battistel2023realtimedecoding,
  title={Real-Time Decoding for Fault-Tolerant Quantum Computing: Progress, Challenges and Outlook},
  author={Battistel, Francesco and Chamberland, Christopher and Johar, Kauser and Overwater, Ramon W. J. and Sebastiano, Fabio and Skoric, Luka and Ueno, Yosuke and Usman, Muhammad},
  journal={Nano Futures},
  volume={7},
  number={3},
  pages={032003},
  year={2023},
  publisher={IOP Publishing},
  doi={10.1088/2399-1984/aceba6},
}

@inproceedings{maurya2025decoderbench,
  title={{decoder-bench}: Benchmarking Decoders for Quantum Error Correction},
  author={Maurya, Satvik and Viszlai, Joshua and Raveendran, Nithin and Das, Poulami and Tannu, Swamit},
  booktitle={Proceedings of the IEEE International Symposium on Workload Characterization (IISWC '25)},
  pages={286--295},
  year={2025},
  publisher={IEEE},
  doi={10.1109/IISWC66894.2025.00032},
}

@inproceedings{liyanage2025deconet,
  title={Network-Integrated Decoding System for Real-Time Quantum Error Correction with Lattice Surgery},
  author={Liyanage, Namitha and Wu, Yue and Houghton, Emmet and Zhong, Lin},
  booktitle={Proceedings of the IEEE International Conference on Quantum Computing and Engineering (QCE '25)},
  volume={1},
  pages={1148--1159},
  year={2025},
  publisher={IEEE},
  doi={10.1109/QCE65121.2025.00129},
}

@article{caune2026realtime,
  title={Demonstrating Real-Time and Low-Latency Quantum Error Correction with Superconducting Qubits},
  author={Caune, Laura and Skoric, Luka and Blunt, Nick S. and Ruban, Archibald and McDaniel, Jimmy and Valery, Joseph A. and Patterson, Andrew D. and Gramolin, Alexander V. and Majaniemi, Joonas and Barnes, Kenton M. and Bialas, Tomasz and Bu{\u{g}}dayc{\i}, Okan and Crawford, Ophelia and Geh{\'e}r, Gy{\"o}rgy P. and Krovi, Hari and Matekole, Elisha and Topal, Canberk and Poletto, Stefano and Bryant, Michael and Snyder, Kalan and Gillespie, Neil I. and Jones, Glenn and Johar, Kauser and Campbell, Earl T. and Hill, Alexander D.},
  journal={Nature Communications},
  year={2026},
  publisher={Springer Nature},
  doi={10.1038/s41467-026-73331-6},
}

@article{roffe2020qldpclandscape,
  title={Decoding Across the Quantum Low-Density Parity-Check Code Landscape},
  author={Roffe, Joschka and White, David R. and Burton, Simon and Campbell, Earl T.},
  journal={Physical Review Research},
  volume={2},
  number={4},
  pages={043423},
  year={2020},
  publisher={American Physical Society},
  doi={10.1103/PhysRevResearch.2.043423},
}

@article{delfosse2021almost,
  title={Almost-linear time decoding algorithm for topological codes},
  author={Delfosse, Nicolas and Nickerson, Naomi H},
  journal={Quantum},
  volume={5},
  pages={595},
  year={2021},
  publisher={Verein zur F{\"o}rderung des Open Access Publizierens in den Quantenwissenschaften},
  doi={10.22331/q-2021-12-02-595},
}

@article{huang2020fault,
  title={Fault-tolerant weighted union-find decoding on the toric code},
  author={Huang, Shilin and Newman, Michael and Brown, Kenneth R},
  journal={Physical Review A},
  volume={102},
  number={1},
  pages={012419},
  year={2020},
  publisher={APS},
  doi={10.1103/physreva.102.012419},
}

@article{delfosse2022toward,
  title={Toward a union-find decoder for quantum LDPC codes},
  author={Delfosse, Nicolas and Londe, Vivien and Beverland, Michael E},
  journal={IEEE Transactions on Information Theory},
  volume={68},
  number={5},
  pages={3187--3199},
  year={2022},
  publisher={IEEE},
  doi={10.1109/tit.2022.3143452},
}

@article{chan2023actis,
  title={Actis: a strictly local union--find decoder},
  author={Chan, Tim and Benjamin, Simon C},
  journal={Quantum},
  volume={7},
  pages={1183},
  year={2023},
  publisher={Verein zur F{\"o}rderung des Open Access Publizierens in den Quantenwissenschaften},
  doi={10.22331/q-2023-11-14-1183},
}

@inproceedings{liyanage2023scalable,
  title={Scalable quantum error correction for surface codes using FPGA},
  author={Liyanage, Namitha and Wu, Yue and Deters, Alexander and Zhong, Lin},
  booktitle={Proceedings of the IEEE International Conference on Quantum Computing and Engineering (QCE '23)},
  volume={1},
  pages={916--927},
  year={2023},
  organization={IEEE},
  doi={10.1109/qce57702.2023.00106},
}

@article{barber2025realtime,
  title={A real-time, scalable, fast and resource-efficient decoder for a quantum computer},
  author={Barber, Ben and Barnes, Kenton M. and Bialas, Tomasz and Bu{\u{g}}dayc{\i}, Okan and Campbell, Earl T. and Gillespie, Neil I. and Johar, Kauser and Rajan, Ram and Richardson, Adam W. and Skoric, Luka and Topal, Canberk and Turner, Mark L. and Ziad, Abbas B.},
  journal={Nature Electronics},
  volume={8},
  pages={84--91},
  year={2025},
  publisher={Nature Publishing Group},
  doi={10.1038/s41928-024-01319-5},
}

@article{poulin2008iterative,
  title={On the iterative decoding of sparse quantum codes},
  author={Poulin, David and Chung, Yeojin},
  journal={arXiv preprint arXiv:0801.1241},
  year={2008}
}

@article{chen2025improved,
  title={Improved belief propagation decoding algorithms for surface codes},
  author={Chen, Jiahan and Yi, Zhengzhong and Liang, Zhipeng and Wang, Xuan},
  journal={IEEE Transactions on Quantum Engineering},
  year={2025},
  publisher={IEEE},
  doi={10.1109/tqe.2025.3577769},
}

@article{muller2025improved,
  title={Improved belief propagation is sufficient for real-time decoding of quantum memory},
  author={M{\"u}ller, Tristan and Alexander, Thomas and Beverland, Michael E and B{\"u}hler, Markus and Johnson, Blake R and Maurer, Thilo and Vandeth, Drew},
  journal={arXiv preprint arXiv:2506.01779},
  year={2025}
}

@inproceedings{yao2024belief,
  title={Belief propagation decoding of quantum {LDPC} codes with guided decimation},
  author={Yao, Hanwen and Laban, Waleed Abu and H{\"a}ger, Christian and i Amat, Alexandre Graell and Pfister, Henry D},
  booktitle={Proceedings of the IEEE International Symposium on Information Theory (ISIT '24)},
  pages={2478--2483},
  year={2024},
  organization={IEEE},
  doi={10.1109/isit57864.2024.10619083},
}

@article{maurer2025real,
  title={Real-time decoding of the gross code memory with FPGAs},
  author={Maurer, Thilo and B{\"u}hler, Markus and Kr{\"o}ner, Michael and Haverkamp, Frank and M{\"u}ller, Tristan and Vandeth, Drew and Johnson, Blake R},
  journal={arXiv preprint arXiv:2510.21600},
  year={2025}
}

@article{valls2021syndrome,
  title={Syndrome-based min-sum vs OSD-0 decoders: FPGA implementation and analysis for quantum LDPC codes},
  author={Valls, Javier and Garcia-Herrero, Francisco and Raveendran, Nithin and Vasi{\'c}, Bane},
  journal={IEEE Access},
  volume={9},
  pages={138734--138743},
  year={2021},
  publisher={IEEE},
  doi={10.1109/access.2021.3118544},
}

@article{bausch2024learning,
  title={Learning high-accuracy error decoding for quantum processors},
  author={Bausch, Johannes and Senior, Andrew W and Heras, Francisco JH and Edlich, Thomas and Davies, Alex and Newman, Michael and Jones, Cody and Satzinger, Kevin and Niu, Murphy Yuezhen and Blackwell, Sam and others},
  journal={Nature},
  volume={635},
  number={8040},
  pages={834--840},
  year={2024},
  publisher={Nature Publishing Group UK London},
  doi={10.1038/s41586-024-08148-8},
}

@article{panteleev2021degenerate,
  title={Degenerate quantum LDPC codes with good finite length performance},
  author={Panteleev, Pavel and Kalachev, Gleb},
  journal={Quantum},
  volume={5},
  pages={585},
  year={2021},
  publisher={Verein zur F{\"o}rderung des Open Access Publizierens in den Quantenwissenschaften},
  doi={10.22331/q-2021-11-22-585},
}

@article{kuo2022exploiting,
  title={Exploiting degeneracy in belief propagation decoding of quantum codes},
  author={Kuo, Kao-Yueh and Lai, Ching-Yi},
  journal={npj Quantum Information},
  volume={8},
  number={1},
  pages={111},
  year={2022},
  publisher={Nature Publishing Group UK London},
  doi={10.1038/s41534-022-00623-2},
}

@article{hu2025efficient,
  title={Efficient and universal neural-network decoder for stabilizer-based quantum error correction},
  author={Hu, Gengyuan and Ouyang, Wanli and Lu, Chao-Yang and Lin, Chen and Zhong, Han-Sen},
  journal={arXiv preprint arXiv:2502.19971},
  year={2025}
}

@article{itogawa2025efficient,
  title={Efficient Magic State Distillation by Zero-Level Distillation},
  author={Itogawa, Tomohiro and Takada, Yugo and Hirano, Yutaka and Fujii, Keisuke},
  journal={PRX Quantum},
  volume={6},
  number={2},
  pages={020356},
  year={2025},
  publisher={APS},
  doi={10.1103/thxx-njr6},
}

@article{gidney2024magic,
  title={Magic state cultivation: growing T states as cheap as CNOT gates},
  author={Gidney, Craig and Shutty, Noah and Jones, Cody},
  journal={arXiv preprint arXiv:2409.17595},
  year={2024}
}

@article{vaknin2025magic,
  title={Magic State Cultivation on the Surface Code},
  author={Vaknin, Yotam and Jacoby, Shoham and Grimsmo, Arne and Retzker, Alex},
  journal={arXiv preprint arXiv:2502.01743},
  year={2025}
}

@article{hirano2025efficient,
  title={Efficient magic state cultivation with lattice surgery},
  author={Hirano, Yutaka and Toshio, Riki and Itogawa, Tomohiro and Fujii, Keisuke},
  journal={arXiv preprint arXiv:2510.24615},
  year={2025}
}

@article{chamberland2020very,
  title={Very low overhead fault-tolerant magic state preparation using redundant ancilla encoding and flag qubits},
  author={Chamberland, Christopher and Noh, Kyungjoo},
  journal={npj Quantum Information},
  volume={6},
  number={1},
  pages={91},
  year={2020},
  publisher={Nature Publishing Group UK London},
  doi={10.1038/s41534-020-00319-5},
}

@article{sahay2025fold,
  title={Fold-Transversal Surface Code Cultivation},
  author={Sahay, Kaavya and Tsai, Pei-Kai and Chang, Kathleen and Su, Qile and Smith, Thomas B and Singh, Shraddha and Puri, Shruti},
  journal={PRX Quantum},
  year={2026},
  doi={10.1103/gpvl-lg4c},
}

@article{claes2025cultivating,
  title={Cultivating {T} states on the surface code with only two-qubit gates},
  author={Claes, Jahan},
  journal={arXiv preprint arXiv:2509.05232},
  year={2025},
}

@inproceedings{ding2018magic,
  title={Magic-state functional units: Mapping and scheduling multi-level distillation circuits for fault-tolerant quantum architectures},
  author={Ding, Yongshan and Holmes, Adam and Javadi-Abhari, Ali and Franklin, Diana and Martonosi, Margaret and Chong, Frederic},
  booktitle={Proceedings of the IEEE/ACM International Symposium on Microarchitecture (MICRO '18)},
  pages={828--840},
  year={2018},
  organization={IEEE},
  doi={10.1109/micro.2018.00072},
}

@article{holmes2019resource,
  title={Resource optimized quantum architectures for surface code implementations of magic-state distillation},
  author={Holmes, Adam and Ding, Yongshan and Javadi-Abhari, Ali and Franklin, Diana and Martonosi, Margaret and Chong, Frederic T},
  journal={Microprocessors and Microsystems},
  volume={67},
  pages={56--70},
  year={2019},
  publisher={Elsevier},
  doi={10.1016/j.micpro.2019.02.007},
}

\end{document}